\pdfoutput=1

\documentclass[cernpreprint,texlive=2011,txfonts,UKenglish]{latex/atlasdoc}

\usepackage[subfigure=true,biblatex=false]{latex/atlaspackage}
\usepackage[BSM,journal,math,misc,other,xref,process=true]{latex/atlasphysics}
\usepackage{multirow}
\usepackage{rotating}
\usepackage{cite}

\usepackage{definitions}

\newcommand{\intlumi}            {\ensuremath{20.3 \pm 0.6}\invfb}

\graphicspath{{logos/}{figures/}}
\newcommand{\papertitle}{
Performance of \pileup mitigation techniques for jets in \pp collisions at 
\ensuremath{\sqrt{s}=8}~TeV using the ATLAS detector
}

\AtlasAbstract{%
The large rate of multiple simultaneous proton--proton interactions, or \pileup, generated by the Large Hadron Collider in \RunOne required the development of many new techniques to mitigate the adverse effects of these conditions. This paper describes the methods employed in the ATLAS experiment to correct for the impact of \pileup on jet energy and jet shapes, and for the presence of spurious additional jets, with a primary focus on the large 20.3\invfb data sample collected at a centre-of-mass energy of \sqseight. The energy correction techniques that incorporate sophisticated estimates of the average \pileup energy density and tracking information are presented. Jet-to-vertex association techniques are discussed and projections of performance for the future are considered. Lastly, the extension of these techniques to mitigate the effect of \pileup on jet shapes using subtraction and grooming procedures is presented.
}

\arXivId{1510.03823}

\PreprintIdNumber{CERN-PH-EP-2015-206}
\AtlasJournalRef{Eur. Phys. J. C (2016) 76:581}
\AtlasDOI{10.1140/epjc/s10052-016-4395-z}

\hypersetup{pdftitle={pileupjet-epjc},pdfauthor={The ATLAS Collaboration},pageanchor=false}
\AtlasTitle{\papertitle}
                 
\begin{document}



\maketitle

\clearpage

\tableofcontents


\section{Introduction}
\label{sec:introduction}

The success of the proton--proton (\pp) operation of the Large Hadron Collider (LHC) at \sqseight led to instantaneous luminosities of up to $7.7\times10^{33}$~\cmSqSec at the beginning of a fill. Consequently, multiple \pp interactions occur within each bunch crossing. Averaged over the full data sample, the mean number of such simultaneous interactions (\pileup) is approximately 21. These additional collisions are uncorrelated with the hard-scattering process that typically triggers the event and can be approximated as contributing a background of soft energy depositions that have particularly adverse and complex effects on jet reconstruction. Hadronic jets are observed as groups of topologically related energy deposits in the ATLAS calorimeters, and therefore \pileup affects  the measured jet energy and jet structure observables. \Pileup interactions can also directly generate additional jets. The production of such \textit{\pileup jets} can occur from additional $2\ra2$ interactions that are independent of the hard-scattering and from contributions due to soft energy deposits that would not otherwise exceed the threshold to be considered a jet. An understanding of all of these effects is therefore critical for precision measurements as well as searches for new physics.

\begin{figure}[!ht]
  \centering
  \includegraphics[width=0.6\textwidth]{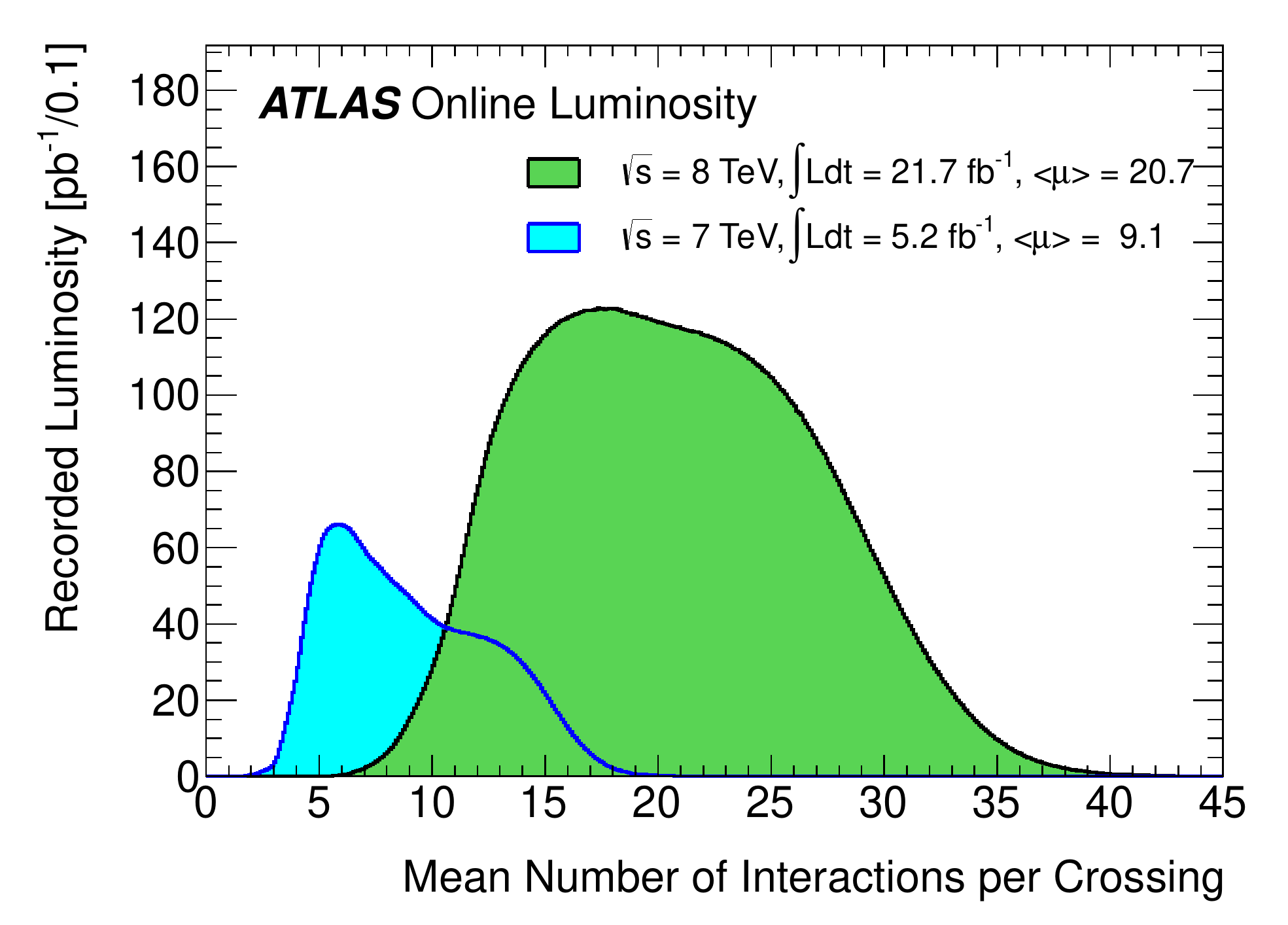}
  \caption{The luminosity-weighted distribution of the mean number of interactions per bunch crossing for the 2011 (\sqsseven) and 2012 (\sqseight) \pp data samples. \label{fig:intro:avgmu}}
\end{figure}

The expected amount of \pileup ($\mu$) in each bunch crossing is related to the instantaneous luminosity ($\lumi_0$) by the following relationship

\begin{equation}
  \mu = \frac{\lumi_0 \sigma_{\mathrm{inelastic}} }{n_{\rm c} \, f_\mathrm{rev}}
  \label{eq:lumi}
\end{equation}

\noindent where $n_{\rm c}$ is the number of colliding bunch pairs in the LHC, $f_\mathrm{rev} = 11.245$~kHz is the revolution frequency~\cite{Bruning:2004ej}, and $\sigma_{\mathrm{inelastic}}$ is the \pp inelastic cross-section. When the instantaneous luminosity is measured by integrating over many bunch crossings, \equref{lumi} yields the average number of interactions per crossing, or \avgmu.  The so-called \textit{in-time \pileup} due to additional \pp collisions within a single bunch crossing can also be accompanied by \textit{out-of-time \pileup} due to signals from collisions in other bunch crossings. This occurs when the detector and/or electronics integration time is significantly larger than the time between crossings, as is the case for the liquid-argon (LAr) calorimeters in the ATLAS detector. The measured detector response as a function of \avgmu in such cases is sensitive to the level of out-of-time \pileup. The distributions of \avgmu for both the \sqsseven and \sqseight runs (collectively referred to as \RunOne) are shown in \figref{intro:avgmu}. The spacing between successive proton bunches was 50~ns for the majority of data collected during \RunOne. This bunch spacing is decreased to 25~ns for LHC \RunTwo. Out-of-time \pileup contributions are likely to increase with this change. However, the LAr calorimeter readout electronics are also designed to provide an optimal detector response for a 25~ns bunch spacing scenario, and so the relative impact of the change to 25~ns may be mitigated, particularly in the case of the calorimeter response (see \secref{Detector}).
    
The different responses of the individual ATLAS subdetector systems to \pileup influence the methods used to mitigate its effects. The sensitivity of the calorimeter energy measurements to multiple bunch crossings, and the LAr EM calorimeter in particular, necessitates correction techniques that incorporate estimates of the impact of both in-time and out-of-time \pileup. These techniques use the average deposited energy density due to \pileup as well as track-based quantities from the inner tracking detector (ID) such as the number of reconstructed primary vertices (\Npv) in an event. Due to the fast response of the silicon tracking detectors, this quantity is not affected by out-of-time \pileup, to a very good approximation.

Resolving individual vertices using the ATLAS ID is a critical task in accurately determining the origin of charged-particle tracks that point to energy deposits in the calorimeter. By identifying tracks that originate in the hard-scatter primary vertex, jets that contain significant contamination from \pileup interactions can be rejected. These approaches provide tools for reducing or even obviating the effects of \pileup on the measurements from individual subdetector systems used in various stages of the jet reconstruction. The result is a robust, stable jet definition, even at very high luminosities.

The first part of this paper describes the implementation of methods to partially suppress the impact of signals from \pileup interactions on jet reconstruction and to directly estimate event-by-event \pileup activity and jet-by-jet \pileup sensitivity, originally proposed in Ref.~\cite{Cacciari:2007fd}.  These estimates allow for a sophisticated \pileup subtraction technique in which the four-momentum of the jet and the jet shape are corrected event-by-event for fluctuations due to \pileup, and whereby jet-by-jet variations in \pileup sensitivity are automatically accommodated.  The performance of these new \pileup correction methods is assessed and compared to previous \pileup corrections based on the number of reconstructed primary vertices and the instantaneous luminosity~\cite{jespaper2010,jespaper2011}. Since the \pileup subtraction is the first step of the jet energy scale (JES) correction in ATLAS, these techniques play a crucial role in establishing the overall systematic uncertainty of the jet energy scale. Nearly all ATLAS measurements and searches for physics beyond the Standard Model published since the end of the 2012 data-taking period utilise these methods, including the majority of the final \RunOne Higgs cross-section and coupling measurements~\cite{Aad:2014tca,Aad:2014eva,Aad:2014eha,Aad:2014xzb,ATLAS:2014aga}.

The second part of this paper describes the use of tracks to assign jets to the hard-scatter interaction. By matching tracks to jets, one obtains a measure of the fraction of the jet energy associated with a particular primary vertex. Several track-based methods allow the rejection of spurious calorimeter jets resulting from local fluctuations in \pileup activity, as well as real jets originating from single \pileup interactions, resulting in improved stability of the reconstructed jet multiplicity against \pileup. Track-based methods to reject \pileup jets are applied after the full chain of JES corrections, as \pileup jet tagging algorithms. 

The discussion of these approaches proceeds as follows. The ATLAS detector is described in \secref{Detector} and the data and Monte Carlo simulation samples are described in \secref{data-mc}. \Secref{topo} describes how the inputs to jet reconstruction are optimised to reduce the effects of \pileup on jet constituents.  Methods for subtracting \pileup from jets, primarily focusing on the impacts on calorimeter-based measurements of jet kinematics and jet shapes, are discussed in \secref{subtraction}. Approaches to suppressing the effects of \pileup using both the subtraction techniques and charged particle tracking information are then presented in \secref{suppression}. Lastly, techniques that aim to correct jets by actively removing specific energy deposits that are due to \pileup, are discussed in \secref{grooming}.

\section{The ATLAS detector}
\label{sec:Detector}
The ATLAS detector~\cite{detPaper,PerfWithData2010} provides nearly full solid angle coverage around the collision point with an inner tracking system covering the pseudorapidity range $|\eta|<2.5$,~\footnote{The ATLAS reference system is a Cartesian right-handed coordinate system, with the nominal collision point at the origin. The anticlockwise beam direction defines the positive $z$-axis, while the positive $x$-axis is defined as pointing from the collision point to the centre of the LHC ring and the positive $y$-axis points upwards. The azimuthal angle $\phi$ is measured around the beam axis, and the polar angle $\theta$ is measured with respect to the $z$-axis. Pseudorapidity is defined as $\eta = - \ln [\tan(\theta/2)]$, rapidity is defined as $y = 0.5\ \ln[(E + p_z)/(E - p_z)]$, where $E$ is the energy and $p_z$ is the $z$-component of the momentum, and transverse energy is defined as $\et = E \sin \theta$.}  electromagnetic and hadronic calorimeters covering $|\eta|<4.9$, and a muon spectrometer covering $|\eta|<2.7$.  

The ID comprises a silicon pixel tracker closest to the beamline, a microstrip silicon tracker, and a straw-tube transition radiation tracker at radii up to $108$ cm. These detectors are layered radially around each other in the central region. A thin superconducting solenoid surrounding the tracker provides an axial 2 T field enabling the measurement of charged-particle momenta. The overall ID acceptance spans the full azimuthal range in $\phi$ for particles originating near the nominal LHC interaction region~\cite{Aad:2010ac, ATLAS-CONF-2012-042, ATLAS-CONF-2014-047}. Due to the fast readout design of the silicon pixel and microstrip trackers, the track reconstruction is only affected by in-time \pileup. The efficiency to reconstruct charged hadrons ranges from 78\% at $\pttrk=500$~MeV to more than 85\% above $10 \GeV$, with a transverse impact parameter ($d_0$) resolution of 10~$\mu$m for high-momentum particles in the central region. For jets with \pt above approximately $500 \GeV$, the reconstruction efficiency for tracks in the core of the jet starts to degrade because these tracks share many clusters in the pixel tracker, creating ambiguities when matching the clusters with track candidates, and leading to lost tracks.


The high-granularity EM and hadronic calorimeters are composed of multiple subdetectors spanning $|\eta|\leq4.9$. The EM barrel calorimeter uses a LAr active medium and lead absorbers. In the region $|\eta| < 1.7$, the hadronic (Tile) calorimeter is constructed from steel absorber and scintillator tiles and is separated into barrel ($|\eta|<1.0$) and extended barrel ($0.8<|\eta|<1.7$) sections. The calorimeter end-cap ($1.375<|\eta|<3.2$) and forward ($3.1<|\eta|<4.9$) regions are instrumented with LAr calorimeters for EM and hadronic energy measurements. The response of the calorimeters to single charged hadrons -- defined as the energy ($E$) reconstructed for a given charged hadron momentum ($p$), or $E/p$ -- ranges from 20\% to 80\% in the range of charged hadron momentum between 1-30~GeV and is well described by Monte Carlo (MC) simulation~\cite{ATL-PHYS-PUB-2014-002}. In contrast to the pixel and microstrip tracking detectors, the LAr calorimeter readout is sensitive to signals from the preceding 12 bunch crossings during 50~ns bunch spacing operation~\cite{Buchanan:2008zzc,Abreu:2010zzc}. For the 25~ns bunch spacing scenario expected during \RunTwo of the LHC, this increases to 24 bunch crossings. The LAr calorimeter uses bipolar shaping with positive and negative output which ensures that the average signal induced by \pileup averages to zero in the nominal 25~ns bunch spacing operation. Consequently, although the LAr detector will be exposed to more out-of-time \pileup in \RunTwo, the signal shaping of the front-end electronics is optimised for this shorter spacing~\cite{ClelandStern1994,Buchanan:2008zzc}, and is expected to cope well with the change. The fast readout of the Tile calorimeter, however, makes it relatively insensitive to out-of-time \pileup~\cite{TileReadiness}. The LAr barrel has three EM layers longitudinal in shower depth (EM1, EM2, EM3), whereas the LAr end-cap has three EM layers (EMEC1, EMEC2, EMEC3) in the range $1.5<|\eta|<2.5$, two layers in the range $2.5<|\eta|<3.2$, and four hadronic layers (HEC1, HEC2, HEC3, HEC4). In addition, there is a pre-sampler layer in front of the LAr barrel and end-cap EM calorimeter (PS). The transverse segmentation of both the EM and hadronic LAr end-caps is reduced in the region between $2.5<|\eta|<3.2$ compared to the barrel layers. The forward LAr calorimeter has one EM layer (FCal1) and two hadronic layers (FCal2, FCal3) with transverse segmentation similar to the more forward HEC region. The Tile calorimeter has three layers longitudinal in shower depth (Tile1, Tile2, Tile3) as well as scintillators in the gap region spanning ($0.85<|\eta|<1.51$) between the barrel and extended barrel sections.

\section{Data and Monte Carlo samples}
\label{sec:data-mc}

This section provides a description of the data selection and definitions of objects used in the analysis (\secref{data_event_selection}) as well as of the simulated event samples to which the data are compared (\secref{MCsimulation}).

\subsection{Object definitions and event selection}\label{sec:data_event_selection}

The full 2012 \pp data-taking period at a centre-of-mass energy of \sqseight is used for these measurements presented here. Events are required to meet baseline quality criteria during stable LHC running periods. The ATLAS data quality (DQ) criteria reject data with significant contamination from detector noise or issues in the read-out~\cite{Aad:2014una} based upon individual assessments for each subdetector. These criteria are established separately for the barrel, end-cap and forward regions, and differ depending on the trigger conditions and reconstruction of each type of physics object (for example jets, electrons and muons). The resulting dataset corresponds to an integrated luminosity of \intlumi following the methodology described in Ref.~\cite{Aad:2013ucp}.

To reject non-collision backgrounds~\cite{Aad:2013zwa}, events are required to contain at least one primary vertex consistent with the LHC beam spot, reconstructed from at least two tracks each with $\pttrk>400$~MeV. The primary hard-scatter vertex is defined as the vertex with the highest $\sum(\pttrk)^2$. To reject rare events contaminated by spurious signals in the detector, all \antikt~\cite{Cacciari200657,Cacciari:2008gp} jets with radius parameter $R=0.4$ and $\ptjet>20 \GeV$ (see below) are required to satisfy the jet quality requirements that are discussed in detail in Ref.~\cite{Aad:2013zwa} (and therein referred to as the ``looser'' selection). These criteria are designed to reject non-collision backgrounds and significant transient noise in the calorimeters while maintaining an efficiency for good-quality events greater than 99.8\% with as high a rejection of contaminated events as possible. In particular, this selection is very efficient in rejecting events that contain fake jets due to calorimeter noise.

Hadronic jets are reconstructed from calibrated three-dimensional \topos~\cite{TopoClusters}. Clusters are constructed from calorimeter cells that are grouped together using a topological clustering algorithm. These objects provide a three-dimensional representation of energy depositions in the calorimeter and implement a nearest-neighbour noise suppression algorithm. The resulting \topos\ are classified as either electromagnetic or hadronic based on their shape, depth and energy density. Energy corrections are then applied to the clusters in order to calibrate them to the appropriate energy scale for their classification. These corrections are collectively referred to as \textit{local cluster weighting}, or LCW, and jets that are calibrated using this procedure are referred to as LCW jets~\cite{jespaper2011}.  

Jets can also be built from charged-particle tracks (track-jets) using the identical \antikt\ algorithm as for jets built from calorimeter clusters.  Tracks used to construct track-jets have to satisfy minimal quality criteria, and are required to be associated with the hard-scatter vertex.

The jets used for the analyses presented here are primarily found and reconstructed using the \antikt\ algorithm with radius parameters $R = 0.4, 0.6$ and $1.0$. In some cases, studies of \textit{groomed} jets are also performed, for which algorithms are used to selectively remove constituents from a jet. Groomed jets are often used in searches involving highly Lorentz-boosted massive objects such as $W/Z$ bosons~\cite{Aad:2014vka} or top quarks~\cite{Aad:2014xra}. Unless noted otherwise, the jet trimming algorithm~\cite{Krohn2010} is used for groomed jet studies in this paper. The procedure implements a \kt algorithm~\cite{Ellis1993, Catani1993} to create small \textit{subjets} with a radius $\drsub=0.3$. The ratio of the \pT\ of these subjets to that of the jet is used to remove constituents from the jet. Any subjets with $\pti/\ptjet < \fcut$ are removed, where \pti\ is the transverse momentum of the $i^{th}$ subjet, and $\fcut=0.05$ is determined to be an optimal setting for improving mass resolution, mitigating the effects of \pileup, and retaining substructure information~\cite{Aad:2013gja}. The remaining constituents form the trimmed jet.

The energy of the reconstructed jet may be further corrected using subtraction techniques and multiplicative jet energy scale correction factors that are derived from MC simulation and validated with the data~\cite{jespaper2010, jespaper2011}. As discussed extensively in \secref{subtraction}, subtraction procedures are critical to mitigating the jet energy scale dependence on \pileup. Specific jet energy scale correction factors are then applied after the subtraction is performed. The same corrections are applied to calorimeter jets in MC simulation and data to ensure consistency when direct comparisons are made between them. 

Comparisons are also made to jets built from particles in the MC generator's event record (``truth particles''). In such cases, the inputs to jet reconstruction are stable particles with a lifetime of at least 10~ps (excluding muons and neutrinos). Such jets are referred to as \textit{generator-level jets} or \textit{truth-particle jets} and are to be distinguished from \textit{parton-level jets}. Truth-particle jets represent the measurement for a hermetic detector with perfect resolution and scale, without \pileup, but including the underlying event. 


Trigger decisions in ATLAS are made in three stages: Level-1, Level-2, and the Event Filter. The Level-1 trigger is implemented in hardware and uses a subset of detector information to reduce the event rate to a design value of at most 75~kHz. This is followed by two software-based triggers, Level-2 and the Event Filter, which together reduce the event rate to a few hundred Hz. The measurements presented in this paper primarily use single-jet triggers. The rate of events in which the highest transverse momentum jet is less than about $400 \GeV$ is too high to record more than a small fraction of them. The triggers for such events are therefore pre-scaled to reduce the rates to an acceptable level in an unbiased manner. Where necessary, analyses compensate for the pre-scales by using weighted events based upon the pre-scale setting that was active at the time of the collision. 

\subsection{Monte Carlo simulation}
\label{sec:MCsimulation}

Two primary MC event generator programs are used for comparison to the data. \Pythia 8.160~\cite{pythia8} with the ATLAS A2 tunable parameter set (tune)~\cite{MC12AU2} and the CT10 NLO parton distribution function (PDF) set~\cite{Lai:2010vv} is used for the majority of comparisons. Comparisons are also made to the \Herwigpp 2.5.2~\cite{Herwigpp} program using the CTEQ6L1~\cite{PDF-CTEQ} PDF set along with the \textsc{UE7-2} tune~\cite{Gieseke:2012ft}, which is tuned to reproduce underlying-event data from the LHC experiments. MC events are passed through the full \geant~\cite{Geant4} detector simulation of ATLAS~\cite{simulation} after the simulation of the parton shower and hadronisation processes. Identical reconstruction and trigger, event, quality, jet, and track selection criteria are then applied to both the MC simulation and to the data.

In some cases, additional processes are used for comparison to data. The \Zboson boson samples used for the validation studies are produced with the \PowhegBox v1.0 generator~\cite{Alioli:2010xd,Frixione:2007vw,Nason:2004rx} and the \Sherpa 1.4.0~\cite{Gleisberg:2008ta} generator, both of which provide NLO matrix elements for inclusive \Zboson boson production. The CT10 NLO PDF set is also used in the matrix-element calculation for these samples. The modelling of the parton shower, multi-parton interactions and hadronisation for events generated using \PowhegBox is provided by \Pythia 8.163 with the AU2 underlying-event tune~\cite{MC12AU2} and the CT10 NLO PDF set. These MC samples are thus referred to as \PowPythia~8 samples. \Pythia is in turn interfaced with {\sc PHOTOS} \cite{Golonka:2005pn} for the modelling of QED final-state radiation.

\Pileup is simulated for all samples by overlaying additional soft \pp collisions which are also generated with \Pythia 8.160 using the ATLAS A2 tune and the MSTW2008LO PDF set~\cite{PDF-MSTW2008}. These additional events are overlaid onto the hard-scattering events according to the measured distribution of the average number \avgmu of \pp interactions per bunch crossing from the luminosity detectors in \ATLAS{}~\cite{Aad:2011dr,Aad:2013ucp} using the full 8 TeV data sample, as shown in \figref{intro:avgmu}. The proton bunches were organised in four trains of 36 bunches with a 50~ns spacing between the bunches. Therefore, the simulation also contains effects from out-of-time \pileup. The effect of this \pileup history for a given detector system is then determined by the size of the readout time window for the relevant electronics. As an example, for the central LAr calorimeter barrel region, which is sensitive to signals from the preceding 12 bunch crossings during 50 ns bunch spacing operation, the digitization window is set to 751 ns before and 101 ns after the simulated hard-scattering interaction.


\section{Topological clustering and cluster-level \pileup suppression}
\label{sec:topo}

The first step for \pileup mitigation in ATLAS is at the level of the constituents used to reconstruct jets. The topological clustering algorithm incorporates a built-in \pileup suppression scheme to limit the formation of clusters produced by \pileup depositions as well as to limit the growth of clusters around highly energetic cells from hard-scatter signals.  The key concept that allows this suppression is the treatment of \pileup as {\it noise}, and the use of cell energy thresholds based on their energy significance relative to the total noise.  

Topological clusters are built using a three-dimensional nearest-neighbour algorithm that clusters  calorimeter cells with energy significance $|E_{\rm cell}|/\sigma^{\rm noise}>4$ for the seed, iterates among all neighbouring cells with $|E_{\rm cell}|/\sigma^{\rm noise}>2$, and finally adds one additional layer of cells $|E_{\rm cell}|/\sigma^{\rm noise}>0$ when no further nearest-neighbours exceed the $2\sigma$ threshold at the boundary (not allowed to extend to next-to-nearest neighbours). The total cell noise, $\sigma^{\rm noise}$, is the sum in quadrature of the cell noise due to the readout electronics and the cell noise that is due to \pileup ($\sigma_{\rm \pileup}^{\rm noise}$).  The \pileup noise for a given cell is evaluated from Monte Carlo simulation and is defined to be the RMS of the energy distribution resulting from \pileup particles for a given number of \pp collisions per bunch crossing (determined by \avgmu) and a given bunch spacing $\Delta t$. It is technically possible to adjust the \pileup noise for specific data-taking periods depending on \avgmu, but it was kept fixed for the entire Run 1 $8~\TeV$ dataset.

By adjusting the \pileup noise value, topological clustering partially suppresses the formation of clusters created by \pileup fluctuations, and reduces the number of cells included in jets.  Raising the \pileup noise value effectively increases the threshold for cluster formation and growth, significantly reducing the effects of \pileup on the input signals to jet reconstruction.  

\begin{figure}[h]
  \centering
    \subfigure[Electronic noise only ($\avgmu=0$)]{
      \includegraphics[width=0.45\textwidth]{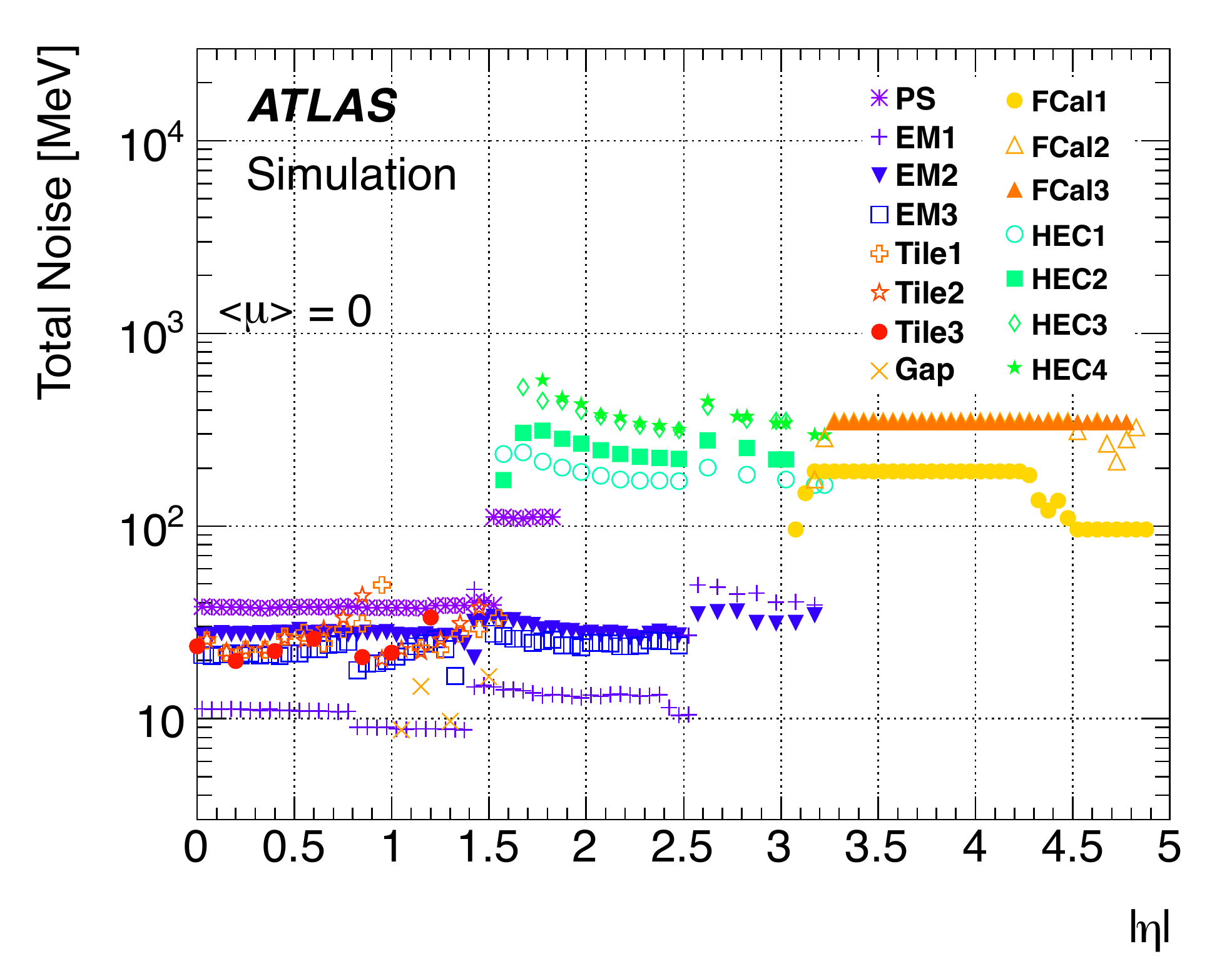}
      \label{fig:noise:elec}
    }
    \subfigure[Electronic + \pileup noise ($\avgmu=30$)]{
      \includegraphics[width=0.45\textwidth]{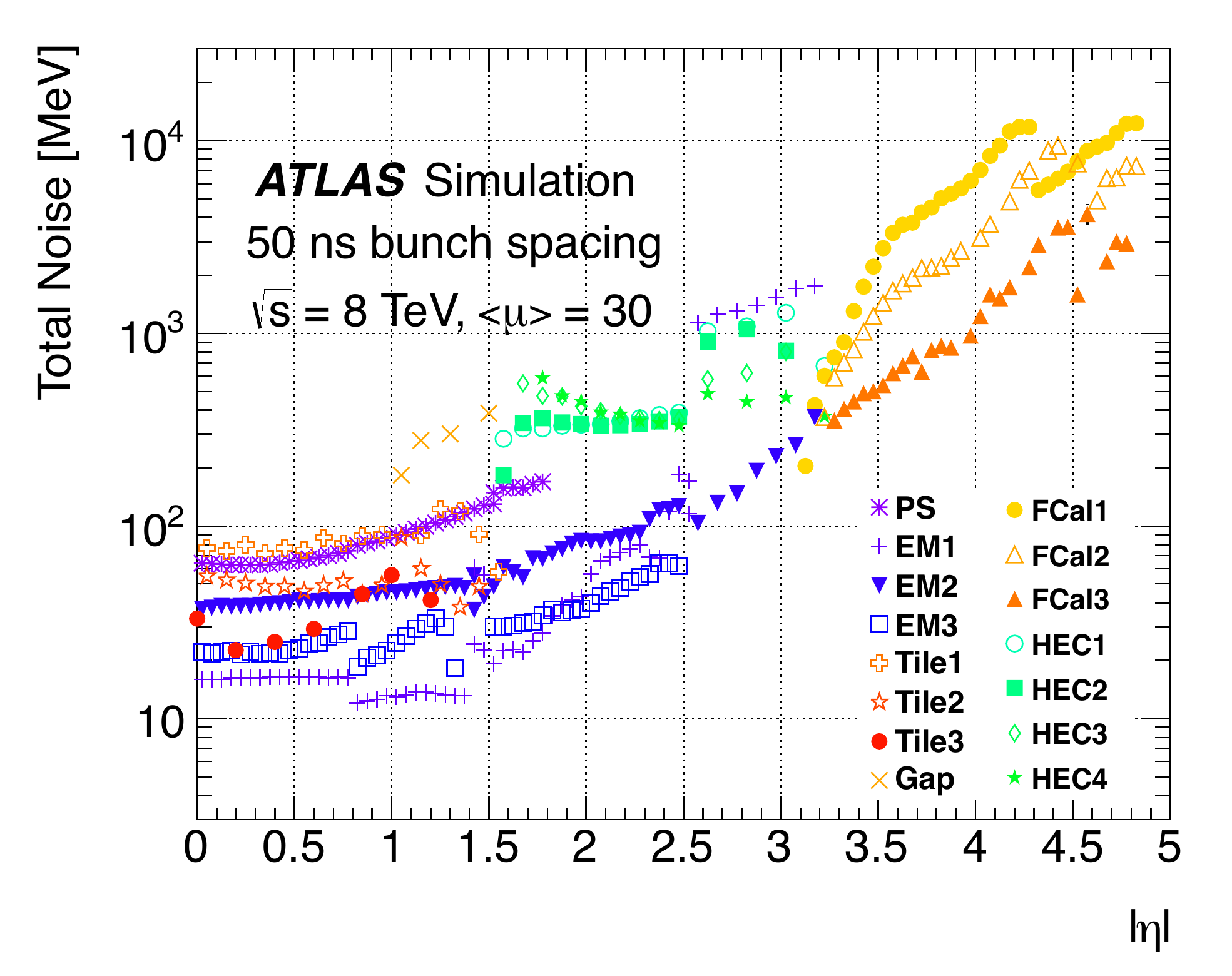}
      \label{fig:noise:pile}
    }
    \caption{
      \subref{fig:noise:elec} Per-cell electronic noise ($\avgmu=0$) and \subref{fig:noise:pile} total noise per cell at high luminosity corresponding to $\avgmu=30$ interactions per bunch crossing with a bunch spacing of $\Delta t = 50$~ns, in MeV, for each calorimeter layer. The different colours indicate the noise in the pre-sampler (PS), the up to three layers of the LAr calorimeter (EM), the up to three layers of the Tile calorimeter (Tile), the four layers of the hadronic end-cap calorimeter (HEC), and the three layers of the forward calorimeter (FCal). The total noise, $\sigma^{\rm noise}$, is the sum in quadrature of electronic noise and the expected RMS of the energy distribution corresponding to a single cell. 
      \label{fig:noise}
    }
\end{figure}

\Figref{noise} shows the electronic and \pileup noise contributions to cells that are used to define the thresholds for the topological clustering algorithm. In events with an average of 30 additional \pileup interactions ($\avgmu=30$), the noise from \pileup depositions is approximately a factor of two larger than the electronic noise for cells in the central electromagnetic calorimeter, and reaches $10\GeV$ in FCal1 and FCal2. This high threshold in the forward region translates into a reduced \topo\ occupancy due to the coarser segmentation of the forward calorimeter, and thus a smaller probability that a given event has a fluctuation beyond $4\sigma$. The implications of this behaviour for the \pileup \pt density estimation are discussed in \secref{rho}.

The value of \avgmu at which $\sigma_{\rm \pileup}^{\rm noise}$ is evaluated for a given data-taking period is chosen to be high enough that the number of clusters does not grow too large due to \pileup and at the same time low enough to retain as much signal as possible.  For a Gaussian noise distribution the actual $4\sigma$ seed threshold leads to an increase in the number of clusters by a factor of five if the noise is underestimated by 10\%.  Therefore $\sigma_{\rm \pileup}^{\rm noise}$ was set to the \pileup noise corresponding to the largest expected \avgmu rather than the average or the lowest expected value. For 2012 (2011) \pileup conditions, $\sigma_{\rm \pileup}^{\rm noise}$ was set to the value of  $\sigma_{\rm \pileup}^{\rm noise}$ corresponding to $\avgmu=30$ ($\avgmu=8$).

The local hadron calibration procedure for clusters depends on the value of $\sigma^{\rm noise}$  since this choice influences the cluster size and thus the shape variables used in the calibration.  Therefore, the calibration constants are re-computed for each $\sigma^{\rm noise}$ configuration. For this reason, a single, fixed value of $\sigma^{\rm noise}$ is used for entire data set periods in order to maintain consistent conditions.

\section{\Pileup subtraction techniques and results}
\label{sec:subtraction}

The independence of the hard-scattering process from additional \pileup interactions in a given event results in positive or negative shifts to the reconstructed jet kinematics and to the jet shape. This motivates the use of subtraction procedures to remove these contributions to the jet. Early subtraction methods~\cite{jespaper2010,jespaper2011} for mitigating the effects of \pileup on the jet transverse momentum in ATLAS relied on an average offset correction ($\langle\ptoffjet\rangle$), 

\begin{equation}
  \ptcorr = \ptjet - \langle\ptoffjet(\avgmu,\Npv,\eta)\rangle.
  \label{eq:offset}
\end{equation}

\noindent In these early approaches, $\langle\ptoffjet\rangle$ is determined from \insitu studies or MC simulation and represents an average offset applied to the jet \pt. This offset is parametrised as a function of \eta, \Npv and \avgmu. Such methods do not fully capture the  fluctuations of the \pileup energy added to the calorimeter on an event-by-event basis; that component is only indirectly estimated from its implicit dependence on \Npv. Moreover, no individual jet's information enters into this correction and thus jet-by-jet fluctuations in the actual offset of that particular jet \pt, \ptoffjet, or the jet shape, cannot be taken into account. Similar methods have also been pursued by the CMS collaboration~\cite{Chatrchyan:2011ds}, as well a much more complex approaches that attempt to mitigate the effects of \pileup prior to jet reconstruction~\cite{Berta:2014eza,Bertolini:2014bba}.

The approach adopted for the final \RunOne ATLAS jet energy scale~\cite{jespaper2011} is to estimate \ptoffjet on an event-by-event basis. To accomplish this, a measure of the jet's susceptibility to soft energy depositions is needed in conjunction with a method to estimate the magnitude of the effect on a jet-by-jet and event-by-event basis. A natural approach is to define a jet \textit{area} (\jetarea)~\cite{Cacciari2008} in $\eta$--$\phi$ space along with a \pileup \pt density, $\rho$. The offset can then be determined dynamically for each jet~\cite{Cacciari:2007fd} using

\begin{equation}
  \ptoffjet = \rho \times \jetarea.
  \label{eq:areacorrection}
\end{equation}

\noindent Nearly all results published by ATLAS since 2012 have adopted this technique for correcting the jet kinematics for \pileup effects. The performance of this approach, as applied to both the jet kinematics and the jet shape, is discussed below.

\subsection{\Pileup event \pt density $\rho$}
\label{sec:rho}

One of the key parameters in the \pileup subtraction methods presented in this paper is the estimated \pileup \pt density characterised by the observable $\rho$. The \pileup \pt density of an event can be estimated as the median of the distribution of the density of many \kt jets, constructed with no minimum \pt threshold~\cite{Ellis1993, Catani1993} in the event. Explicitly, this is defined as

\begin{equation}
  \rho = \mathrm{median}\left\{ \frac{\ptjeti}{\jetareai}\right \},
  \label{eq:rho}
\end{equation}

\noindent where each \kt\ jet $i$ has transverse momentum \ptjeti, and area \jetareai, and is defined with a nominal radius parameter $\rkt=0.4$.  The chosen radius parameter value is the result of a dedicated optimisation study, balancing two competing effects:  the sensitivity to biases from hard-jet contamination in the $\rho$ calculation when $\rkt$ is large, and statistical fluctuations when $\rkt$ is small. The sensitivity to the chosen radius value is not large, but measurably worse performance was observed for radius parameters larger than 0.5 and smaller than 0.3. 

\begin{figure}[!ht]
  \centering
  \includegraphics[width=0.60\textwidth]{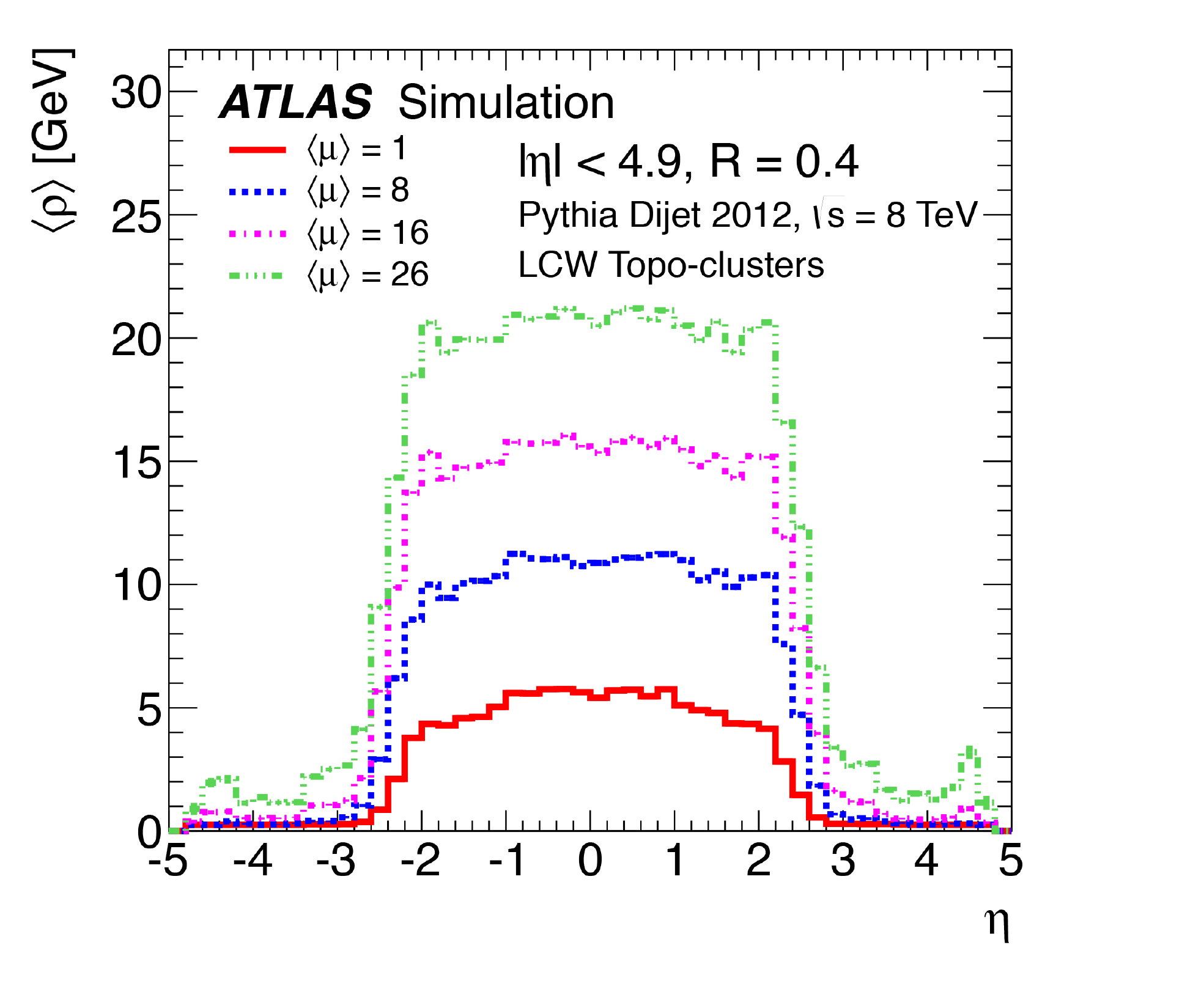}
  \caption{The mean estimated \pileup \pt density, $\rho$ as a function of $\eta$, in simulated \Pythia8 dijet events\label{fig:rho:distribution:eta}}
\end{figure}

The use of the \kt algorithm in \equref{rho} is motivated by its sensitivity to soft radiation and thus no minimum $\pt$ selection is applied to the \kt jets that are used. In ATLAS, the inputs to the \kt jets used in the $\rho$ calculation are positive-energy calorimeter \topos\ within $|\eta| \leq 2.0$. The $\eta$ range chosen for calculating $\rho$ is motivated by the calorimeter occupancy, which is low in the forward region relative to the central region.  The cause of the low occupancy in the forward region is complex and is intrinsically related to the calorimeter segmentation and response. The coarser calorimeter cell size at higher $|\eta|$~\cite{detPaper}, coupled with the noise suppression inherent in topological clustering, plays a large role. Since \topos\ are seeded according to significance relative to (electronic and \pileup) noise rather than an absolute threshold, having a larger number of cells (finer segmentation) increases the probability that the energy of one cell fluctuates up to a significant value due to (electronic or \pileup) noise. With the coarser segmentation in the end-cap and forward regions beginning near $|\eta|= 2.5$ (see \figref{noise}), this probability becomes smaller, and clusters are predominantly seeded only by the hard-scatter signal. In addition, the likelihood that hadronic showers overlap in a single cell increases along with the probability that fluctuations in the calorimeter response cancel, which affects the energy deposited in the cell. The mean $\rho$ measured as a function of $\eta$ is shown in \figref{rho:distribution:eta}. The measurements are made in narrow strips in $\eta$ which are $\Delta\eta = 0.7$ wide and shifted in steps of $\delta\eta = 0.1$ from $\eta = - 4.9$ to 4.9. The $\eta$ reported in \figref{rho:distribution:eta} is the central value of each strip. The measured $\rho$ in each strip quickly drops to nearly zero beyond $|\eta|\simeq2$. Due to this effectively stricter suppression in the forward region, a calculation of $\rho$ in the central region gives a more meaningful measure of the \pileup activity than the median over the entire $\eta$ range, or an $\eta$-dependent $\rho$ calculated in slices across the calorimeter. 

\begin{figure}[!ht]
  \centering
  \includegraphics[width=0.60\textwidth]{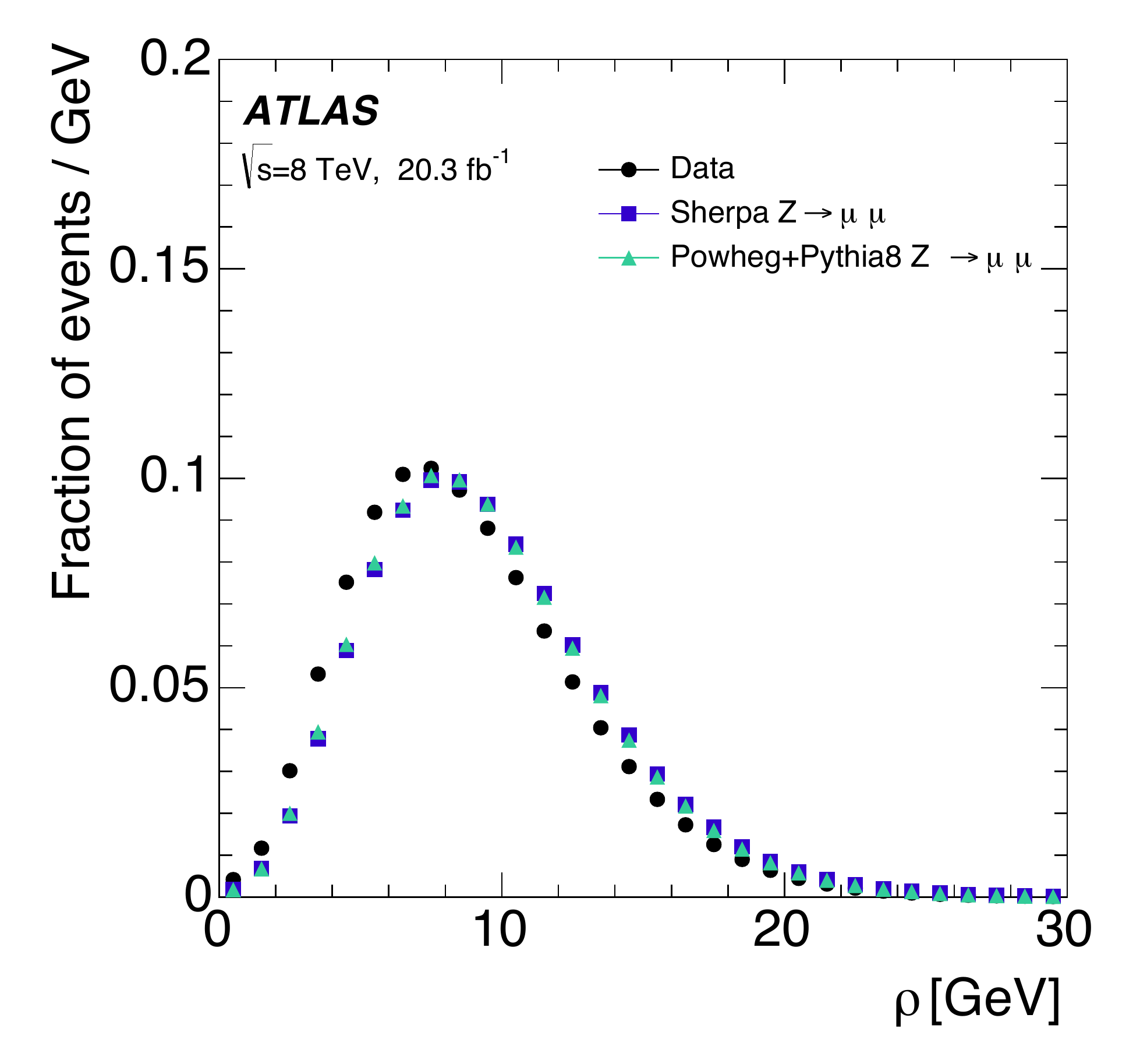}
  \caption{The distribution of estimated \pileup \pt density, $\rho$, in $\Zboson(\to\mu\mu)$+jets events using data and two independent MC simulation samples (\Sherpa and \PowPythia~8). Both MC generators use the same \pileup simulation model (\Pythia~8.160), and this model uses the \avgmu distribution for $8 \TeV$ data shown in \figref{intro:avgmu}. $\rho$ is calculated in the central region using \topos\ with positive energy within $|\eta| \leq 2.0$. \label{fig:rho:distribution:dataMC}}
\end{figure}

Distributions of $\rho$ in both data and MC simulation are presented in \figref{rho:distribution:dataMC} for \Sherpa and \PowPythia~8. Both MC generators use the same \pileup simulation model. The event selection used for these distributions corresponds to $\Zboson(\to\mu\mu)$+jets events where a \Zboson boson ($\Zpt>30$~\GeV) and a jet ($|\eta|<2.5$ and $\pt>20$~\GeV) are produced back-to-back ($\Delta\phi(Z,{\rm leading~jet})>2.9$). Both MC simulations slightly overestimate $\rho$, but agree well with each other. Small differences between the MC simulations can be caused by different modelling of the soft jet spectrum associated with the hard-scattering and the underlying event.

Since $\rho$ is computed event-by-event, separately for data and MC, a key advantage of the jet area subtraction is that it reduces the \pileup uncertainty from detector mismodelling effects.  This is because different values of $\rho$ are determined in data and simulation depending on the measured \pileup activity rather than using a predicted value for $\rho$ based on MC simulations.



\subsection{\Pileup energy subtraction}
\label{sec:subtraction:energy}

The median \pt density $\rho$ provides a direct estimate of the global \pileup activity in each event, whereas the jet area provides an estimate of an individual jet's susceptibility to \pileup.  \Equref{offset} can thus be expressed on a jet-by-jet basis using \equref{areacorrection} instead of requiring an average calculation of the offset, $\langle\ptoff\rangle$. This yields the following \pileup subtraction scheme:

\begin{equation}
  \ptcorr = \ptjet - \rho \times \jetarea.
  \label{eq:areasubtraction}
\end{equation}

There are two ways in which \pileup can contribute energy to an event: either by forming new clusters, or by overlapping with signals from the hard-scattering event. Because of the noise suppression inherent in topological clustering, only \pileup signals above a certain threshold can form separate clusters. Low-energy \pileup deposits can thus only contribute measurable energy to the event if they overlap with other deposits that survive noise suppression. The probability of overlap is dependent on the transverse size of EM and hadronic showers in the calorimeter, relative to the size of the calorimeter cells. Due to fine segmentation, \pileup mainly contributes extra clusters in the central regions of the calorimeter where $\rho$ is calculated ($|\eta|\lesssim 2$). 

As discussed in \secref{Detector}, the details of the readout electronics for the LAr calorimeter  can result in signals associated with out-of-time \pileup activity. If out-of-time signals from earlier bunch crossings are isolated from in-time signals, they may form negative energy clusters, which are excluded from jet reconstruction and the calculation of $\rho$. However, overlap between the positive jet signals and out-of-time activity results in both positive and negative modulation of the jet energy. Due to the long negative component of the LAr pulse shape, the probability is higher for an earlier bunch crossing to negatively contribute to signals from the triggered event than a later bunch crossing to contribute positively. This feature results in a negative dependence of the jet \pt on out-of-time \pileup.  Such overlap is more probable at higher $|\eta|$, due to coarser segmentation relative to the transverse shower size. In addition, the length of the bipolar pulse is shorter in the forward calorimeters, which results in larger fluctuations in the out-of-time energy contributions to jets in the triggered event since the area of the pulse shape must remain constant. As a result, forward jets have enhanced sensitivity to out-of-time \pileup due to the larger impact of fluctuations of \pileup energy depositions in immediately neighbouring bunch crossings.

Since the $\rho$ calculation is dominated by lower-occupancy regions in the calorimeter, the sensitivity of $\rho$ to \pileup does not fully describe the \pileup sensitivity of the high-occupancy region at the core of a high-\pt jet. The noise suppression provided by the topological clustering procedure has a smaller impact in the dense core of a jet where significant nearby energy deposition causes a larger number of small signals to be included in the final clusters than would otherwise be possible. Furthermore, the effects of \pileup in the forward region are not well described by the median \pt density as obtained from positive clusters in the central region. 
A residual correction is therefore necessary to obtain an average jet response that is insensitive to \pileup across the full \pt range.

\begin{figure*}
  \centering
    \subfigure[]{
    \includegraphics[width=0.48\textwidth]{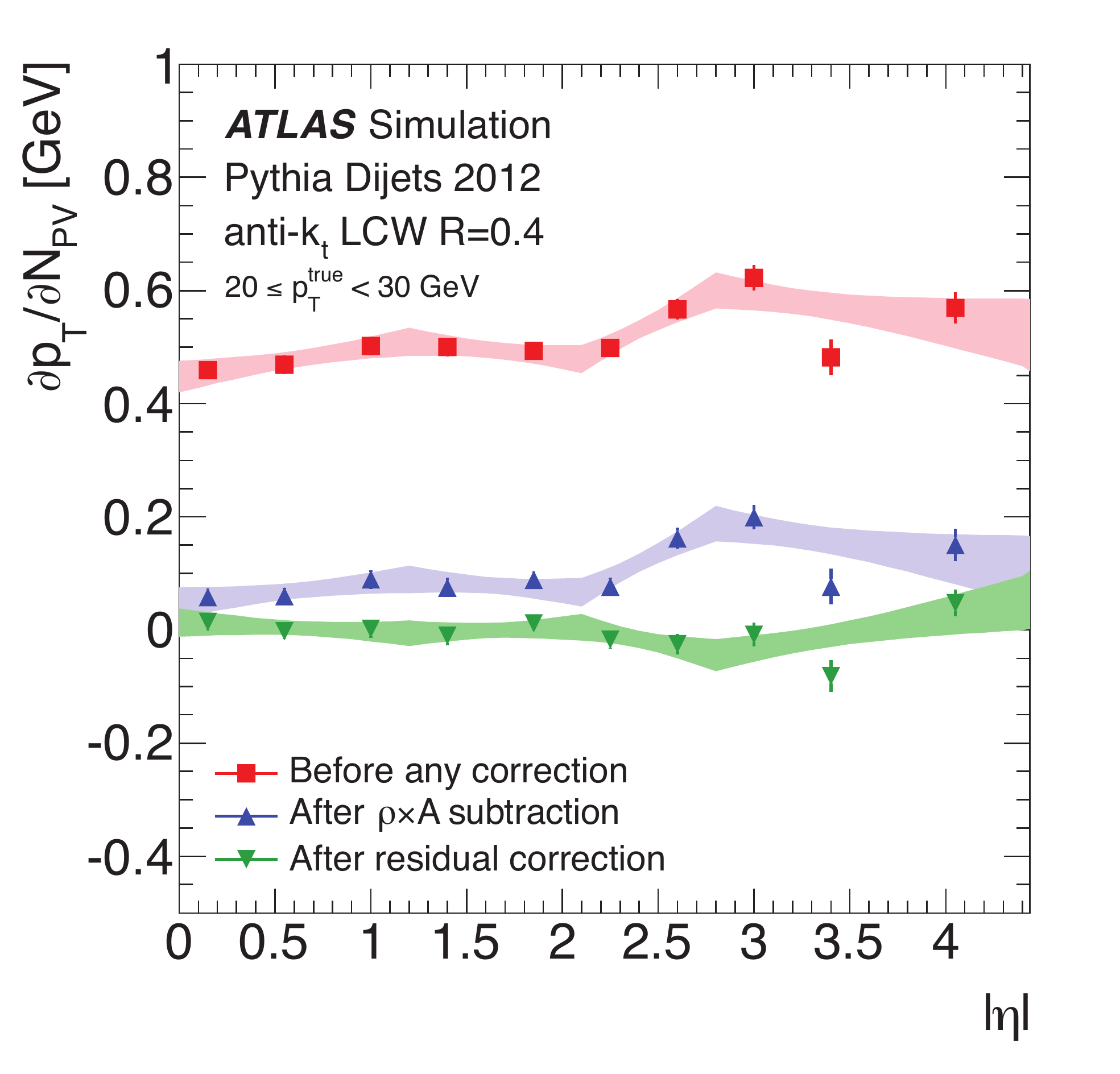}
    \label{fig:residCorr_npvTerm_akt4lc}}
  \subfigure[]{
    \includegraphics[width=0.48\textwidth]{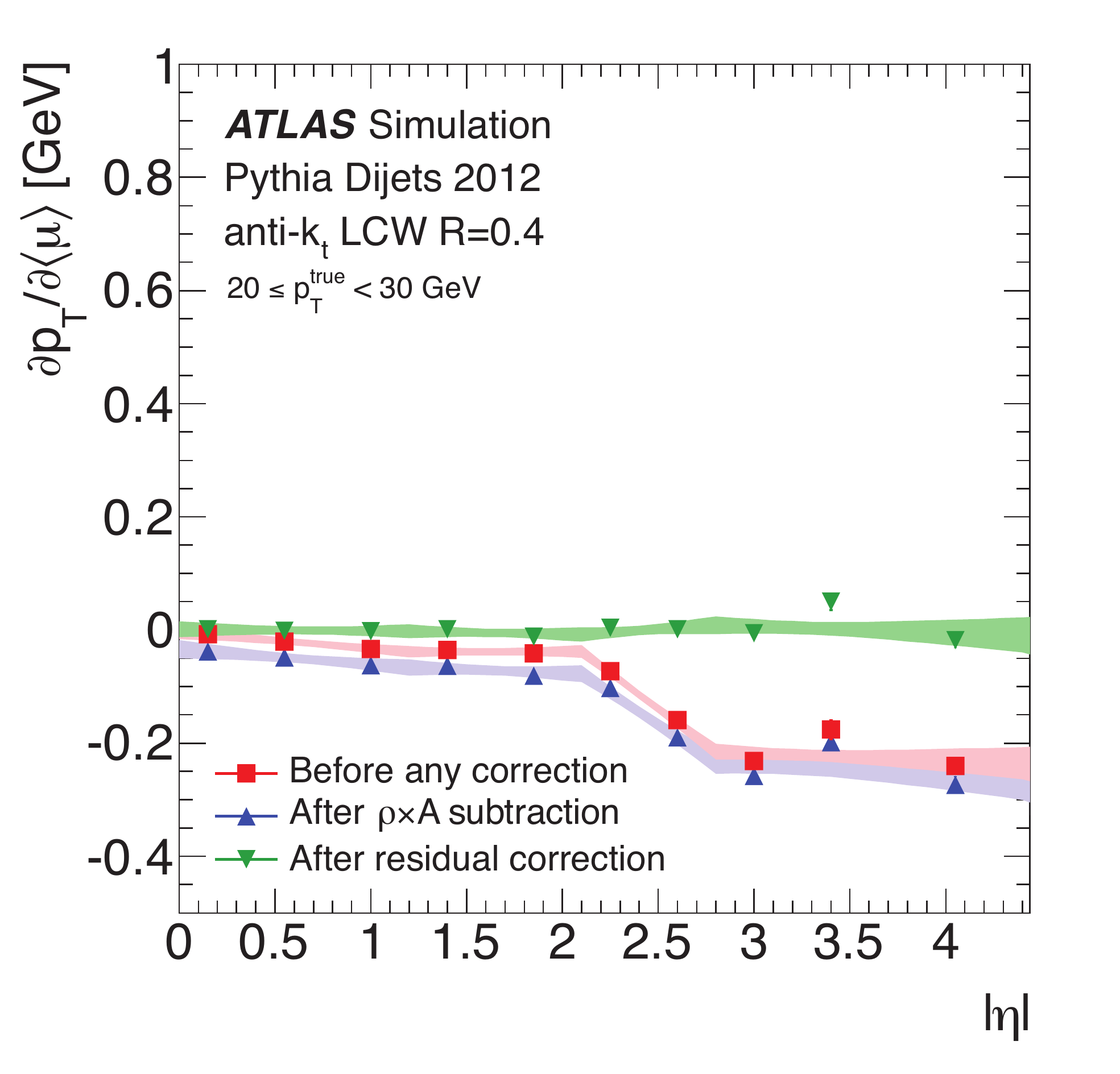}
    \label{fig:residCorr_muTerm_akt4lc}} 
\caption{Dependence of the reconstructed jet \pt (\antikt, $R = 0.4$, LCW scale) on \subref{fig:residCorr_npvTerm_akt4lc} in-time \pileup measured using \Npv and \subref{fig:residCorr_muTerm_akt4lc} out-of-time \pileup measured using \avgmu. In each case, the dependence of the jet \pT is shown for three correction stages: before any correction, after the $\rho\times\jetarea$ subtraction, and after the residual correction.  The error bands show the 68\% confidence intervals of the fits. The dependence was obtained by comparison with truth-particle jets in simulated dijet events, and corresponds to a truth-jet \pt range of 20--30\GeV.}
              
  \label{fig:residCorr_akt4lc}
  
\end{figure*}

\Figref{residCorr_akt4lc} shows the $\eta$ dependence of the transverse momentum of \akt $R = 0.4$ jets on \Npv (for fixed \avgmu) and on \avgmu (for fixed \Npv). Separating these dependencies probes the effects of in-time and out-of-time \pileup, respectively, as a function of $\eta$. These results were obtained from linear fits to the difference between the reconstructed and the true jet \pt (written as $\ptreco - \pttrue$) as a function of both \Npv and \avgmu. The subtraction of $\rho\times\jetarea$  removes a significant fraction of the sensitivity to in-time \pileup. In particular, the dependence decreases from nearly $0.5 \GeV$ per additional vertex to $\lesssim 0.2 \GeV$ per vertex, or a factor of 3--5 reduction in \pileup sensitivity. This reduction in the dependence of the $\pT$ on \pileup does not necessarily translate into a reduction of the \pileup dependence of other jet observables. Moreover, some residual dependence on \Npv remains. \Figref{residCorr_muTerm_akt4lc} shows that $\rho\times\jetarea$ subtraction has very little effect on the sensitivity to out-of-time \pileup, which is particularly significant in the forward region. The dependence on \Npv is evaluated in bins of \avgmu, and vice versa. Both dependencies are evaluated in bins of \pttrue and $\eta$ as well.  The slope of the linear fit as a function of \Npv does not depend significantly on \avgmu, or vice versa, within each $(\pttrue, \eta)$ bin.  In other words, there is no statistically significant evidence for non-linearity or cross-terms in the sensitivity of the jet \pt to in-time or out-of-time \pileup for the values of \avgmu seen in 2012 data. A measurable effect of such non-linearities may occur with the shorter bunch spacing operation, and thus increased out-of-time \pileup effects, expected during \RunTwo of the LHC. Measurements and validations of this sort are therefore important for establishing the sensitivity of this correction technique to such changes in the operational characteristics of the accelerator.

After subtracting $\rho\times\jetarea$ from the jet \pt, there is an additional subtraction of a residual term proportional to the number ($\Npv - 1$) of reconstructed \pileup vertices, as well as a residual term proportional to \avgmu (to account for out-of-time \pileup). This residual correction is derived by comparison to truth particle jets in simulated dijet events, and it is completely analogous to the average \pileup offset correction used previously in \ATLAS~\cite{jespaper2011}. Due to the preceding $\rho\times\jetarea$ subtraction, the residual correction is generally quite small for jets with $|\eta| < 2.1$.  In the forward region, the negative dependence of jets on out-of-time \pileup results in a significantly larger residual correction. The \avgmu-dependent term of the residual correction is approximately the same size as the corresponding term in the average offset correction of \equref{offset}, but the \Npv-dependent term is significantly smaller.  This is true even in the forward region, which shows that $\rho$ is a useful estimate of in-time \pileup activity even beyond the region in which it is calculated.

Several additional jet definitions are also studied, including larger nominal jet radii and alternative jet algorithms. Prior to the jet area subtraction, a larger sensitivity to in-time \pileup is observed for larger-area jets, as expected.  Following the subtraction procedure in \equref{areasubtraction} similar results are obtained even for larger-area jet definitions. These results demonstrate that $\rho\times\jetarea$ subtraction is able to effectively reduce the impact of in-time \pileup regardless of the jet definition, although a residual correction is required to completely remove the dependence on \Npv and \avgmu.

\begin{figure}
  \centering
  
  \subfigure[]{
    \label{fig:subtraction:resolution:pt}
    \includegraphics[width=0.48\textwidth]{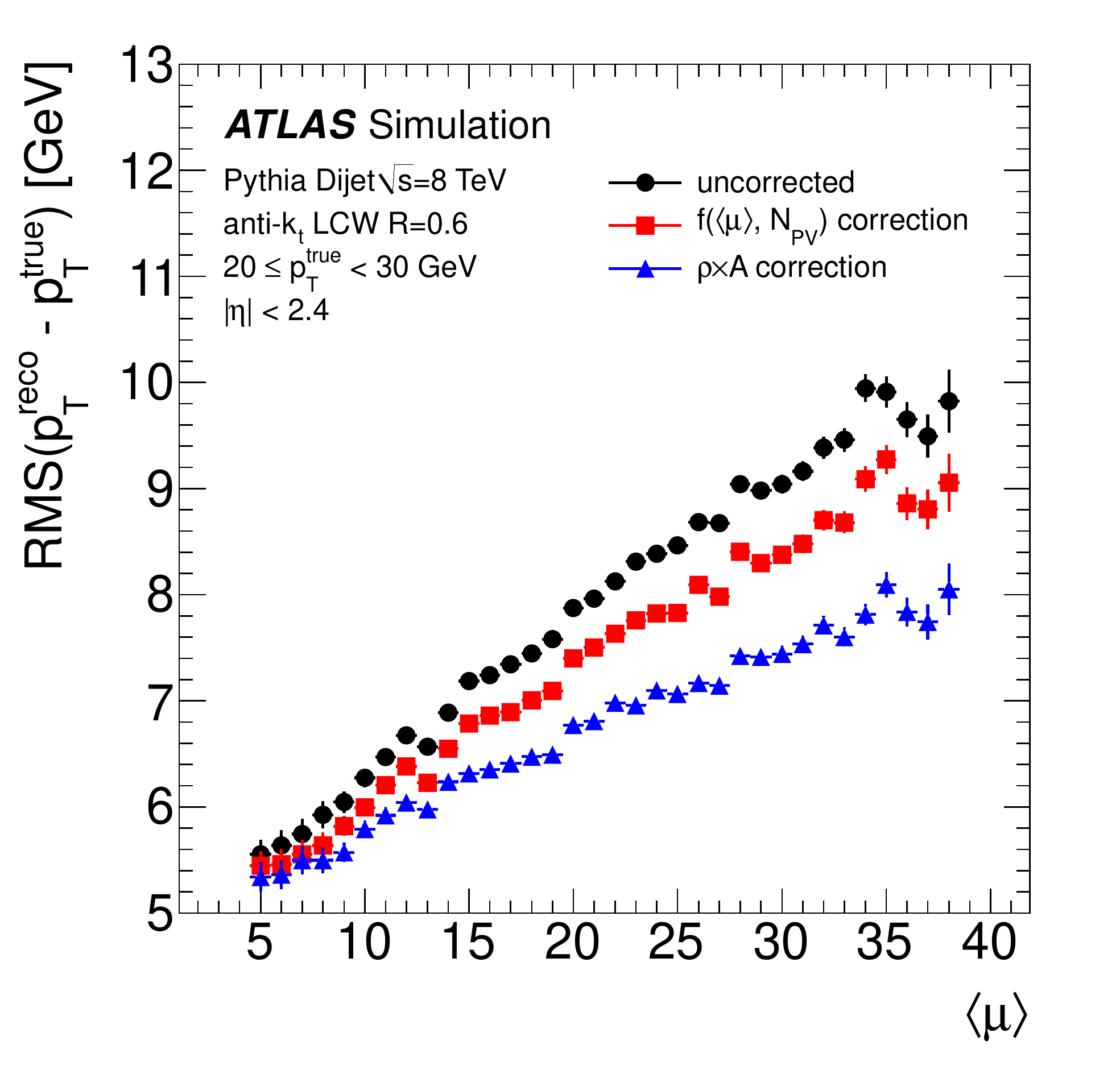}}
  \subfigure[]{
    \label{fig:subtraction:resolution:eta}
    \includegraphics[width=0.48\textwidth]{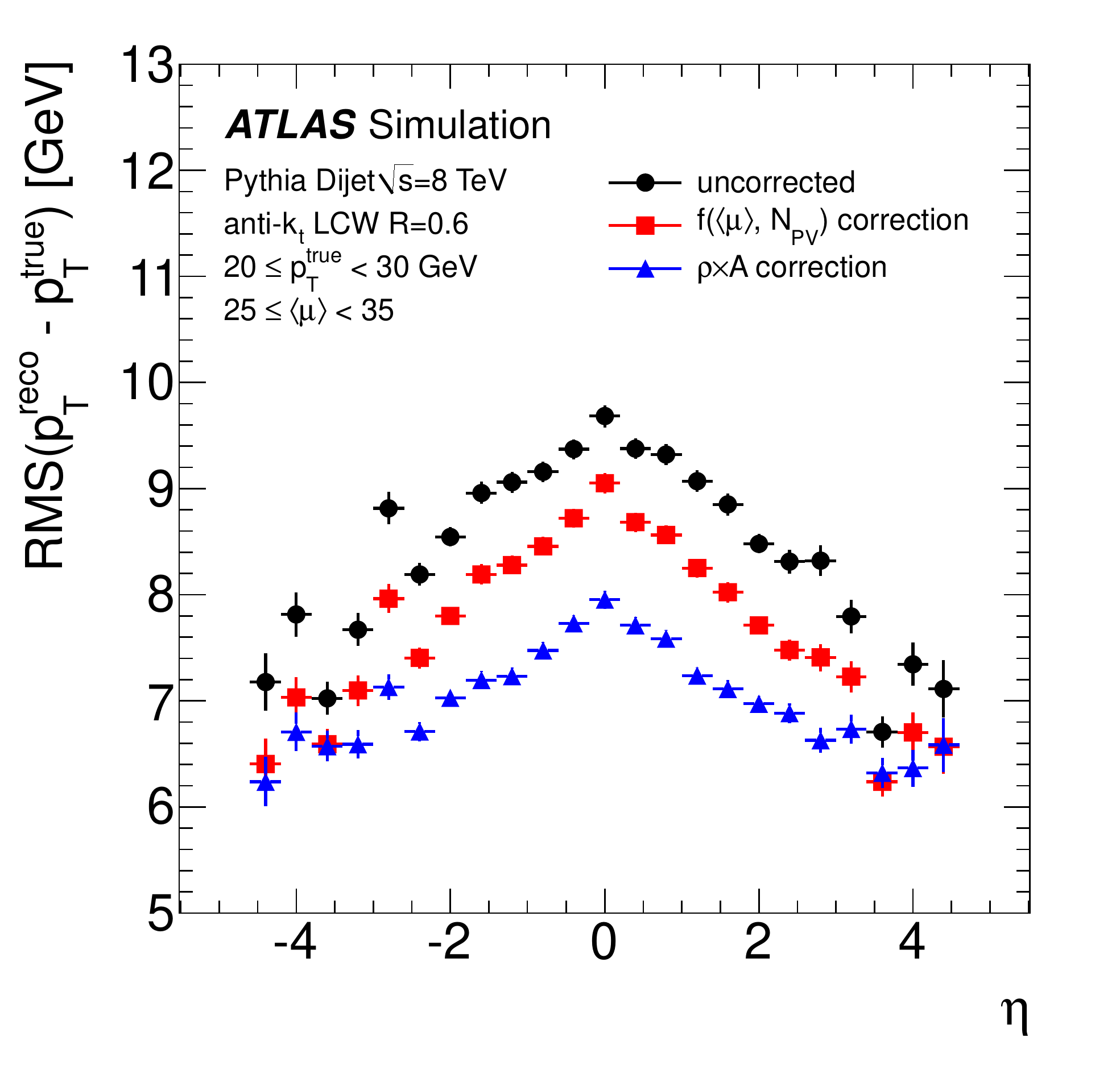}}

\caption{\subref{fig:subtraction:resolution:pt} RMS width of the $\ptreco - \pttrue$ distribution versus \avgmu and \subref{fig:subtraction:resolution:eta} versus pseudorapidity $\eta$, for \antikt $R=0.6$ jets at the LCW scale matched to truth-particle jets satisfying $20 < \pttrue < 30 \GeV$, in simulated dijet events.  A significant improvement is observed compared to the previous subtraction method (shown in red)~\cite{jespaper2011}. \label{fig:subtraction:resolution}}

\end{figure}

In addition to the slope of the \pt dependence on \Npv, the RMS of the $\ptreco - \pttrue$ distribution is studied as a function of \avgmu and $\eta$ in \figref{subtraction:resolution}. For this result, \antikt $R=0.6$ jets are chosen due to their greater susceptibility to \pileup and the greater challenge they therefore pose to \pileup correction algorithms. The RMS width of this distribution is an approximate measure of the jet \pt resolution for the narrow truth-particle jet \pt ranges used in \figref{subtraction:resolution}. These results show that the area subtraction procedure provides an approximate 20\% reduction in the magnitude of the jet-by-jet fluctuations introduced by \pileup relative to uncorrected jets and approximately a 10\% improvement over the simple offset correction. Smaller radius $R=0.4$ jets exhibit a similar relative improvement compared to the simple offset correction.  It should be noted that the \pileup activity in any given event may have significant local fluctuations similar in angular size to jets, and a global correction such as that provided by the area subtraction procedure defined in \equref{areacorrection} cannot account for them. Variables such as the \textit{jet vertex fraction} \JVF, \textit{corrected \JVF} or \cJVF, or the \textit{jet vertex tagger} \JVT may be used to reject jets that result from such fluctuations in \pileup \pt density, as described in \secref{suppression}.

Two methods of \insitu validation of the \pileup correction are employed to study the dependence of jet \pt on \Npv and \avgmu.  The first method uses track-jets to provide a  measure of the jet \pt that is \pileup independent. This requires the presence of track-jets and so can only be used in the most central region of the detector for $|\eta| < 2.1$. It is not statistically limited. The second method exploits the \pt balance between a reconstructed jet and a \Zboson boson, using the \Zpt as a measure of the jet \pt. This enables an analysis over the full ($|\eta|<4.9$) range of the detector, but the extra selections applied to the jet and \Zboson boson reduce its  statistical significance.  The \Npv dependence must therefore be evaluated inclusively in \avgmu and vice versa. This results in a degree of correlation between the measured \Npv and \avgmu dependence.

While the \pileup residual correction is derived from simulated dijet events, the \insitu validation is done entirely using \Zjets events. In the track-jet validation, although the kinematics of the \Zboson boson candidate are not used directly, the dilepton system is relied upon for triggering, thus avoiding any potential bias from jet triggers. 

\begin{figure}
  \centering
    \subfigure[]{
    \includegraphics[width=0.475\textwidth]{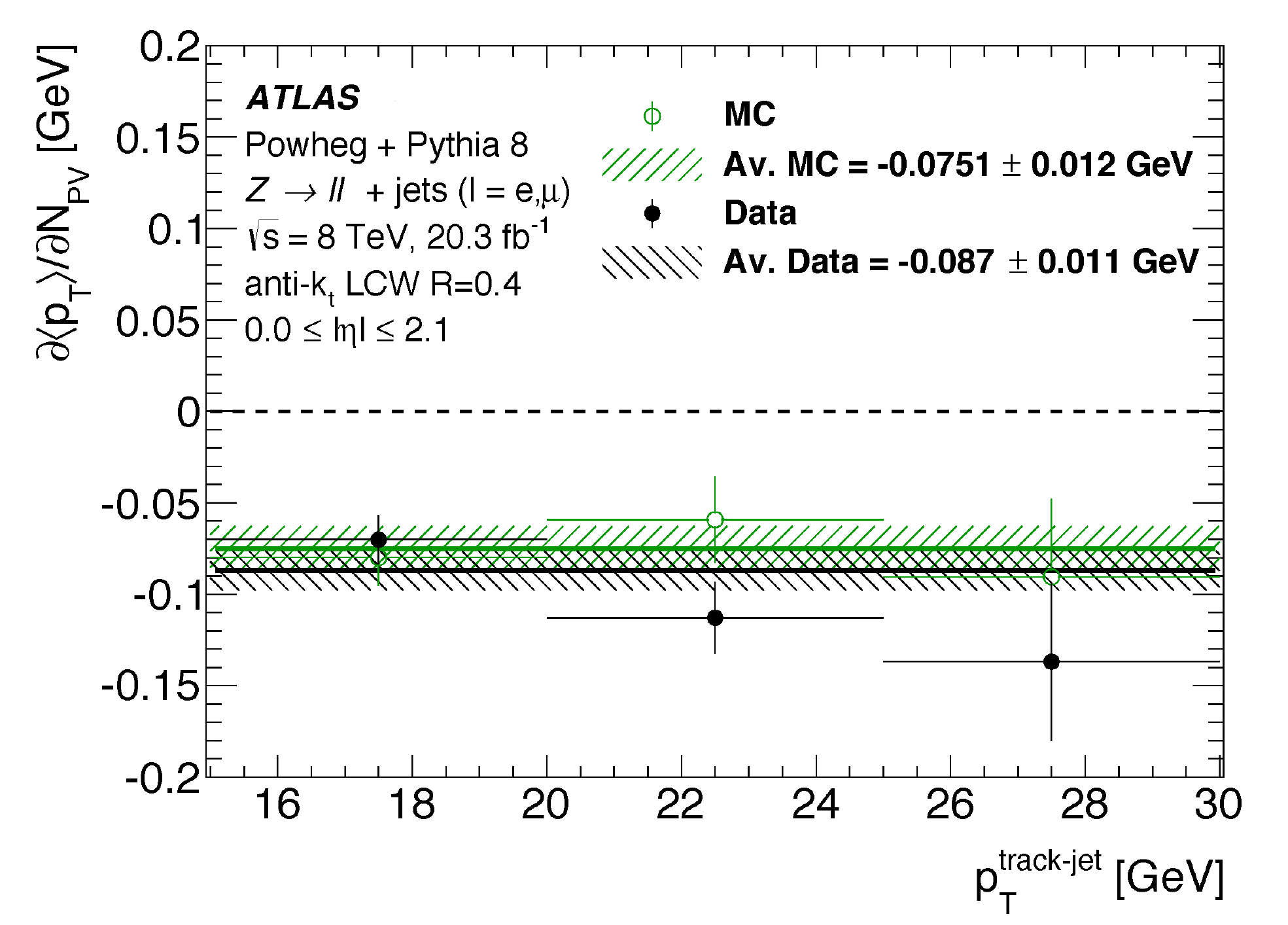}
    \label{fig:subtraction:energy:datamc:pttrk}} 
  \subfigure[]{
    \includegraphics[width=0.48\textwidth]{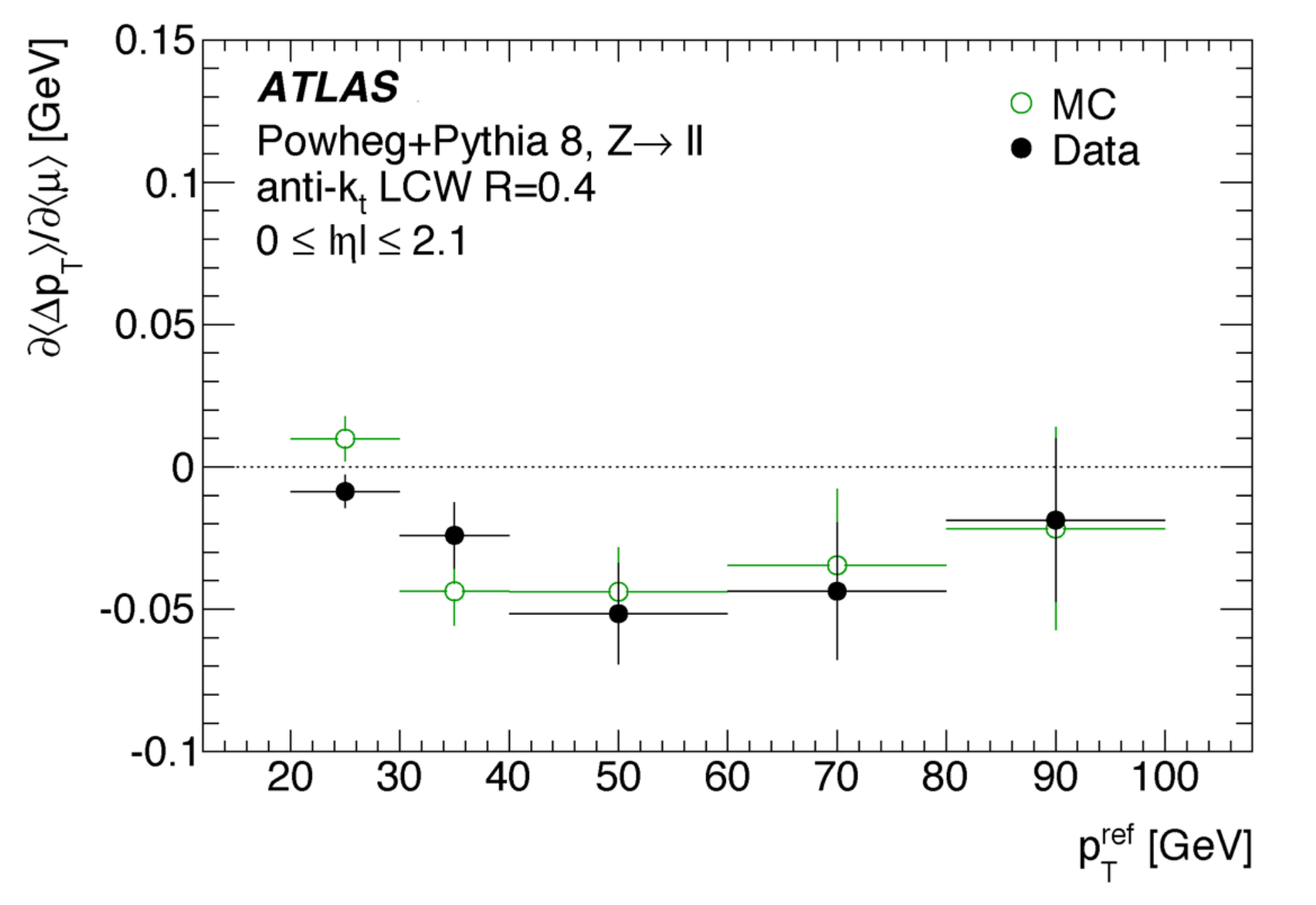}
    \label{fig:subtraction:energy:datamc:ptref}}
  \caption{
     \subref{fig:subtraction:energy:datamc:pttrk} \Npv dependence of the reconstructed \pt of \antikt $R=0.4$ LCW jets after the area subtraction as a function of track-jet \pT. 
     \subref{fig:subtraction:energy:datamc:ptref} Validation results from \Zjets events showing the \avgmu dependence as a function of the \Zboson boson \pt, denoted by \ptref, for \antikt $R=0.4$ LCW jets in the central region after the area subtraction. The points represent central values of the linear fit to \ptmuslope and the error bars correspond to the associated fitting error. \label{fig:subtraction:energy:datamc}}
  
\end{figure}

\Figref{subtraction:energy:datamc:pttrk} shows the results obtained when matching \antikt $R=0.4$, LCW reconstructed jets to \antikt $R=0.4$ track-jets. No selection is applied based on the calorimeter-based jet \pt. Good agreement is observed between data and MC simulation; however, a small overcorrection is observed in the \Npv dependence of each. For the final uncertainties on the method, this non-closure of the correction is taken as an uncertainty in the jet \pt dependence on \Npv. 

In events where a \Zboson boson is produced in association with one jet, momentum conservation ensures balance between the \Zboson boson and the jet in the transverse plane. In the direct \pt balance method, this principle is exploited by using \Zpt as a proxy for the true jet \pt. In the case of a perfect measurement of lepton energies and provided that all particles recoiling against the \Zboson boson are included in the jet cone, the jet is expected to balance the \Zboson boson. Therefore the estimated \Zboson boson \pt is used as the reference scale, denoted by \ptref.

Taking the mean, $\langle\Delta \pt\rangle$, of the ($\Delta \pt = \pt - \ptref$) distribution, the slope \ptmuslope is extracted and plotted as a function of \ptref, as shown in \figref{subtraction:energy:datamc:ptref}. A small residual slope is observed after the jet-area correction, which is well modelled by the MC simulation, as can be seen in \figref{subtraction:energy:datamc:ptref}. The mismodelling is quantified by the maximum differences between data and MC events for both \ptnpvslope and \ptmuslope. These differences (denoted by \ptnpvunc and \ptmuunc) are included in the total systematic uncertainty.

The systematic uncertainties are obtained by combining the measurements from \Zboson--jet balance and track-jet \insitu validation studies. In the central region ($|\eta| < 2.1$) only the track-jet measurements are used whereas \Zboson--jet balance is used for $2.1 < |\eta| < 4.5$. In the case of the \Zboson--jet balance in the forward region, the effects of in-time and out-of-time \pileup cannot be fully decoupled. Therefore, the \Npv uncertainty is assumed to be $\eta$-independent and is thus extrapolated from the central region. In the forward region, the uncertainty on the \avgmu dependence, \ptmuunc, is taken to be the maximum difference between \ptmuslope in the central region and \ptmuslope in the forward region. In this way, the forward region \ptmuunc uncertainty implicitly includes any $\eta$ dependence.

\subsection{\Pileup shape subtraction}
\label{sec:subtraction:shape}

The jet shape subtraction method~\cite{Soyez2012} determines the sensitivity of jet shape observables, such as the jet width or substructure shapes, to \pileup by evaluating the sensitivity of that shape to variations in infinitesimally soft energy depositions. This variation is evaluated numerically for each jet in each event and then extrapolated to zero to derive the correction. 

The procedure uses a uniform distribution of infinitesimally soft particles, or \textit{ghosts}, that are added to the event. These ghost particles are distributed with a number density \ghostdensity per unit in $y$--$\phi$ space, yielding an individual ghost area $\Aghost=1/\ghostdensity$. The four-momentum of ghost $i$ is defined as
%
\begin{equation}
  \ghostfmi = \ghostpt \cdot [\cos{\phi_i}, \sin{\phi_i}, \sinh{y_i}, \cosh{y_i}],
  \label{eq:ghost4mom}
\end{equation}

\noindent where \ghostpt is the ghost transverse momentum (initially set to $10^{-30} \GeV$), and the ghosts are defined to have zero mass. This creates a uniform ghost density given by $\ghostpt/\Aghost$ which is used as a proxy for the estimated \pileup contribution described by \equref{rho}. These ghosts are then incorporated into the jet finding and participate in the jet clustering. By varying the amount of ghost \pt density incorporated into the jet finding and determining the sensitivity of a given jet's shape to that variation, a numerical correction can be derived. A given jet shape variable $\shapeV$ is assumed to be a function of ghost \pt, $\shapeV(\ghostpt)$. The reconstructed (uncorrected) jet shape is then $\shapeV(\ghostpt=0)$. The corrected jet shape can be obtained by extrapolating to the value of \ghostpt which cancels the effect of the \pileup \pt density, namely $\ghostpt = -\rho \cdot \Aghost$. The corrected shape is then given by $\shapeV_{\mathrm{corr}} = \shapeV(\ghostpt = -\rho \cdot \Aghost)$. This solution can be achieved by using the Taylor expansion:
%
\begin{equation}
  \shapeV_{\mathrm{corr}}= \sum\limits_{k=0}^{\infty} \left(-\rho \cdot \Aghost\right)^{k} \frac{\partial^k \shapeV(\rho,\ghostpt)}{\partial \ghostpt^k}\bigg|_{\ghostpt=0}.
  \label{eq:corrShape_expansion}
\end{equation}

\noindent The derivatives are obtained numerically by evaluating several values of $\shapeV(\ghostpt)$ for $\ghostpt \geq 0$. Only the first three terms in \equref{corrShape_expansion} are used for the studies presented here.

One set of shape variables which has been shown to significantly benefit from the correction defined by the expansion in \equref{corrShape_expansion} is the set of $N$-subjettiness observables \tauN~\cite{Thaler:2010tr, Thaler:2011gf}. These observables measure the extent to which the constituents of a jet are clustered around a given number of axes denoted by $N$ (typically with $N=1,2,3$) and are related to the corresponding subjet multiplicity of a jet. The ratios $\tau_2/\tau_1$ (\tauTwoOne) and $\tau_3/\tau_2$ (\tauThrTwo) can be used to provide discrimination between Standard Model jet backgrounds and boosted $\W/\Z$ bosons~\cite{Thaler2008,Thaler:2010tr,Aad:2013gja}, top quarks~\cite{Thaler2008,Thaler:2010tr,BOOST2011,Aad:2013gja}, or even gluinos~\cite{RPVGluino7TeV}. For example, $\tauTwoOne \simeq 1$ corresponds to a jet that is very well described by a single subjet whereas a lower value implies a jet that is much better described by two subjets rather than one.

Two approaches are tested for correcting the $N$-subjettiness ratios \tauTwoOne and \tauThrTwo. The first approach is to use the individually corrected \tauN for the calculation of the numerators and denominators of the ratios. 
A second approach is also tested in which the full ratio is treated as a single observable
and corrected directly. The resulting agreement between data and MC simulation is very similar in the two cases. However, for very high \pt jets ($600\GeV\leq\ptjet<800 \GeV$) the first approach is preferable since it yields final ratios that are closer to the values obtained for truth-particle jets and a mean $\langle\tauThrTwo\rangle$ that is more stable against \avgmu. On the other hand, at lower jet \pt ($200\GeV\leq\ptjet<300 \GeV$), applying the jet shape subtraction to the ratio itself
performs better than the individual \tauN corrections according to the same figures of merit. Since substructure studies and the analysis of boosted hadronic objects typically focus on the high jet \pt regime, all results shown here use the individual corrections for \tauN in order to compute the corrected \tauTwoOne and \tauThrTwo.

\begin{figure}
  \centering
  
  \subfigure[]{
    \includegraphics[width=0.48\textwidth]{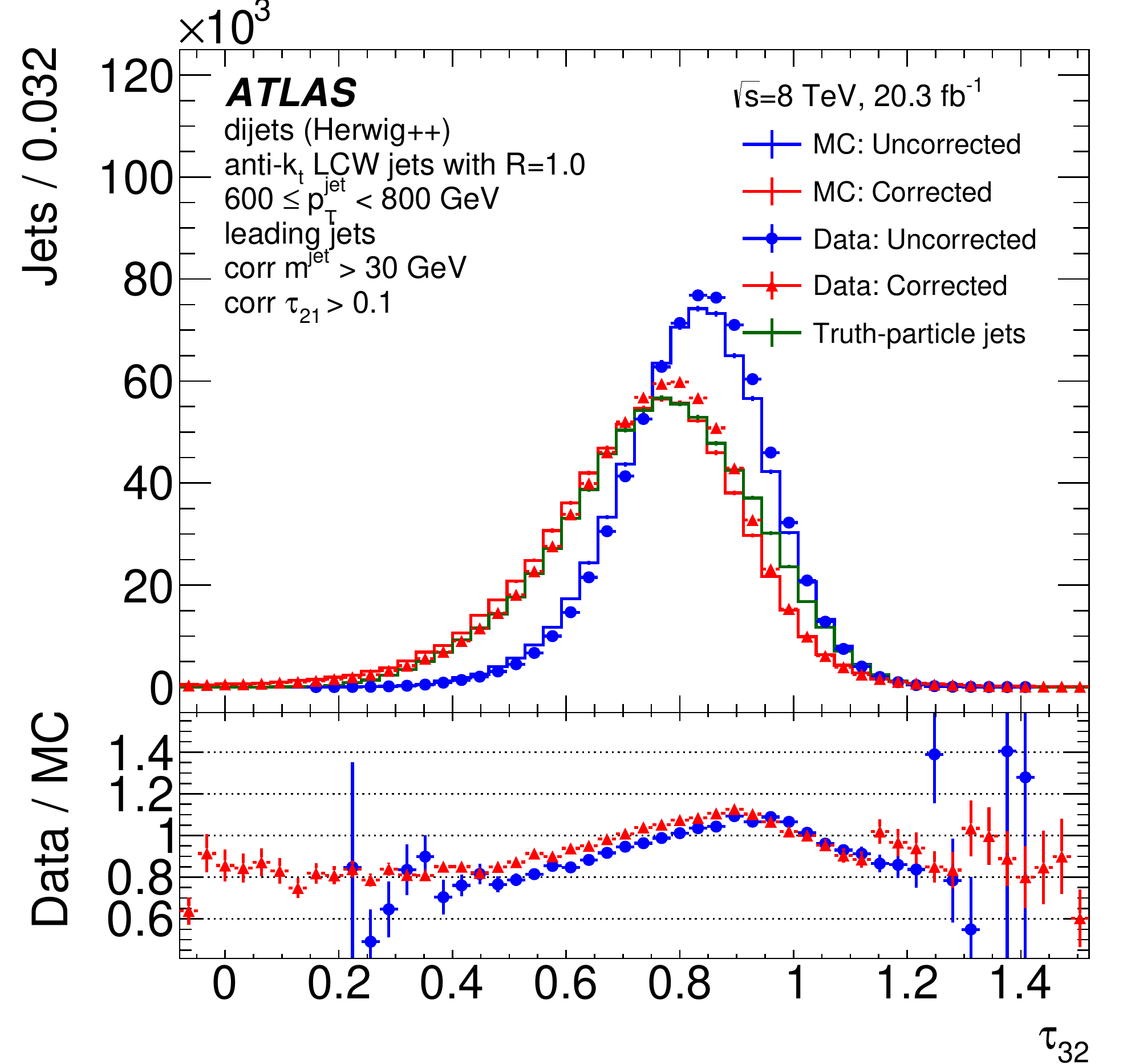}
    \label{fig:subtraction:shape:tau32}}
  \subfigure[]{
    \includegraphics[width=0.48\textwidth]{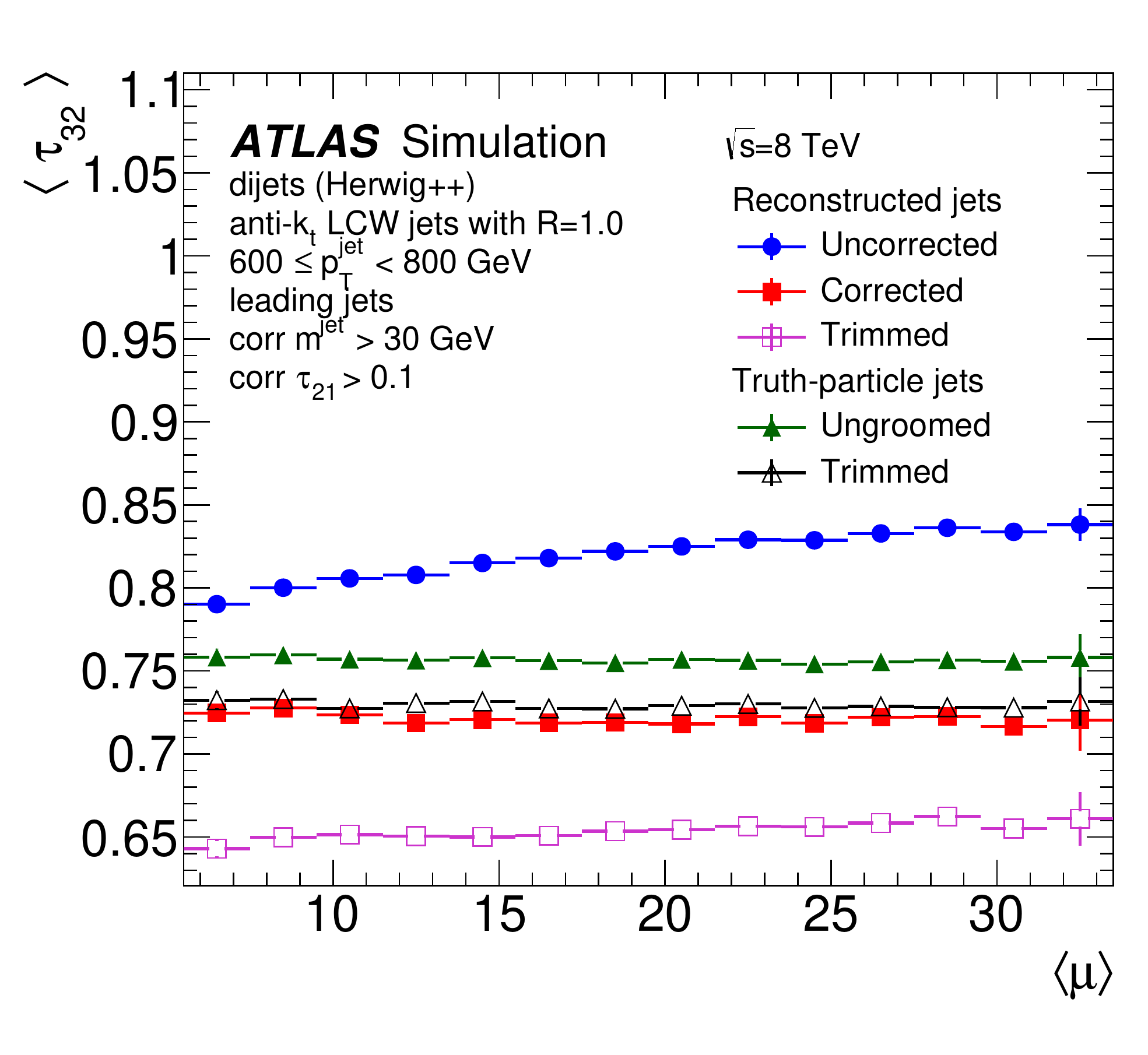}
    \label{fig:subtraction:shape:mc:avgtau32}} 
    
  \caption{\subref{fig:subtraction:shape:tau32} Comparisons of the uncorrected (filled blue circles), corrected (red) distributions of the ratio of $3$-subjettiness to $2$-subjettiness (\tauThrTwo) for data (points) and for MC simulation (solid histogram) for leading jets in the range $600 \leq \pT < 800 \GeV$. The distribution of \tauThrTwo computed using stable truth particles (filled green triangles) is also included. The lower panel displays the ratio of the data to the MC simulation.
  \subref{fig:subtraction:shape:mc:avgtau32} Dependence of \tauThrTwo on \avgmu for the uncorrected (filled blue circles), corrected (filled red squares), and trimmed (open purple squares) distributions for reconstructed jets in MC simulation for leading jets in the range $600 \leq \pT < 800 \GeV$. The mean value of \tauThrTwo computed using stable truth particles (green) is also included.
          }
           
  \label{fig:subtraction:shapes}
\end{figure}

\Figref{subtraction:shapes} presents the uncorrected and corrected distributions of \tauThrTwo, in both the observed data and MC simulation, as well as the truth-particle jet distributions. In the case of \figref{subtraction:shape:mc:avgtau32}, the mean value of \tauThrTwo is also presented for trimmed jets, using both the reconstructed and truth-particle jets. This comparison allows for a direct comparison of the shape subtraction method to trimming in terms of their relative effectiveness in reducing the \pileup dependence of the jet shape. Additional selections are applied to the jets used to study \tauThrTwo in this case: $\tauTwoOne>0.1$ (after correction) and jet mass $\massjet>30 \GeV$ (after correction). These selections provide protection against the case where $\tau_2$ becomes very small and small variations in $\tau_3$ can thus lead to large changes in the ratio. The requirement on \tauTwoOne rejects approximately 1\% of jets, whereas the mass requirement removes approximately 9\% of jets. As discussed above, the default procedure adopted here is to correct the ratio \tauTwoOne by correcting \tauOne and \tauTwo separately. In cases where both the corrected \tauOne and \tauTwo are negative, the sign of the corrected \tauTwoOne is set to negative. 

The corrected $N$-subjettiness ratio \tauThrTwo shows a significant reduction in \pileup dependence, as well as a much closer agreement with the distribution expected from truth-particle jets. \Figref{subtraction:shape:mc:avgtau32} provides comparisons between the shape subtraction procedure and jet trimming. Trimming is very effective in removing the \pileup dependence of jet substructure variables (see Ref.~\cite{Aad:2013gja} and \secref{grooming}). However, jet shape variables computed after jet trimming are considerably modified by the removal of soft subjets and must be directly compared to truth-level jet shape variables constructed with trimming at the truth level as well. Comparing the mean trimmed jet \tauThrTwo at truth level to the reconstructed quantity in \Figref{subtraction:shape:mc:avgtau32} (open black triangles and open purple square markers, respectively), and similarly for the shape correction method (filled green triangles and filled red square markers, respectively) it is clear that the shape expansion correction obtains a mean value closer to the truth.

\section{\Pileup jet suppression techniques and results}
\label{sec:suppression}

The suppression of \pileup jets is a crucial component of many physics analyses in ATLAS.  \Pileup jets arise from two sources: hard QCD jets originating from a \pileup vertex, and local fluctuations of \pileup activity. The \pileup QCD jets are genuine jets and must be tagged and rejected using the vertex-pointing information of charged-particle tracks (out-of-time QCD jets have very few or no associated tracks since the ID reconstructs tracks only from the in-time events). \Pileup jets originating from local fluctuations are a superposition of random combinations of particles from multiple \pileup vertices, and are generically referred to here as {\it stochastic} jets.  Stochastic jets are preferentially produced in regions of the calorimeter where the global $\rho$ estimate is smaller than the actual \pileup activity.  Tracking information also plays a key role in tagging and rejecting stochastic jets.  Since tracks can be precisely associated with specific vertices, track-based observables can provide information about the \pileup structure and vertex composition of jets within the tracking detector acceptance ($|\eta|<2.5$) that can be used for discrimination. The composition of \pileup jets depends on both \avgmu and \pt. Stochastic jets have a much steeper \pt spectrum than \pileup QCD jets. Therefore, higher-\pt jets that are associated with a primary vertex which is not the hard-scatter vertex are more likely to be \pileup QCD jets, not stochastic jets.  On the other hand, while the number of QCD \pileup jets increases linearly with \avgmu, the rate of stochastic jets increases more rapidly such that at high luminosity the majority of \pileup jets at low \pt are expected to be stochastic in nature~\cite{BOOST2012}.  

\subsection{\Pileup jet suppression from subtraction}
\label{sec:suppression:subtraction}

\begin{figure}[!ht]
  \centering

   \includegraphics[width=0.6\textwidth]{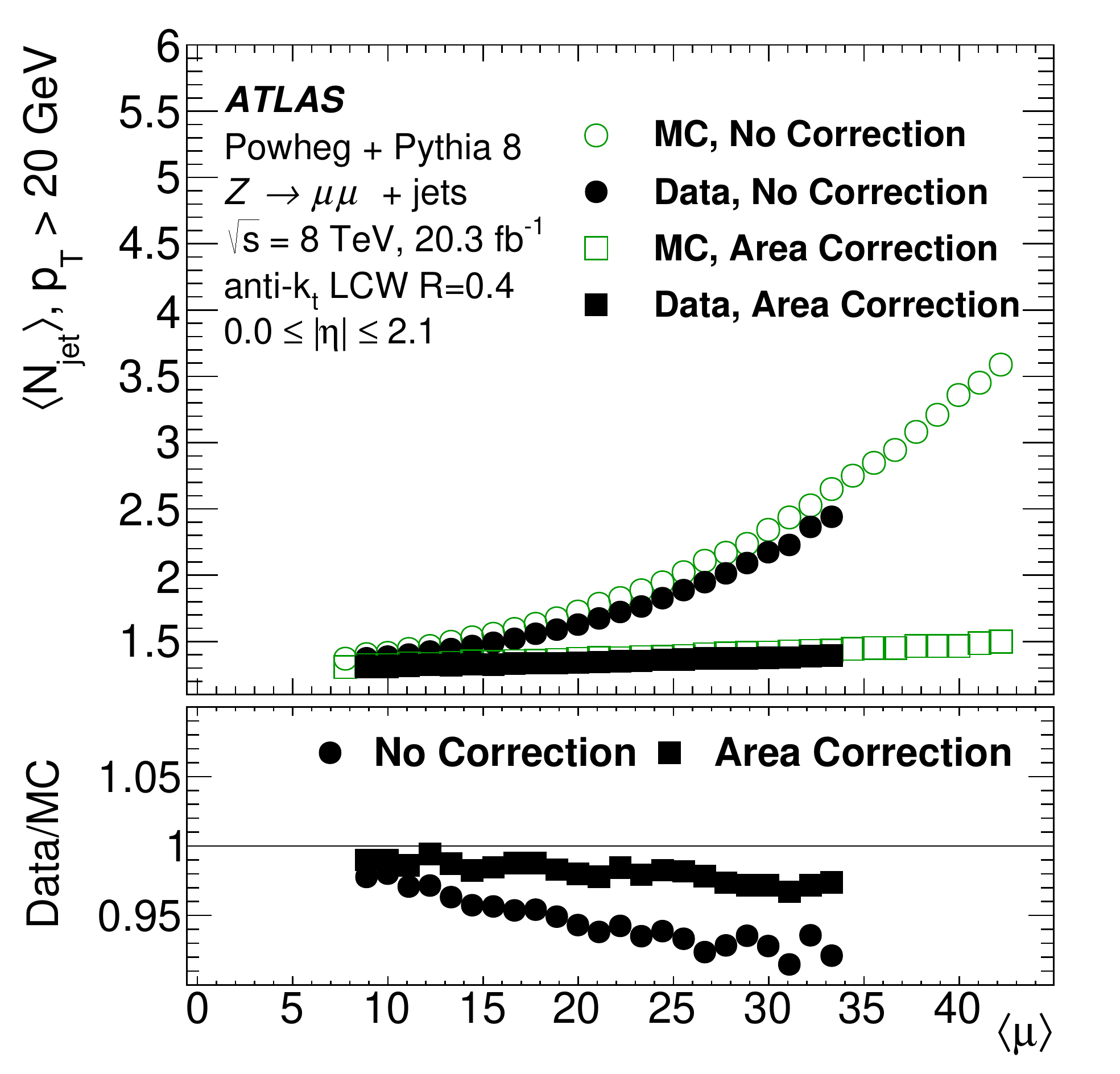}

  \caption{The mean \antikt\ $R=0.4$ LCW jet multiplicity as a function of \avgmu in \Zjets events for jets with $\pt > 20\GeV$ and $|\eta| < 2.1$. Events in this plot are required to have at least 1 jet both before and after the application of the jet-area based \pileup correction.}
           
  \label{fig:suppression:areasubt}
\end{figure}

The number of reconstructed jets increases with the average number of \pileup interactions, as shown in \figref{suppression:areasubt} using the \Zjets event sample described in \secref{rho}. Event-by-event \pileup subtraction based on jet areas, as described in \secref{subtraction:energy}, removes the majority of \pileup jets by shifting their \pt below the \pt threshold of $20\GeV$.  This has the effect of improving the level of agreement between data and MC simulation. The phenomenon of \pileup jets is generally not well-modelled, as shown in the ratio plot of \figref{suppression:areasubt}.

\subsection{\Pileup jet suppression from tracking}
\label{sec:suppression:tracking}

Some \pileup jets remain even after \pileup subtraction mainly due to localised fluctuations in \pileup activity which are not fully corrected by $\rho$ in \equref{areasubtraction}. Information from the tracks matched to each jet may be used to further reject any jets not originating from the hard-scatter interaction. ATLAS has developed three different track-based tagging approaches for the identification of \pileup jets: The jet vertex fraction (\JVF) algorithm, used in almost all physics analyses in \RunOne,  a set of two new variables (\cJVF, and \RpT) for improved performance, and a new combined discriminant, the jet vertex tagger (\JVT) for optimal performance. While the last two approaches were developed using \RunOne data, most analyses based on \RunOne data were completed before these new algorithms for \pileup suppression were developed. Their utility is already being demonstrated for use in high-luminosity LHC upgrade studies, and will be available to all ATLAS analyses at the start of \RunTwo.

\subsubsection{Jet vertex fraction}
\label{sec:suppression:JVF}

The jet vertex fraction (\JVF) is a variable used in ATLAS to identify the primary vertex from which the jet originated. A cut on the \JVF variable can help to remove jets which are not associated with the hard-scatter primary vertex. Using tracks reconstructed from the ID information, the \JVF variable can be defined for each jet with respect to each identified primary vertex (PV) in the event, by identifying the PV associated with each of the charged-particle tracks pointing towards the given jet. Once the hard-scatter PV is identified, the \JVF variable can be used to select jets having a high likelihood of originating from that vertex. Tracks are assigned to calorimeter jets following the ghost-association procedure~\cite{Cacciari2008}, which consists of assigning tracks to jets by adding tracks with infinitesimal \pt\ to the jet clustering process. Then, the \JVF is calculated as the ratio of the scalar sum of the \pt\ of matched tracks that originate from a given PV to the scalar sum of \pt\ of all matched tracks in the jet, independently of their origin.

\JVF is defined for each jet with respect to each PV. For a given jet$_{i}$, its \JVF with respect to the primary vertex PV$_{j}$ is given by:

\begin{equation}
 \JVF({\rm jet}_{i},{\rm PV}_{j})=\frac{\sum_m \pT ({\rm track}_{m}^{\rm jet_i},{\rm PV}_j)}{\sum_n\sum_l  \pT ({\rm track}_{l}^{\rm jet_i},{\rm PV}_n)},
 \label{eq:jvf}
\end{equation} 

\noindent where $m$ runs over all tracks originating from PV$_{j}$\footnote{Tracks are assigned to vertices by requiring $| \Delta z \times \sin \theta | < 1$~mm.  In cases where more than one vertex satisfies this criterion, ambiguity is resolved by choosing the vertex with the largest summed $\pT^2$ of tracks.} matched to jet$_{i}$, $n$ over all primary vertices in the event and $l$ over all tracks originating from PV$_{n}$ matched to jet$_{i}$. Only tracks with $\pt > 500\MeV$ are considered in the \JVF calculation.  \JVF is bounded by 0 and 1, but a value of $-1$ is assigned to jets with no associated tracks. 

\begin{figure}[h]
  \centering
      \subfigure[]{
         \includegraphics[width=0.48\textwidth]{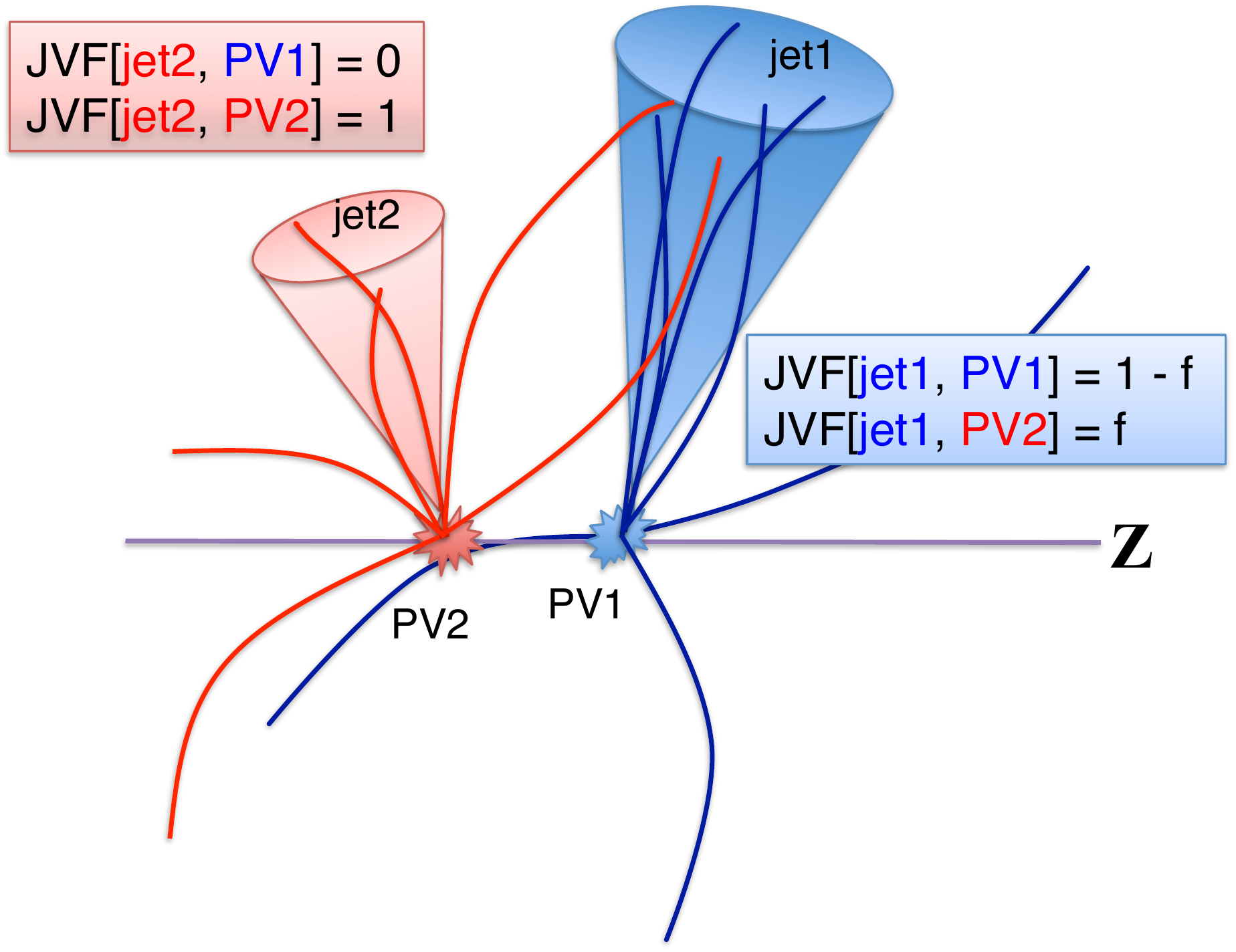}
         \label{fig:JVFschema:JVF1}}
     \subfigure[]{
         \includegraphics[width=0.48\textwidth]{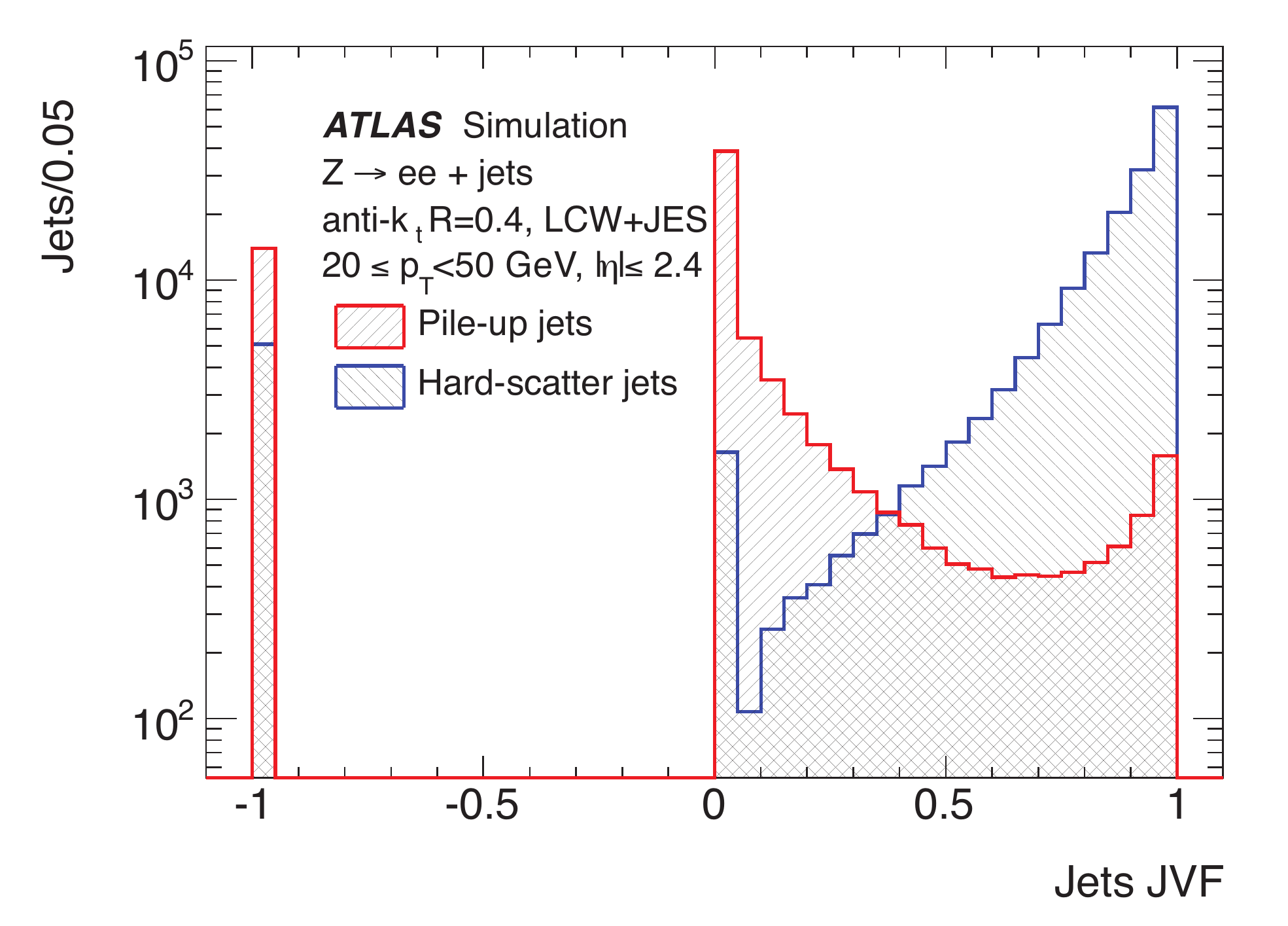}
         \label{fig:JVFschema:JVF2}}
  \caption{\subref{fig:JVFschema:JVF1} Schematic representation of the jet vertex fraction \JVF principle where $f$  denotes the fraction of track \pt\ contributed to jet 1 due to the second vertex (PV2). \subref{fig:JVFschema:JVF2} \JVF distribution for hard-scatter (blue) and \pileup (red) jets with $20 < \pT < 50\GeV$ and $|\eta|<2.4$ after \pileup subtraction and jet energy scale correction in simulated $Z+$jets events.}
  \label{fig:JVFschema}
\end{figure}

For the purposes of this paper, \JVF is defined from now on with respect to the hard-scatter primary vertex. In the \Zjets events used for these studies of \pileup suppression, this selection of the hard-scatter primary vertex is found to be correct in at least 98\% of events.  \JVF may then be interpreted as an estimate of the fraction of \pt in the jet that can be associated with the hard-scatter interaction.  The principle of the \JVF variable is shown schematically in \figref{JVFschema:JVF1}. \Figref{JVFschema:JVF2} shows the \JVF distribution in MC simulation for hard-scatter jets and for \pileup jets with $\pt > 20\GeV$ after \pileup subtraction and jet energy scale correction in a $Z(\rightarrow ee)+$jets sample with the \avgmu distribution shown in \figref{intro:avgmu}. Hard-scatter jets are calorimeter jets that have been matched to truth-particle jets from the hard-scatter with an angular separation of $\Delta R\leq0.4$, whereas \pileup jets are defined as calorimeter jets with an angular separation to the nearest truth-particle jet of $\Delta R>0.4$. The thresholds for truth-particle jets are $\pt>10$~GeV for those originating from the hard-scatter, and $\pt>4$~GeV for those originating in \pileup interactions. This comparison demonstrates the discriminating power of the \JVF variable.

While \JVF is highly correlated with the actual fraction of hard-scatter activity in a reconstructed calorimeter jet, it is important to note that the correspondence is imperfect.  For example, a jet with significant neutral \pileup contributions may receive $\JVF = 1$, while $\JVF = 0$ may result from a fluctuation in the fragmentation of a hard-scatter jet such that its charged constituents all fall below the track \pt threshold.  \JVF also relies on the hard-scatter vertex being well-separated along the beam axis from all of the \pileup vertices.  In some events, a \pileup jet may receive a high value of \JVF because its associated primary vertex is very close to the hard-scatter primary vertex.  While this effect is very small for 2012 \pileup conditions, it will become more important at higher luminosities, as the average distance between interactions decreases as $1/\avgmu$. For these reasons, as well as the lower probability for producing a \pileup QCD jet at high \pt, \JVF selections are only applied to jets with $\pt\leq50\GeV$.

\begin{figure*}
  \centering
  \subfigure[$\pT > 20 \GeV$]{\includegraphics[width=0.44\textwidth]{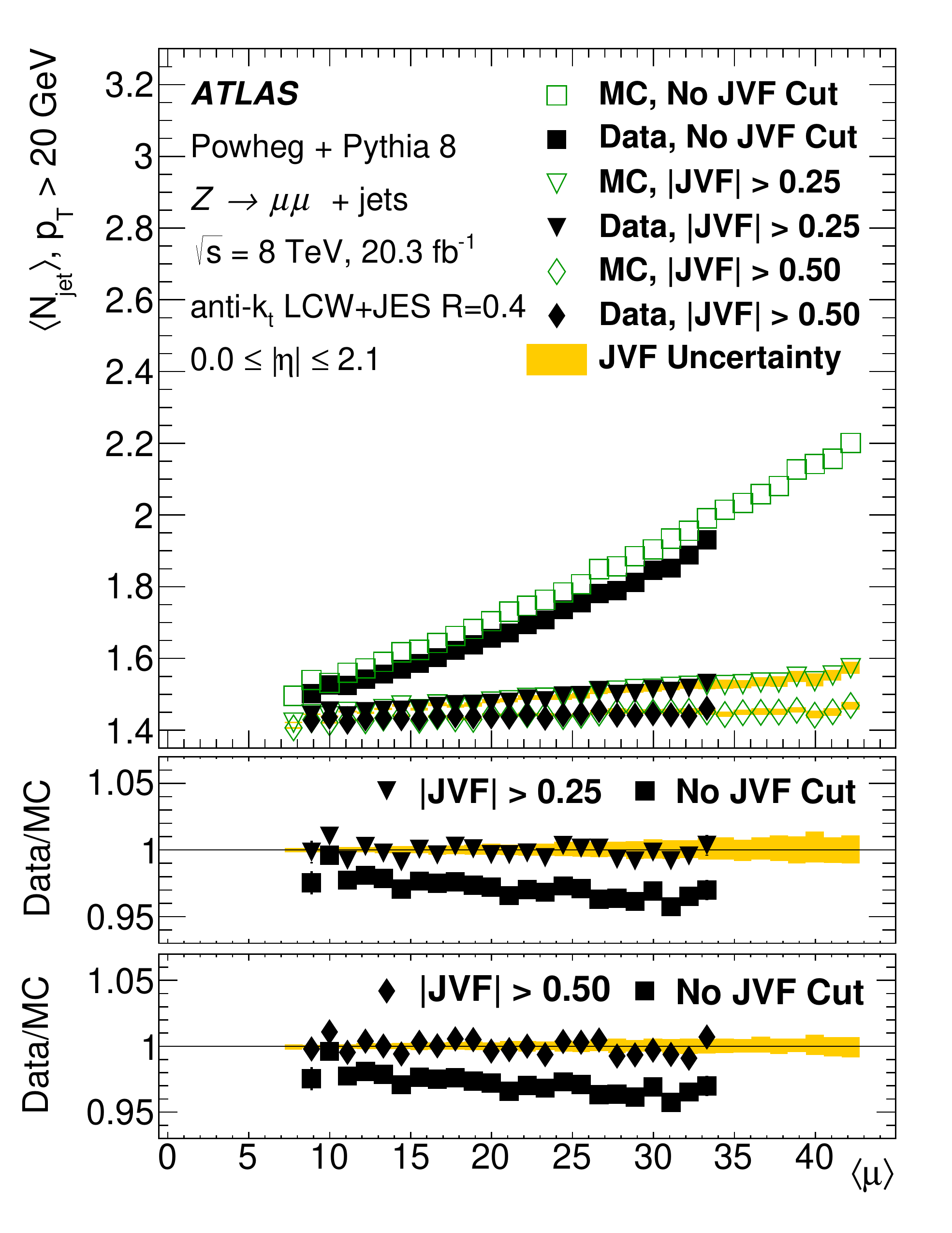}\label{fig:njetsJVF:20}}
  \subfigure[$\pT > 30 \GeV$]{\includegraphics[width=0.44\textwidth]{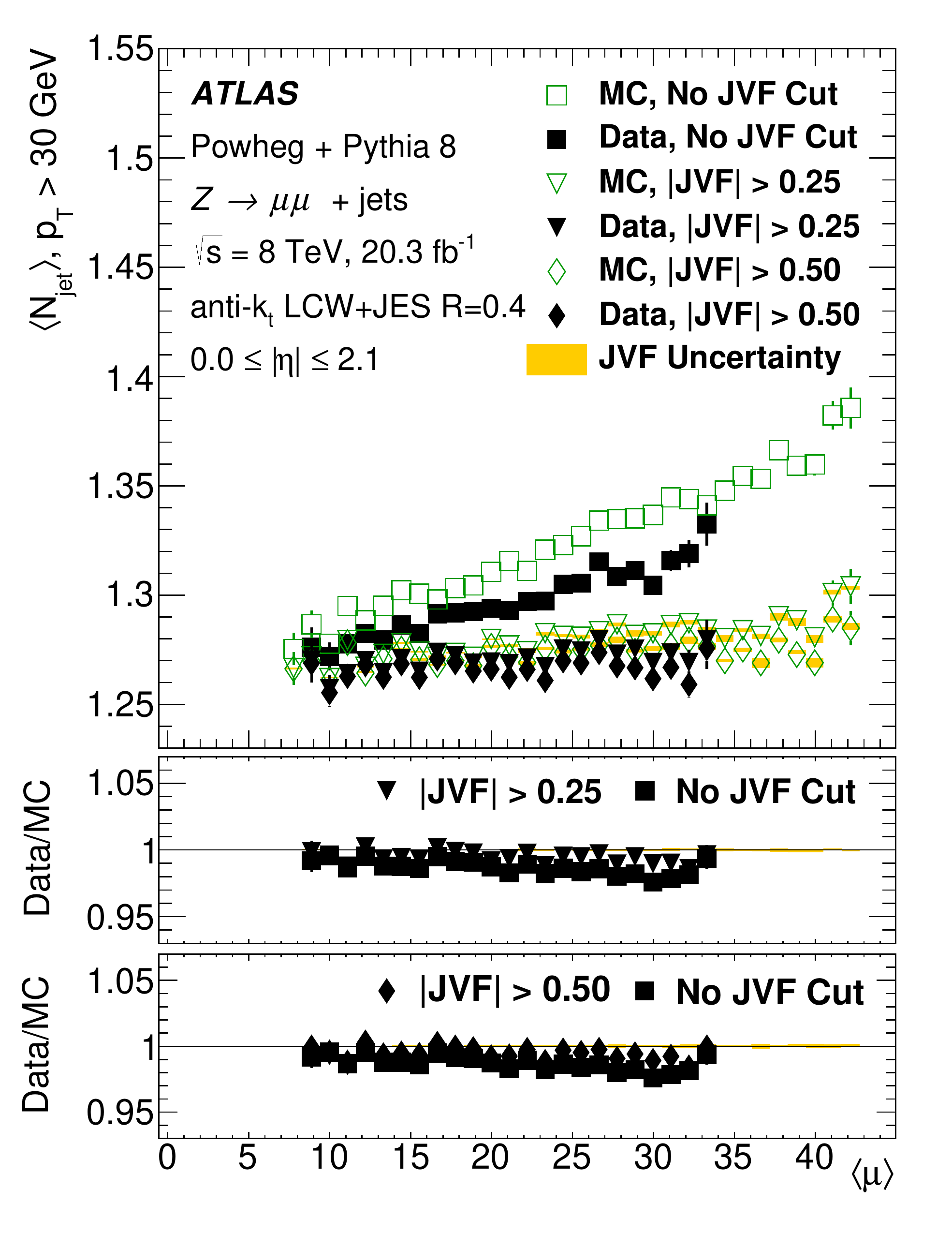}\label{fig:njetsJVF:30}}\\
  \subfigure[$\pT > 40 \GeV$]{\includegraphics[width=0.44\textwidth]{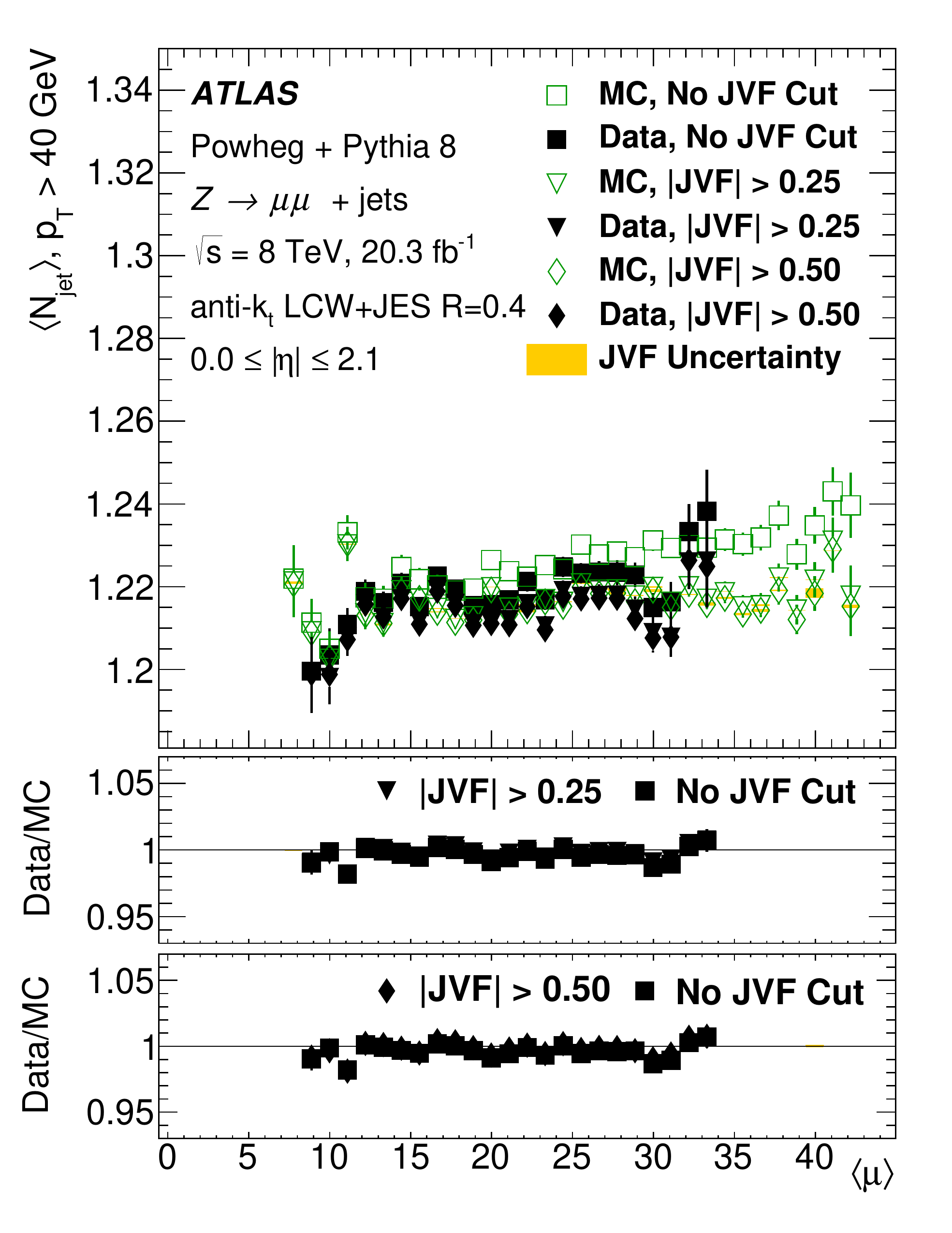}\label{fig:njetsJVF:40}}
  \caption{The mean \antikt\ $R=0.4$ LCW+JES jet multiplicity as a function of \avgmu in \Zjets events for jets with $|\eta| < 2.1$, back-to-back with the \Zboson boson, before and after several $|\JVF|$ cuts were applied to jets with $\pT < 50 \GeV$. Results for jets with \subref{fig:njetsJVF:20} $\pT > 20\GeV$, \subref{fig:njetsJVF:30} $\pT > 30\GeV$, and \subref{fig:njetsJVF:40} $\pT > 40\GeV$ are shown requiring at least one jet of that \pT. To remove effects of hard-scatter modelling the dependence on \avgmu was fit and the MC simulation shifted so that data and simulation agree at zero \pileup, $\avgmu = 0$. The upper ratio plots show before and after applying a $|\JVF|$ cut of 0.25 and the lower ratio plots show the same for a cut of 0.50. The \JVF uncertainty is very small when counting jets with $\pT > 40 \GeV$.} 
  \label{fig:njetsJVF}
\end{figure*}

The modelling of \JVF is investigated in $\Zboson(\to\mu\mu)$+jets events using the same selection as discussed in \secref{rho}, which yields a nearly pure sample of hard-scatter jets. By comparison to truth-particle jets in MC simulation, it was found that the level of \pileup jet contamination in this sample is close to 2\% near $20\GeV$ and almost zero at the higher end of the range near $50\GeV$. The \JVF distribution for the jet balanced against the $Z$ boson in these events is well-modelled for hard-scatter jets. However, the total jet multiplicity in these events is overestimated in simulated events, due to mismodelling of \pileup jets. This is shown in \figref{njetsJVF}, for several different choices of the minimum \pt cut applied at the fully calibrated jet energy scale (including jet-area-based \pileup subtraction).  The application of a \JVF cut significantly improves the data/MC agreement because the majority of \pileup jets fail the \JVF cut: across all \pt bins, data and MC simulation are seen to agree within 1\% following the application of a \JVF cut. It is also observed that the application of a \JVF cut results in stable values for the mean jet multiplicity as a function of \avgmu.

\Figref{njetsJVF} also shows the systematic uncertainty bands, which are only visible for the lowest \pt selection of $20\GeV$.  These uncertainties are estimated by comparing the \JVF distributions for hard-scatter jets in data and MC simulation.  The efficiency of a nominal \JVF cut of $X$ is defined as the fraction of jets, well balanced against the $Z$ boson, passing the cut, denoted by $\mathcal{E}_{\rm MC}^{\rm nom}$ and $\mathcal{E}_{\rm data}^{\rm nom}$ for MC events and data, respectively. The systematic uncertainty is derived by finding two \JVF cuts with $\mathcal{E}_{\rm MC}$ differing from $\mathcal{E}_{\rm MC}^{\rm nom}$ by $\pm (\mathcal{E}_{\rm MC}^{\rm nom}-\mathcal{E}_{\rm data}^{\rm nom})$. The \JVF uncertainty band is then formed by re-running the analysis with these up and down variations in the \JVF cut value. Systematic uncertainties vary between 2 and 6\% depending on jet \pt and $\eta$. 

\subsubsection{Improved variables for \pileup jet vertex identification}
\label{sec:suppression:cJVFRpT}

While a \JVF selection is very effective in rejecting \pileup jets, it has limitations when used in higher (or varying) luminosity conditions.  As the denominator of \JVF increases with the number of reconstructed primary vertices in the event, the mean \JVF for signal jets is shifted to smaller values.  This explicit \pileup dependence of \JVF results in an \Npv-dependent jet efficiency when a minimum \JVF criterion is imposed to reject \pileup jets.  This \pileup sensitivity is addressed in two different ways.  First, by correcting \JVF for the explicit \pileup dependence in its denominator (\cJVF), and second, by introducing a new variable defined entirely from hard-scatter observables (\RpT). 

The quantity \cJVF is a variable similar to \JVF, but corrected for the \Npv dependence. It is defined as
 
\begin{equation}
\cJVF = \frac{\sum_m{p_{{\rm T}, m}^{{\rm track}}({\rm PV}_0)}}{\sum_l p_{{\rm T}, l}^{{\rm track}}({\rm PV}_0) + \frac{\sum_{n\geq1}\sum_l p_{{\rm T}, l}^{{\rm track}}({\rm PV}_n)}{ (k \cdot \nPUtrk)}}.
\label{eq:corrJVF}
\end{equation}

\noindent where $\sum_m{p_{{\rm T}, m}^{{\rm track}}({\rm PV}_0)}$ is the scalar sum of the \pt of the tracks that are associated with the jet and originate from the hard-scatter vertex. The term $\sum_{n\geq1}\sum_l p_{{\rm T}, l}^{{\rm track}}({\rm PV}_n)=\pT^{\mathrm{PU}}$ denotes the scalar sum of the \pt of the associated tracks that originate from any of the \pileup interactions. 

The \cJVF variable uses a modified track-to-vertex association method that is different from the one used for \JVF. The new selection aims to improve the efficiency for $b$-quark jets and consists of two steps.  In the first step, the vertex reconstruction is used to assign tracks to vertices. If a track is attached to more than one vertex, priority is given to the vertex with higher $\sum(\pttrk)^2$. In the second step, if a track is not associated with any primary vertex after the first step but satisfies $|\Delta z|<3$~mm with respect to the hard-scatter primary vertex, it is assigned to the hard-scatter primary vertex. The second step targets tracks from decays in flight of hadrons that originate from the hard-scatter but are not likely to be attached to any vertex. The $|\Delta z|<3$~mm criterion was chosen based on the longitudinal impact parameter distribution of tracks from $b$-hadron decays, but no strong dependence of the performance on this particular criterion was observed when the cut value was altered within 1~mm. The new 2-step track-to-vertex association method results in a significant increase in the hard-scatter jet efficiency at fixed rate of fake \pileup jets, with a large performance gain for jets initiated by $b$-quarks.

To correct for the linear increase of $\langle\pT^{\mathrm{PU}}\rangle$ with  the total number of \pileup tracks per event (\nPUtrk), $\pT^{\mathrm{PU}}$ is divided by $(k \cdot \nPUtrk)$, with $k=0.01$, in the \cJVF definition. The total number of \pileup tracks per event is computed from all tracks associated with vertices other than the hard-scatter vertex. The scaling factor $k$ is approximated by the slope of $\langle\pT^{\mathrm{PU}}\rangle$ with $\nPUtrk$, but the resulting discrimination between hard-scatter and \pileup jets is insensitive to the choice of $k$.\footnote{With this particular choice of $k$, the resulting \cJVF shapes for hard-scatter and \pileup jets are similar to the corresponding ones for \JVF.}

\begin{figure}[!htbp]
  \centering
  \subfigure[]{
      \includegraphics[width= 0.48\textwidth]{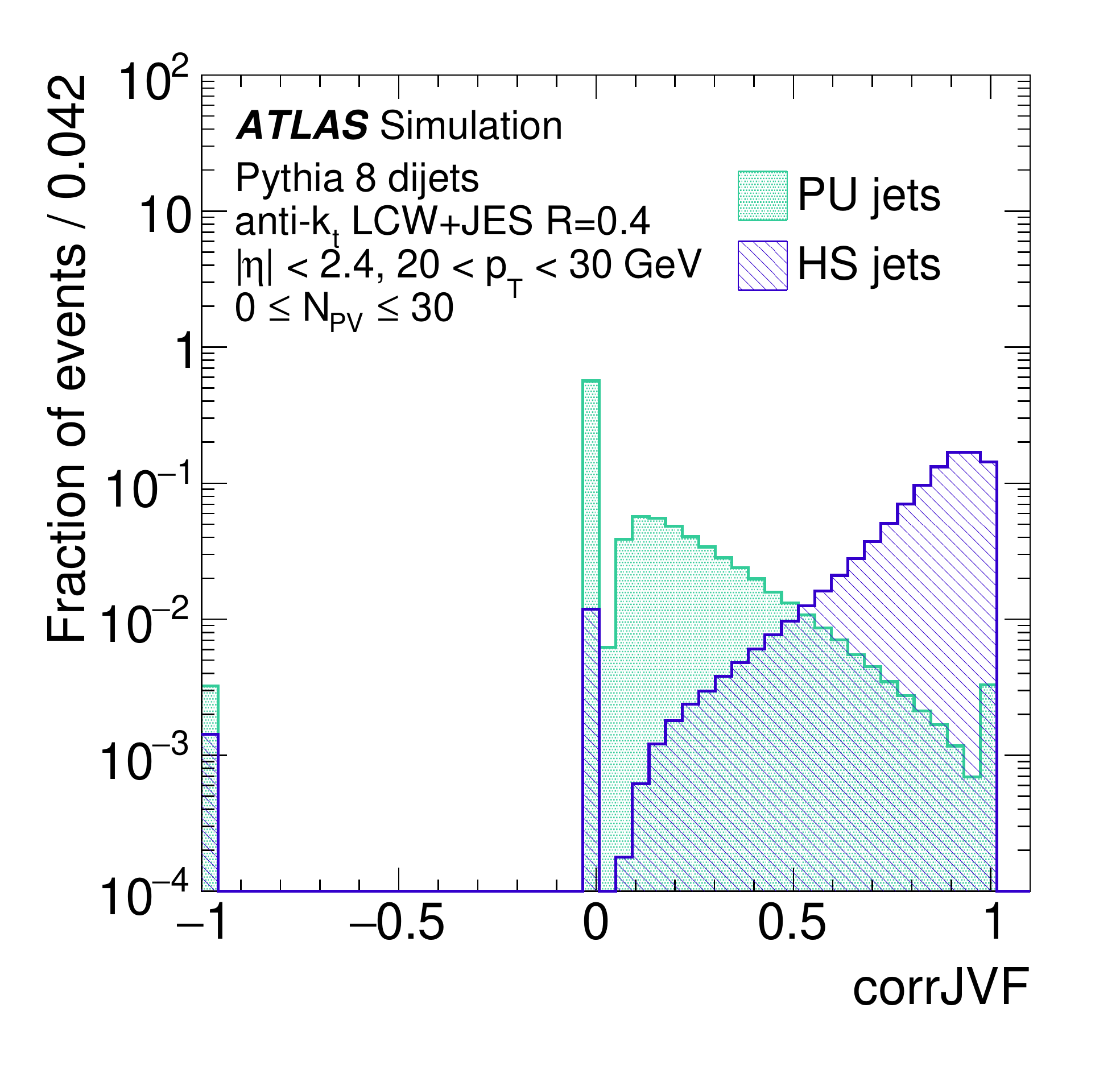}
    \label{fig:corrJVF_Dist}
  }
  \subfigure[]{
      \includegraphics[width=0.48\textwidth]{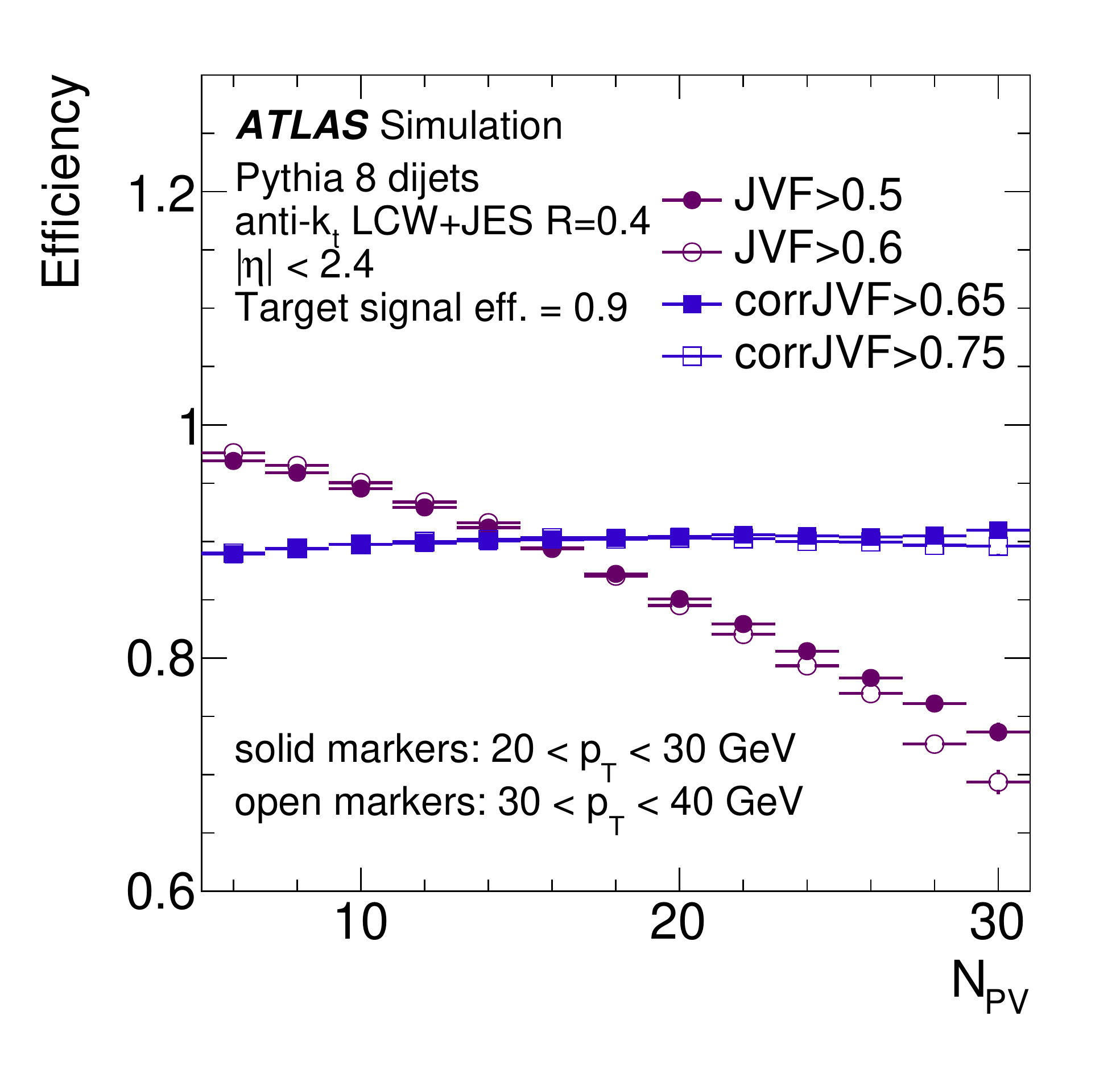}
    \label{fig:corrJVF_NPVdependence}
  }
  \caption{\subref{fig:corrJVF_Dist} Distribution of \cJVF for \pileup (PU) and hard-scatter (HS) jets with $20 < \pT < 30\GeV$. 
           \subref{fig:corrJVF_NPVdependence} Primary-vertex dependence of the hard-scatter jet efficiency for $20 < \pT < 30\GeV$ (solid markers) and $30 < \pt < 40\GeV$ (open markers) jets for fixed cuts of \cJVF (blue square) and \JVF (violet circle) such that the inclusive efficiency is $90\%$. The selections placed on \cJVF and \JVF, which depend on the \pt bin, are specified in the legend.
    }
\label{fig:corrJVF}
\end{figure}

\Figref{corrJVF_Dist} shows the \cJVF distribution for \pileup and hard-scatter jets in simulated dijet events. A value $\cJVF=-1$ is assigned to jets with no associated tracks.  Jets with $\cJVF = 1$ are not included in the studies that follow due to use of signed \cJVF selections. About $1\%$ of hard-scatter jets with $20 < \pt < 30 \GeV$ have no associated hard-scatter tracks and thus $\cJVF=0$. 

\Figref{corrJVF_NPVdependence} shows the hard-scatter jet efficiency as a function of the number of reconstructed primary vertices in the event when imposing a minimum \cJVF or \JVF requirement such that the efficiency measured across the full range of \Npv is 90\%.  For the full range of \Npv considered, the hard-scatter jet efficiency after a selection based on \cJVF is stable at $90\%\pm1\%$, whereas for \JVF the efficiency degrades by about 20\%, from 97\% to 75\%. The choice of scaling factor $k$ in the \cJVF distribution does not affect the stability of the hard-scatter jet efficiency with \Npv. 

The variable \RpT is defined as the scalar sum of the \pt of the tracks that are associated with the jet and originate from the hard-scatter vertex divided by the fully calibrated jet \pt, which includes \pileup subtraction:

\begin{equation}
\RpT = \frac{\sum_k{p_{{\rm T}, k}^{{\rm track}}({\rm PV}_0)}}{\ptjet}.
\label{eq:RpT}
\end{equation}

\begin{figure}[!htbp]
  \centering
  \subfigure[]{
      \includegraphics[width= 0.48\textwidth]{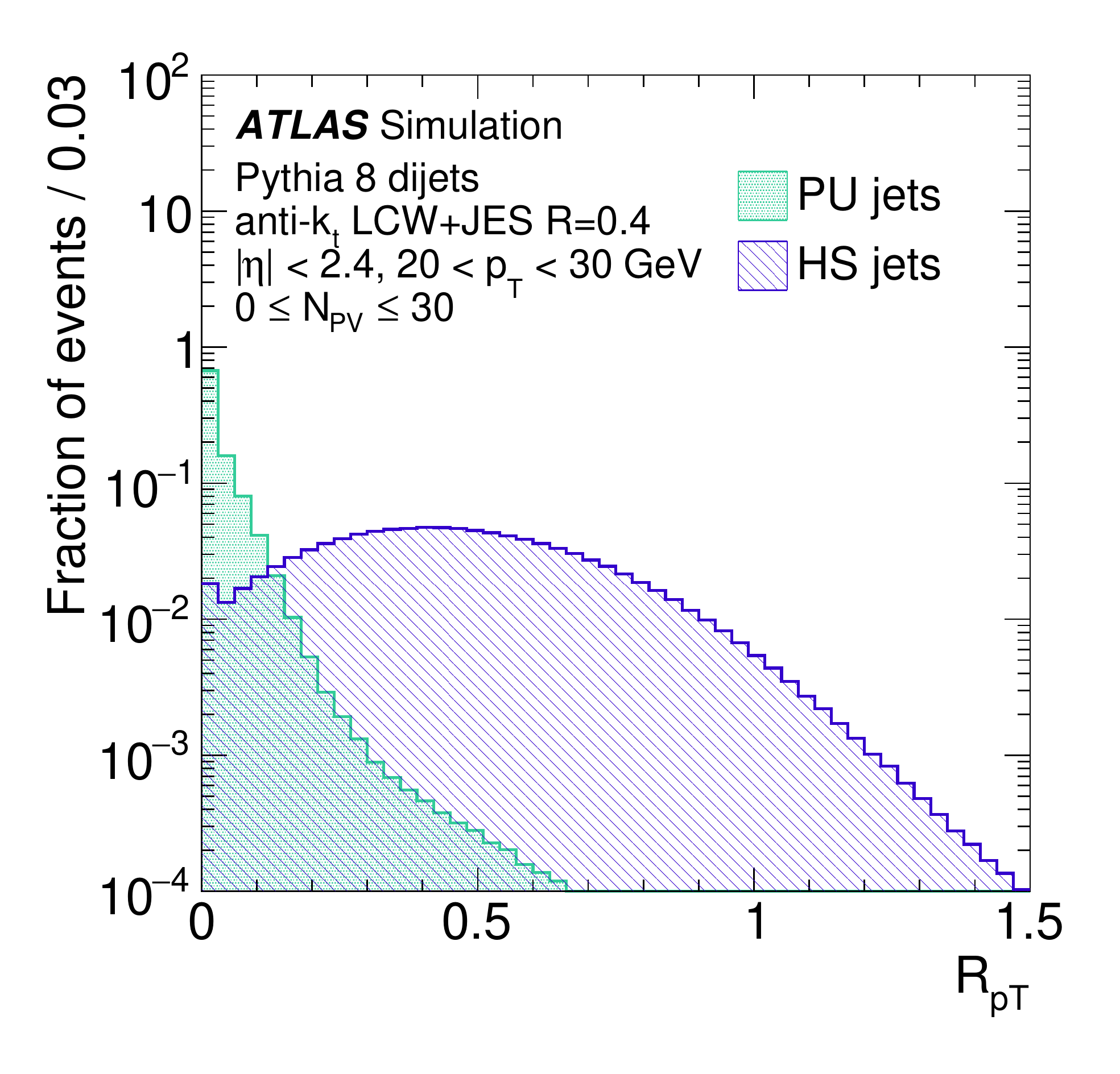}
    \label{fig:RpT_Dist}
  }
  \subfigure[]{
      \includegraphics[width=0.48\textwidth]{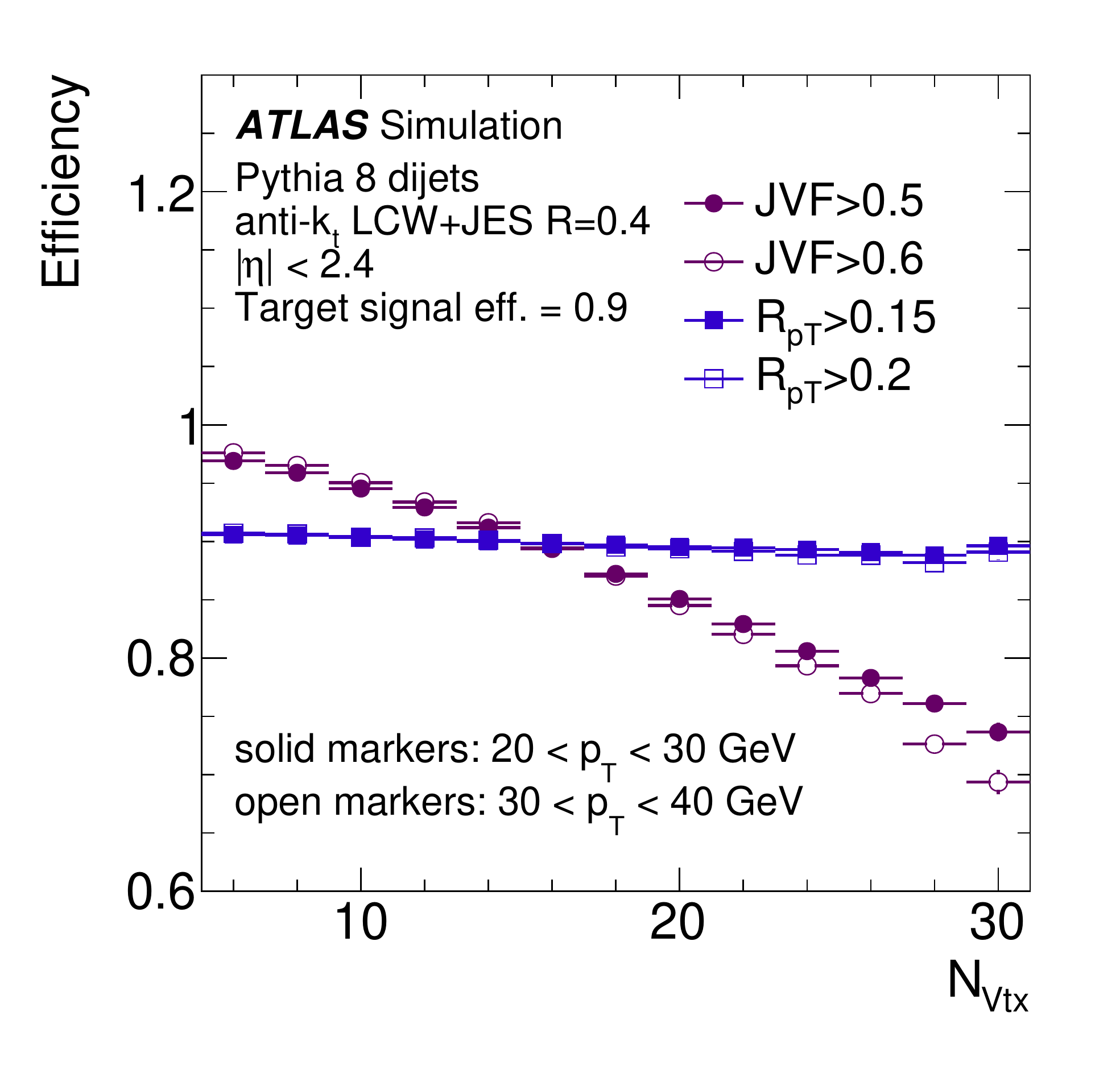}
    \label{fig:RpT_NPVdependence}
  }
  \caption{\subref{fig:RpT_Dist} Distribution of \RpT for \pileup (PU) and hard-scatter (HS) jets with $20 < \pt < 30\GeV$.
           \subref{fig:RpT_NPVdependence} Primary-vertex dependence of the hard-scatter jet efficiency for $20 < \pt < 30\GeV$ (solid markers) and $30 < \pt < 40\GeV$ (open markers) jets for fixed cuts of \RpT (blue square) and \JVF (violet circle) such that the inclusive efficiency is $90\%$. The cut values imposed on \RpT and \JVF, which depend on the \pt bin, are specified in the legend.}

\label{fig:RpT}
\end{figure}

The \RpT  distributions for \pileup and hard-scatter jets are shown in \figref{RpT_Dist}. \RpT is peaked at 0 and is steeply falling for \pileup jets, since tracks from the hard-scatter vertex rarely contribute. For hard-scatter jets, however, \RpT has the meaning of a charged \pt fraction and its mean value and spread are larger than for \pileup jets. Since \RpT involves only tracks that are associated with the hard-scatter vertex, its definition is at first order independent of \Npv. \Figref{RpT_NPVdependence} shows the hard-scatter jet efficiency as a function of \Npv when imposing a minimum \RpT and \JVF requirement such that the \Npv inclusive efficiency is $90\%$. For the full range of \Npv considered, the hard-scatter jet efficiency after a selection based on \RpT is stable at $90\%\pm1\%$.

\subsubsection{Jet vertex tagger}
\label{sec:suppression:JVT}

A new discriminant called the jet-vertex-tagger (\JVT) is constructed using \RpT and \cJVF as a two-dimensional likelihood derived using simulated dijet events and based on a k-nearest neighbour (kNN) algorithm~\cite{Hocker:2007ht}. For each point in the two-dimensional $\cJVF$--$\RpT$ plane, the relative probability for a jet at that point to be of signal type is computed as the ratio of the number of hard-scatter jets to the number of hard-scatter plus \pileup jets found in a local neighbourhood around the point using a training sample of signal and \pileup jets with $20 < \pt < 50\GeV$ and $|\eta| < 2.4$. The local neighbourhood is defined dynamically as the 100 nearest neighbours around the test point using a Euclidean metric in the $\RpT$--$\cJVF$ space, where \cJVF and \RpT are rescaled so that the variables have the same range. 
 
\begin{figure}[!htbp]
  \centering
  \subfigure[]{
  \includegraphics[width= 0.48\textwidth]{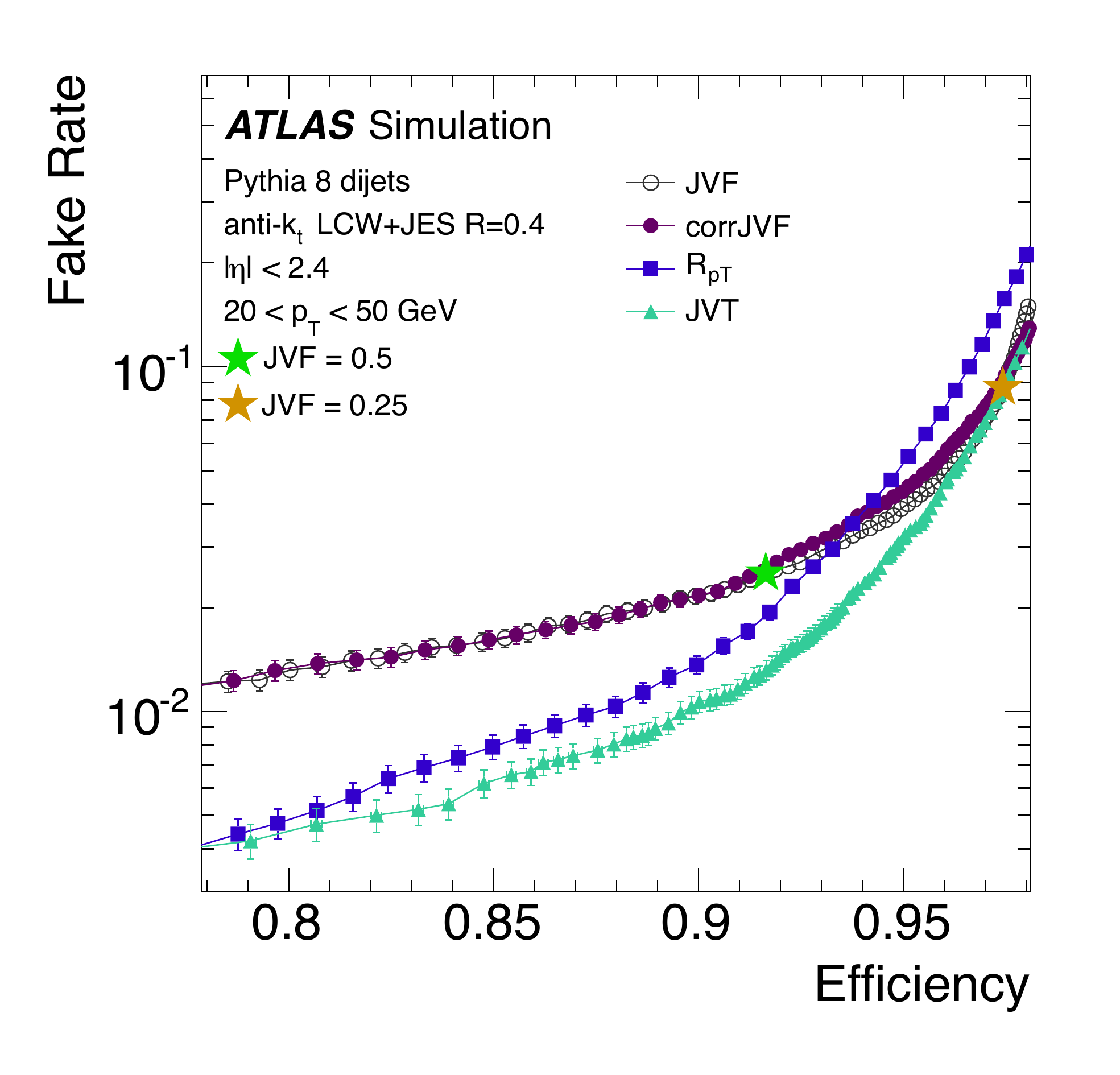}
  \label{fig:ROC}
  }
  \subfigure[]{
  \includegraphics[width= 0.48\textwidth]{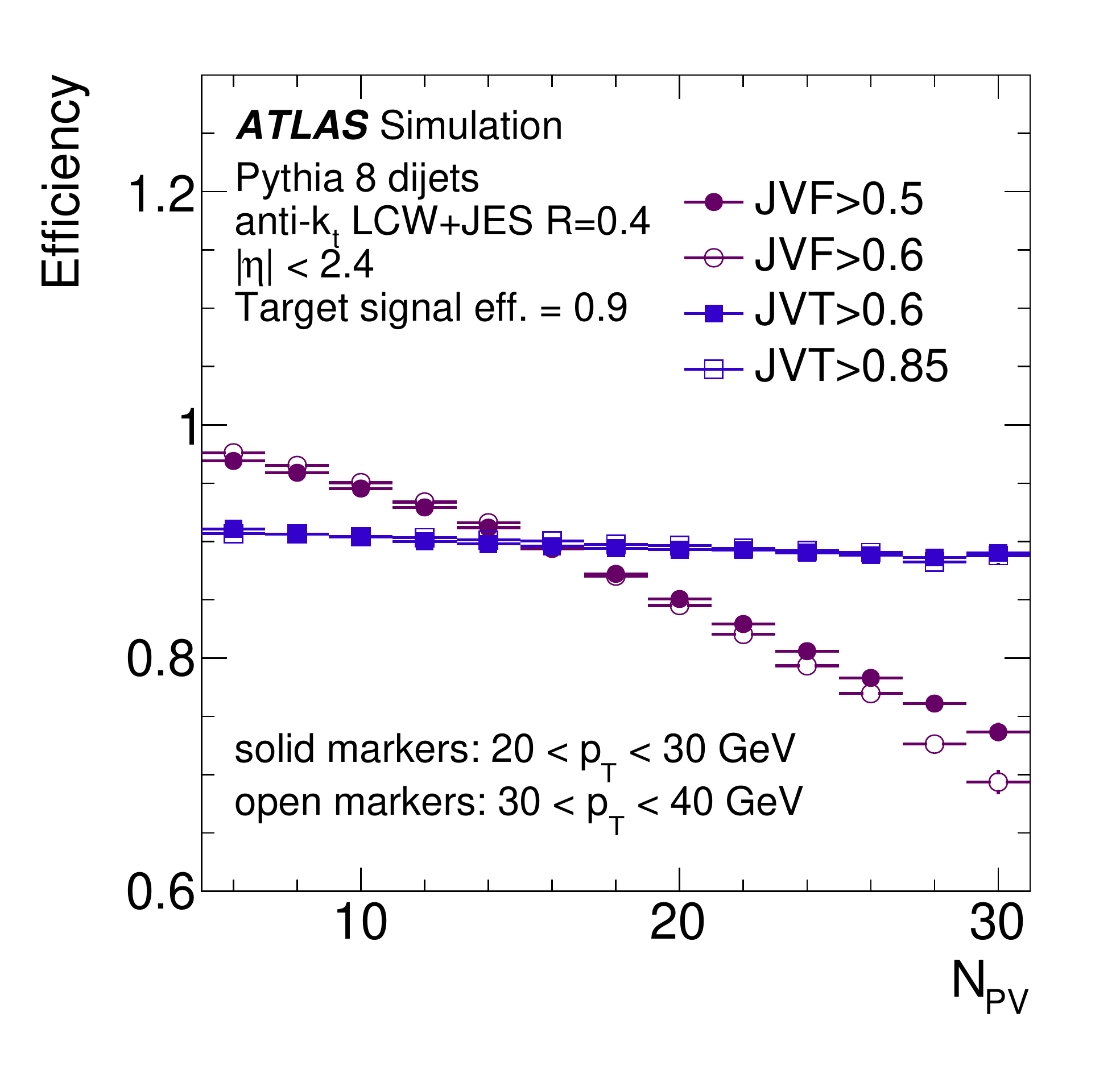}
  \label{fig:JVT_stab_95}
  }
  \caption{\subref{fig:ROC} Fake rate from \pileup jets versus hard-scatter jet efficiency curves for \JVF, \cJVF, \RpT, and \JVT. The widely used \JVF working points with cut values $0.25$ and $0.5$ are indicated with gold and green stars. 
  \subref{fig:JVT_stab_95} Primary vertex dependence of the hard-scatter jet efficiency for $20 < \pt < 30\GeV$ (solid markers) and $30 < \pt < 40\GeV$ (open markers) jets for fixed cuts of \JVT (blue square) and \JVF (violet circle) such that the inclusive efficiency is $90\%$.
  } 
\end{figure}

\Figref{ROC} shows the fake rate versus efficiency curves comparing the performance of the four variables \JVF, \cJVF, \RpT, and \JVT when selecting a sample of jets with $20 < \pt < 50\GeV$, $|\eta|<2.4$ in simulated dijet events.

The figure shows the fraction of \pileup jets  passing a minimum \JVF, \cJVF, \RpT or \JVT requirement as a function of the signal-jet efficiency resulting from the same requirement. The \JVT performance is driven by \cJVF (\RpT) in the region of high signal-jet efficiency (high \pileup rejection). Using \JVT, signal jet efficiencies of $80\%$, $90\%$ and $95\%$ are achieved for \pileup fake rates of respectively $0.4\%$, $1.0\%$ and $3\%$. When imposing cuts on \JVF that result in the same jet efficiencies, the \pileup fake rates are $1.3\%$, $2.2\%$ and $4\%$.

The dependence of the hard-scatter jet efficiencies on \Npv is shown in \figref{JVT_stab_95}.  For the full range of \Npv considered, the hard-scatter jet efficiencies after a selection based on \JVT are stable within $1\%$.

The differences in fragmentation and showering between jets initiated by gluons and light quarks affect the shapes of the \cJVF and \RpT distributions and thus the performance of the \JVT-based \pileup jet suppression.  Jets initiated by light quarks ($u,d,s$) have on average a lower number of associated hard-scatter tracks but a slightly higher jet energy response~\cite{Aad:2014gea} and both effects lead towards an increase in the number of jets with no associated tracks from the hard-scatter primary vertex relative to gluon-initiated jets.    

\begin{figure*}[!htbp]
  \centering
  \subfigure[]{
    \includegraphics[width= 0.45\textwidth]{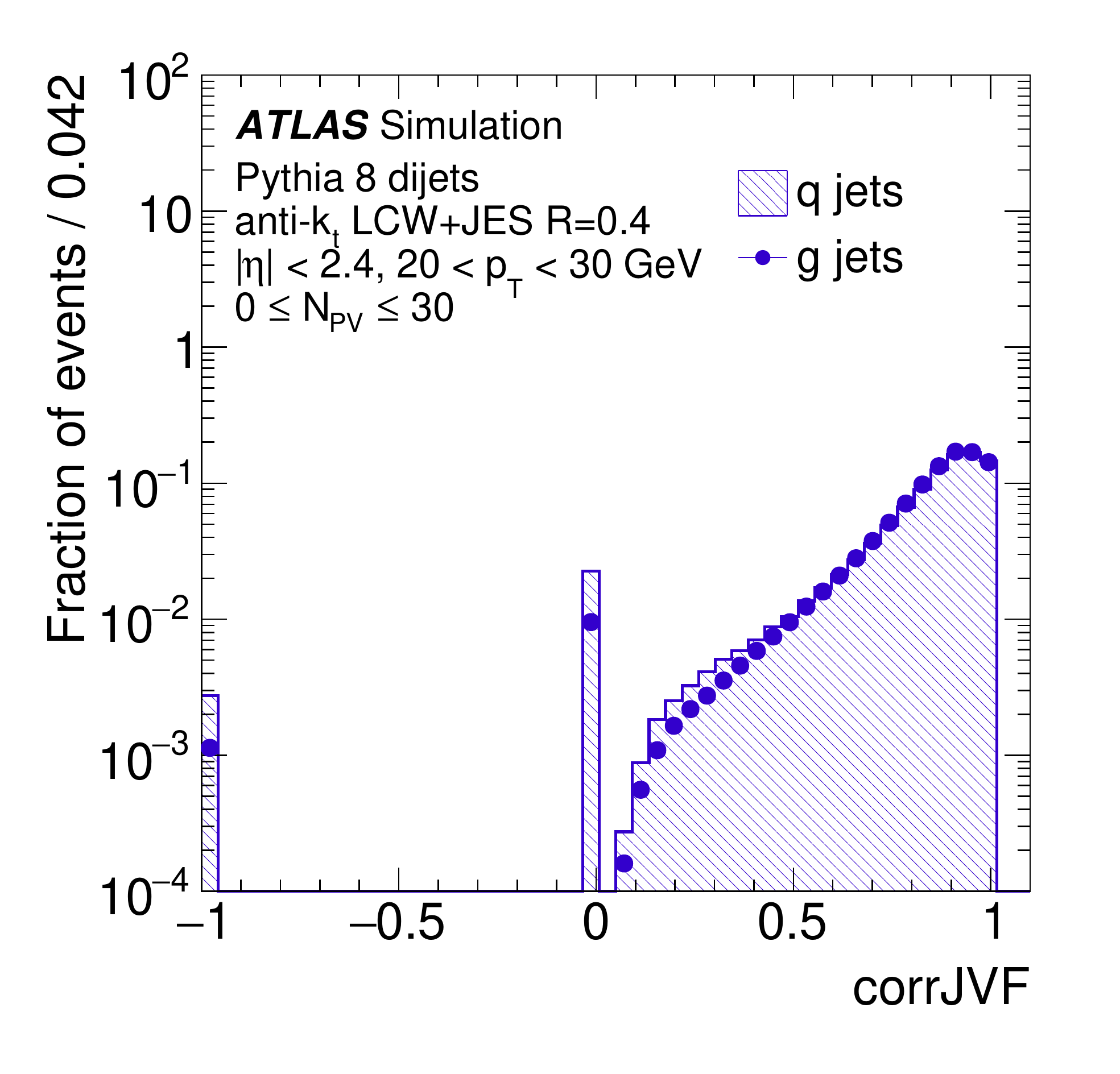}
    \label{fig:corrJVF_flavor_dep_1}
  }
  \subfigure[]{
    \includegraphics[width= 0.45\textwidth]{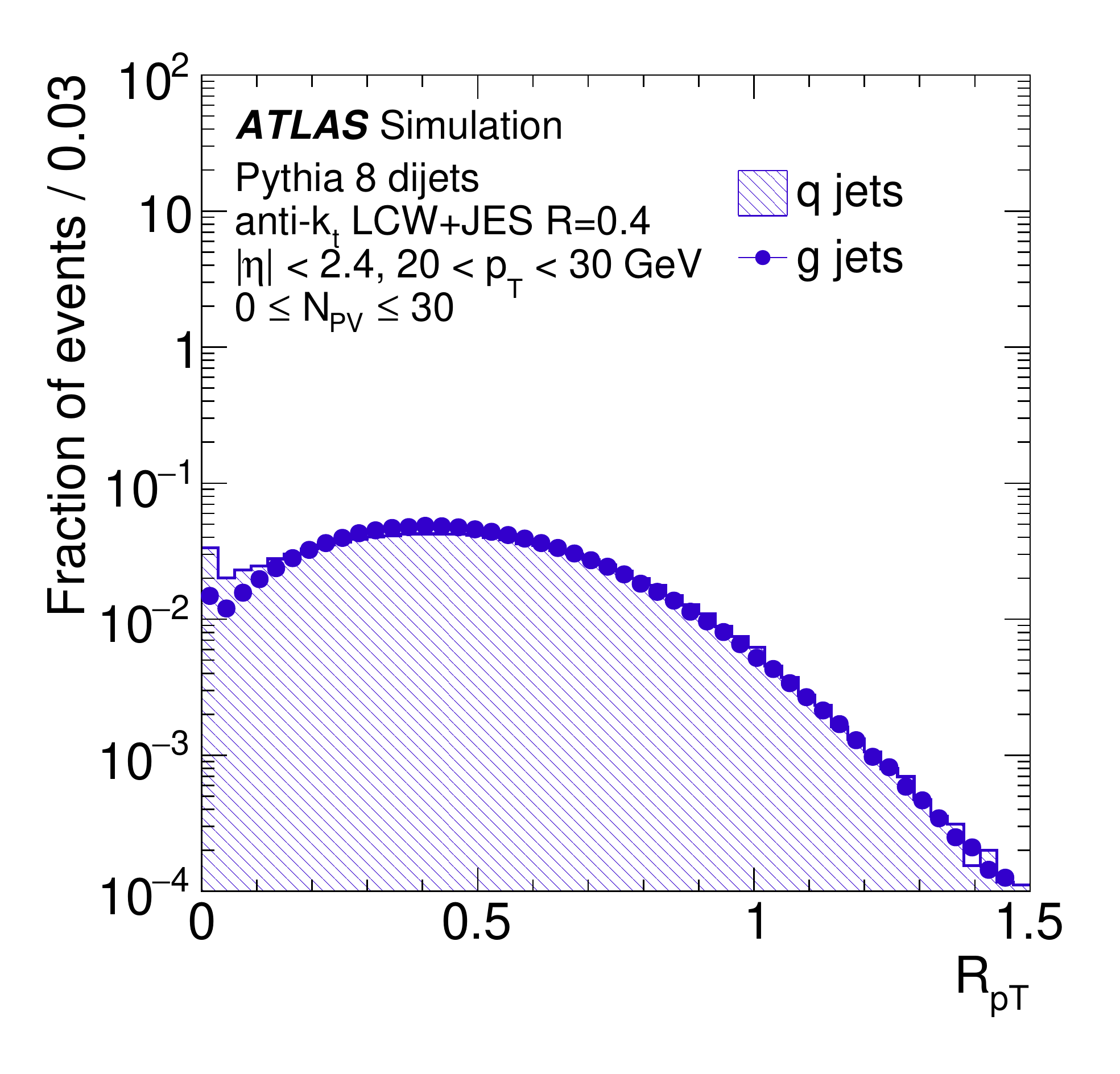}
    \label{fig:RpT_flavor_dep_2}
  }\\
  \subfigure[]{
    \includegraphics[width= 0.45\textwidth]{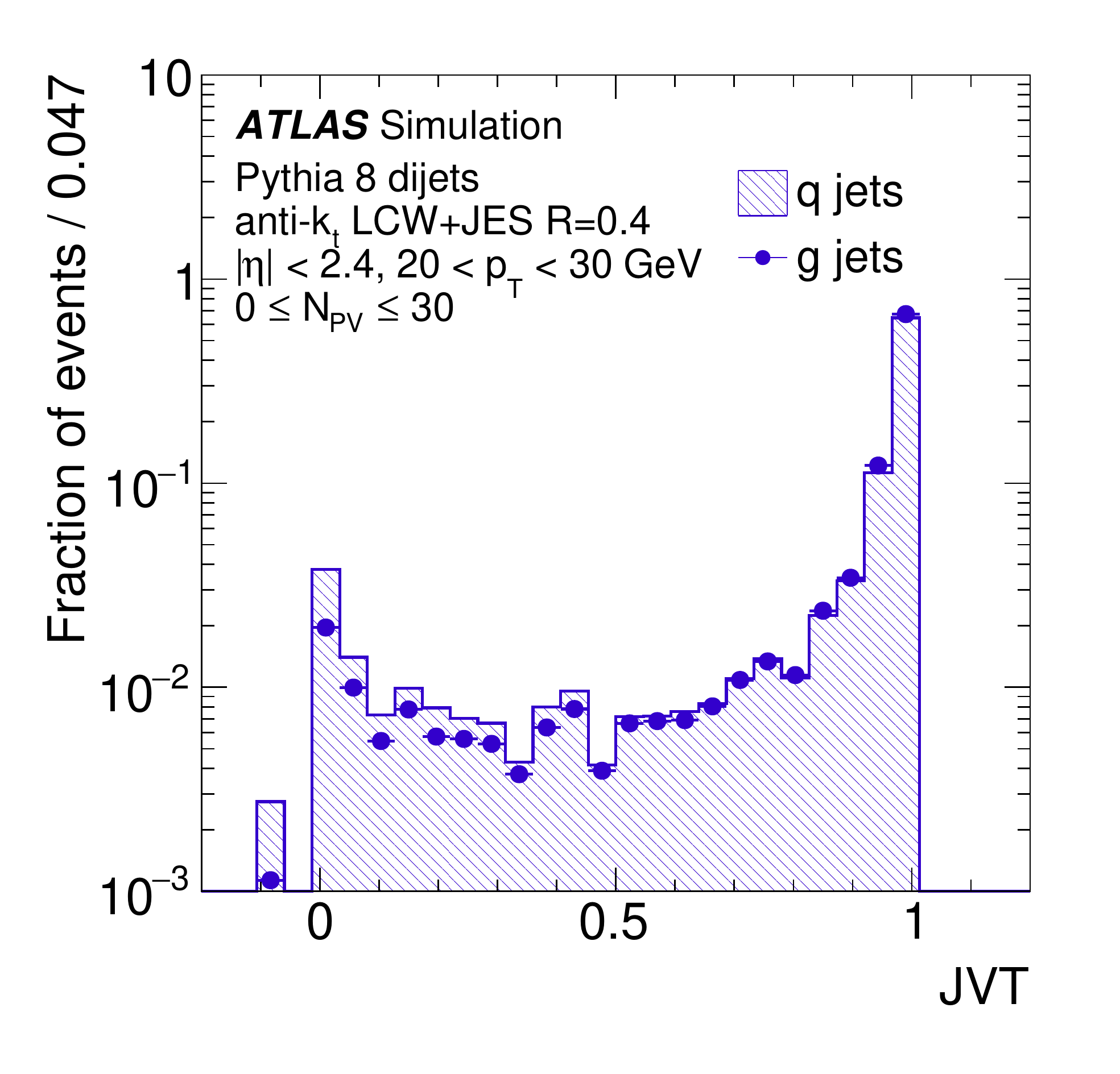}
    \label{fig:JVT_flavor_dep_3}
  }
  \caption{The distributions of \subref{fig:corrJVF_flavor_dep_1} \cJVF, \subref{fig:RpT_flavor_dep_2} \RpT and \subref{fig:JVT_flavor_dep_3} \JVT for light-quark and gluon initiated hard-scatter jets.\label{fig:flavor_dep}}
\end{figure*}

\Figref{flavor_dep} shows the \cJVF, \RpT and \JVT distributions for hard-scatter jets with $20<\pT<30\GeV$ initiated by gluons and light quarks. Using a leading-order notion of jet flavour, the parton-level flavour labelling refers to the highest-energy parton within a narrow cone of size $\Delta R = 0.3$ around the jet axis. The distributions for jets initiated by light quarks have more entries at low \cJVF, \RpT and \JVT values and consequently a worse separation from \pileup jets.  Most notably, about twice as many light-quark jets have no associated tracks from the hard-scatter primary vertex, that is $\cJVF=\JVT=0$. 

\begin{figure}[!htbp]
  \centering
  \includegraphics[width= 0.60\textwidth]{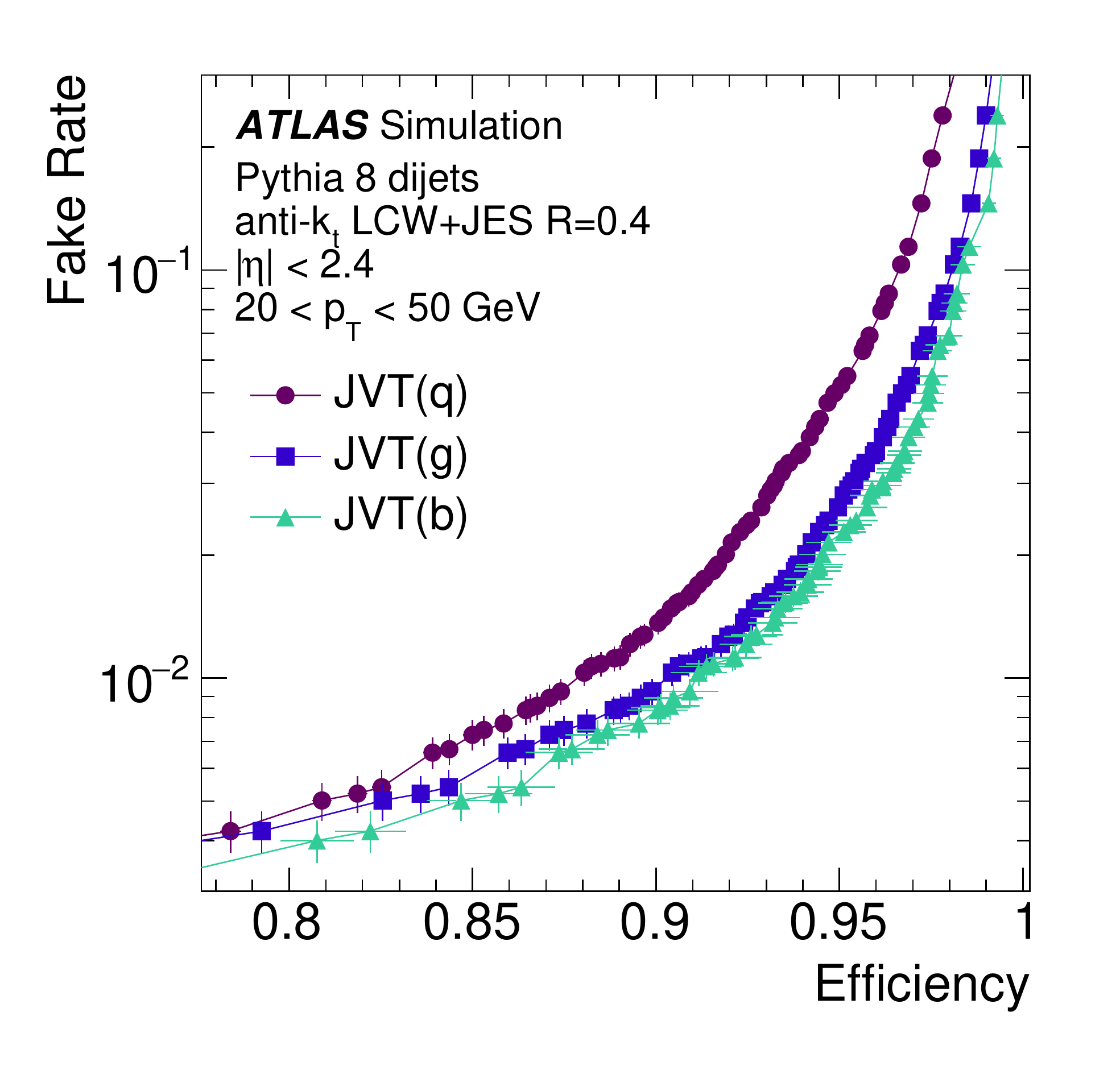}
  \caption{The fake rate from \pileup jets versus hard-scatter jet efficiency curves for \JVT separating jets initiated by light quarks, gluons and $b$-quarks.}
  \label{fig:ROC_q_vs_g_vs_b}
\end{figure}

\Figref{ROC_q_vs_g_vs_b} shows the efficiency versus fake-rate curve for \JVT for jets initiated by light quarks, gluons and $b$-quarks. As expected from \figref{flavor_dep}, the performance is worse for jets initiated by light quarks. The \pileup versus hard-scatter jet discrimination performs best for hard-scatter jets initiated by $b$-quarks. The efficiency versus fake-rate curve for jets initiated by $c$-quarks is similar to that of gluon jets. 

The stability of the hard-scatter efficiencies as a function of \Npv is found to be independent of the flavour of the parton initiating the jet. 

To test the sample dependence of \JVT, the likelihood was also derived using a sample of $20 < \pT < 50\GeV$ jets in simulated $\Zboson(\to\mu\mu)$+jets events. The performance of the \JVT-based \pileup jet suppression (evaluated in terms of fake rate versus efficiency curves) was found to be independent of the sample from which the likelihood is derived.

The hard-scatter jet efficiency for \JVT in data was measured using the tag-and-probe method in $\Z(\to\mu\mu)$+jets events, using a procedure similar to that described in \secref{suppression:JVF} (see also Ref.~\cite{CMS-PAS-JME-13-005}). Using the leading jet recoiling against the \Z boson as a probe, a signal region for hard-scatter jets is defined as the back-to-back region specified by the requirement $|\Delta \phi (\Z, {\rm jet})| > 2.8$.  The \pileup contamination in the signal region is estimated from a \pileup control region, based on the assumption that the $|\Delta \phi (\Z, {\rm jet})|$ distribution is flat for \pileup jets.

\begin{figure*}
  \centering
  \subfigure[]{
      \includegraphics[width= 0.45\textwidth]{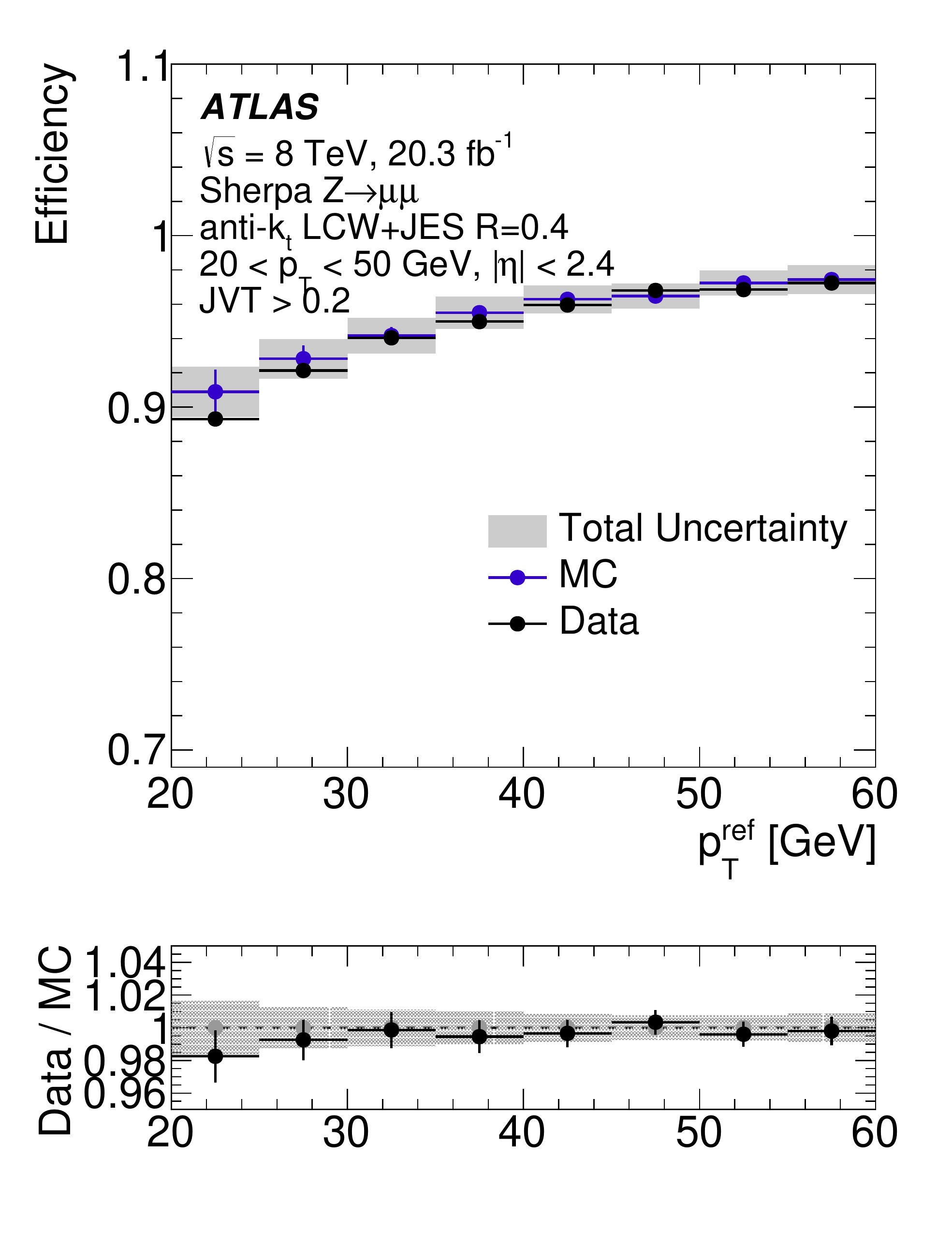}
      \label{fig:JVTCal_0p2}
  }
  \subfigure[]{
      \includegraphics[width= 0.45\textwidth]{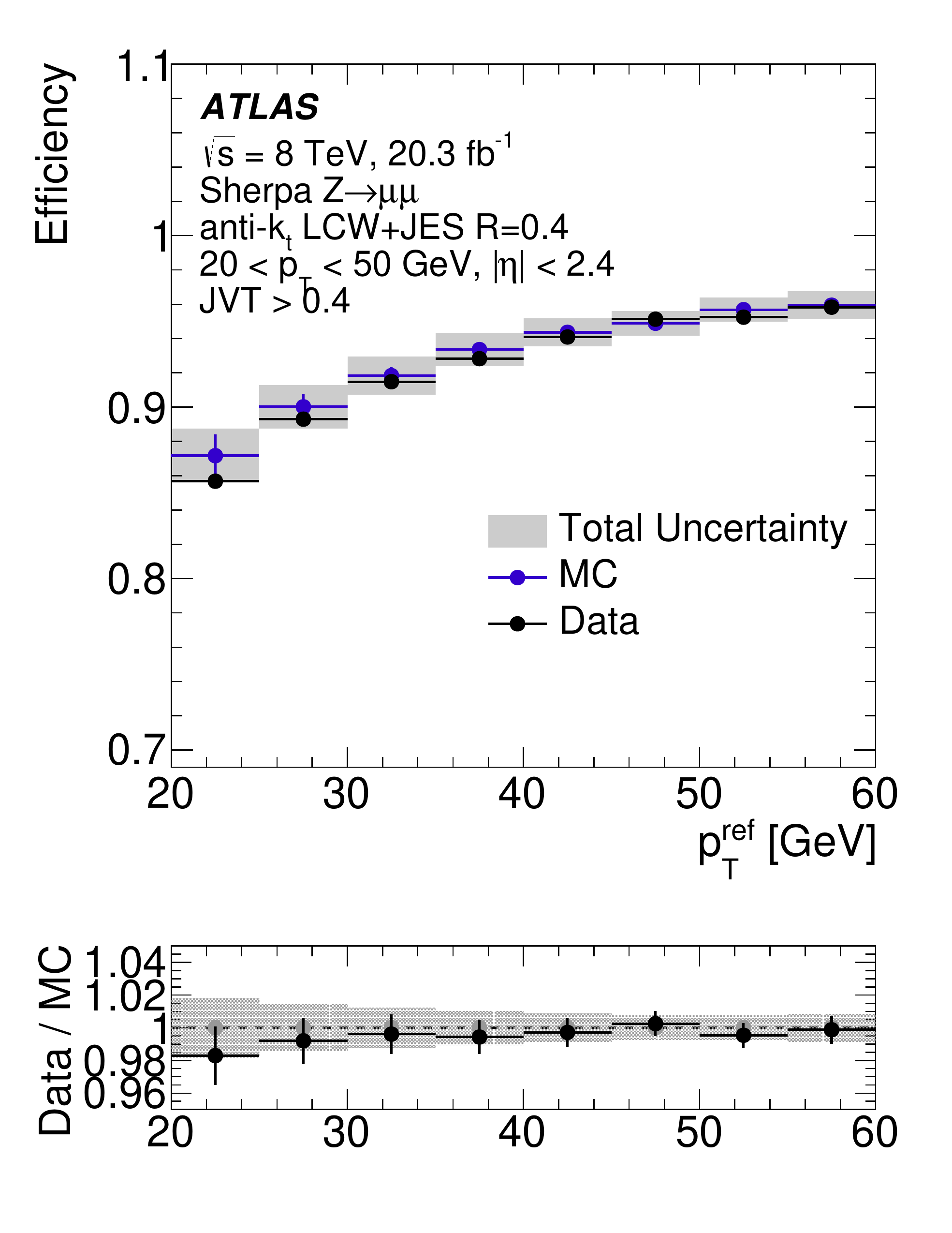}
      \label{fig:JVTCal_0p4}
  }\\
  \subfigure[]{
      \includegraphics[width= 0.45\textwidth]{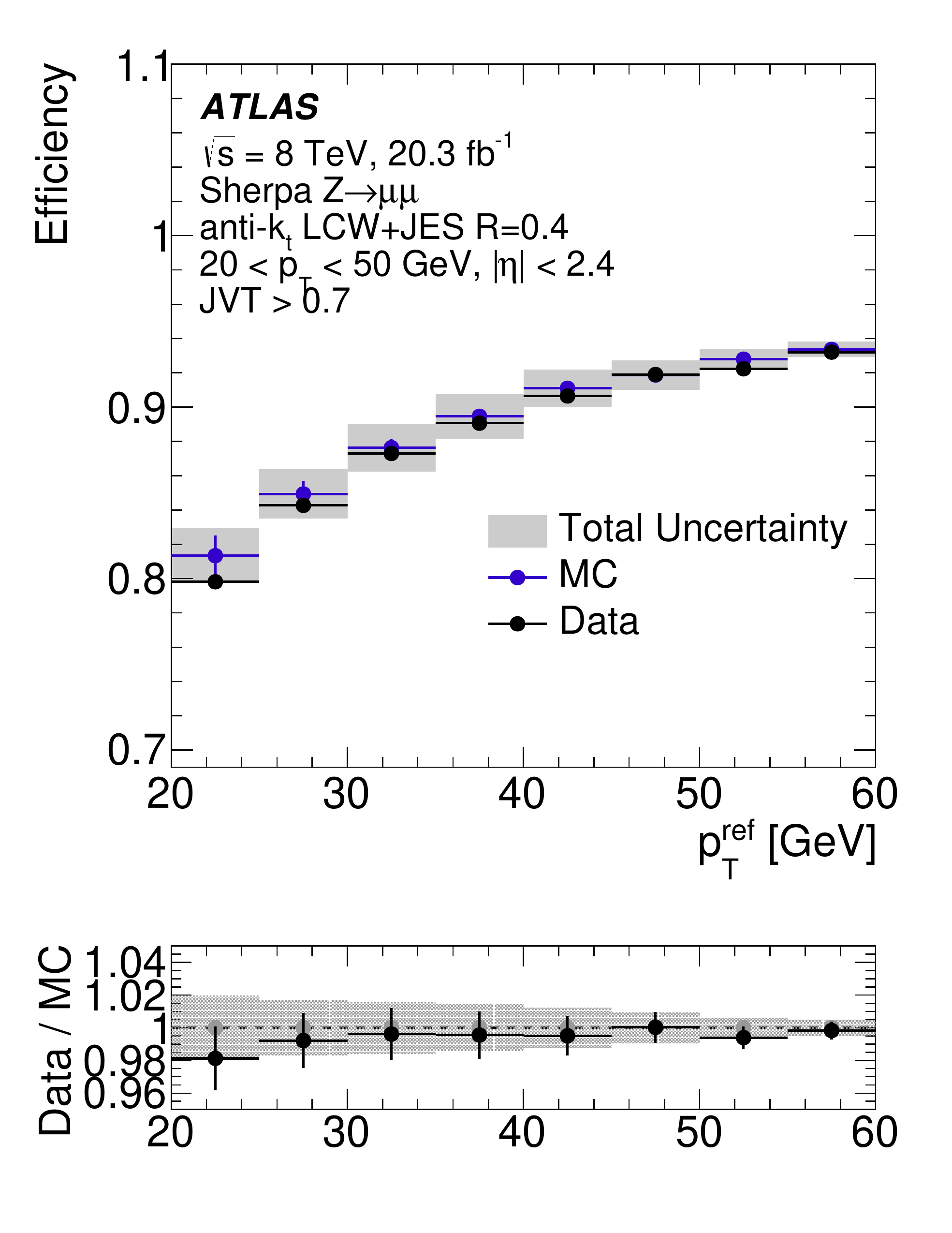}
  \label{fig:JVTCal_0p7}
  }
  \caption{Efficiency measured in $\Zboson(\to\mu\mu)$+jets events as a function of \ptref in data and MC simulation for  \subref{fig:JVTCal_0p2} $\JVT>0.2$, \subref{fig:JVTCal_0p4} $\JVT>0.4$ and \subref{fig:JVTCal_0p7} $\JVT>0.7$, where $\ptref = \Zpt$. The bottom panels of each figure show the ratio of efficiencies measured in data and MC simulation.}
  \label{fig:JVTCalibration}
\end{figure*}

\figrange{JVTCal_0p2}{JVTCal_0p7} show the jet efficiencies for  minimum \JVT requirements of 0.2, 0.4 and 0.7 respectively, measured in bins of $\ptref = \Zpt$.

Good agreement is observed between data and simulation, although there is a very slight tendency for the MC simulation to predict an efficiency higher than that found in data at low \ptref, but this difference is within the statistical uncertainty. The simulation-to-data scale factors are consistent with unity within the uncertainties. The grey band reflects the total uncertainty on the efficiency in simulation, adding the statistical and the systematic uncertainties in quadrature. The systematic uncertainty is determined by accounting for the differences in efficiency observed between the \Sherpa and the \PowPythia~8 $\Z(\to\mu\mu)$+jets MC samples, and by the mismodelling in the simulation of the $\Delta \phi(\Z, {\rm jet})$ shape for hard-scatter jets. The total uncertainty ranges from 2\% to 1\% when \ptref varies from 20 to $60\GeV$.


\section{Jet grooming for \pileup mitigation and suppression}
\label{sec:grooming}

The algorithmic removal of substructures within a jet based on kinematic criteria is generally referred to as \textit{jet grooming}. Several types of jet grooming have been explored in ATLAS~\cite{Aad:2013gja} for their ability to reduce the backgrounds to boosted-object selection while maintaining high efficiencies for signal processes. Improving the individual jet mass resolution and mitigating the effects of \pileup are critical issues in these studies. Indeed, these measures of performance are used as some of the primary figures of merit in determining a subset of groomed-jet algorithms on which to focus for physics analysis in \ATLAS.

Previous studies show that trimming and filtering both significantly reduce the dependence of the jet mass on \pileup~\cite{Aad:2013gja}. As described in \secref{data_event_selection}, trimming removes subjets with $\pti/\ptjet < \fcut$, where \pti\ is the transverse momentum of the $i^{th}$ subjet, and $\fcut=0.05$. Filtering proceeds similarly, but utilises the relative masses of the subjets defined and the original jet. For at least one of the configurations tested, trimming and filtering are both able to approximately eliminate the \pileup dependence of the jet mass. Building upon the success of calorimeter-based grooming methods and track-based \pileup suppression of small-radius jets, a new, track-based, grooming technique can be designed by vetoing individual subjets of \largeR jets that are associated with \pileup interactions using tracking information. 

The implementations of track-based grooming in ATLAS have so far focused on \cJVF and so-called \textit{jet cleansing} methods~\cite{Krohn:2013lba}. The algorithm which uses \cJVF relies on the application of \cJVF to the individual subjets of \largeR jets wherein tracks matched to each subjet are used in the calculation of \cJVF for that subjet. In particular, track-based trimming is implemented by replacing the \fcut criterion with a requirement on the \cJVF of subjets.  
\begin{figure}[!htbp]
  \centering
  \includegraphics[width= 0.60\textwidth]{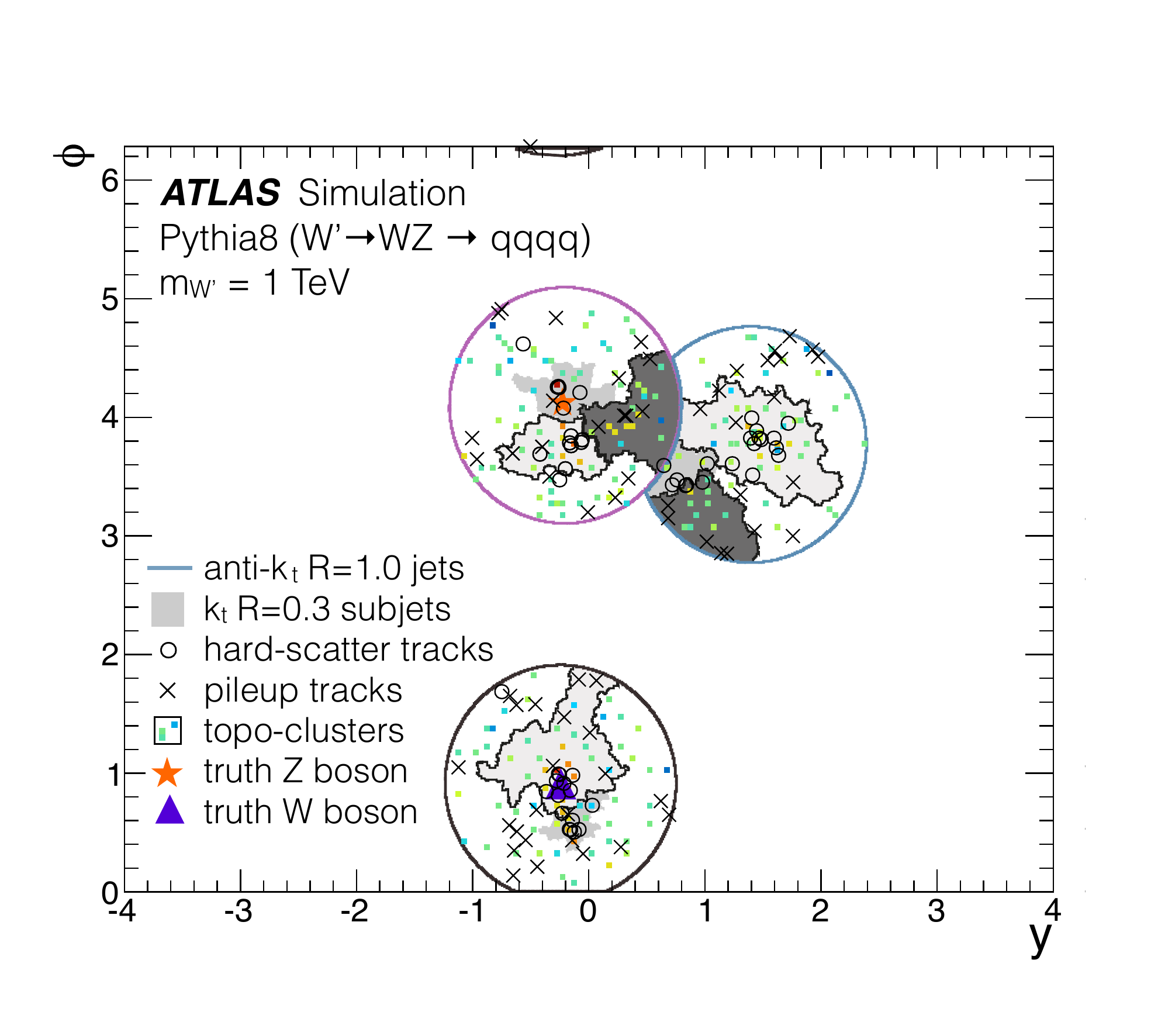}
  \caption{Rapidity--$\phi$ view of a simulated event of a $\Wboson^{\prime}$ boson with a mass of 1~\TeV\ decaying to a \Wboson boson and a \Zboson boson, both of which decay to jet pairs.}
\label{fig:EventDisp}
\end{figure}

The concept of track-based grooming can be illustrated in an event display. \Figref{EventDisp} shows both calorimeter and tracking information in the rapidity ($y$) versus azimuthal angle ($\phi$) plane of a simulated event where a $\Wboson^{\prime}$ boson with a mass of 1~TeV decays to a \Wboson boson and a \Zboson boson, which decay hadronically. The orange star and blue triangle indicate the $y$--$\phi$ positions of the generated \Wboson and \Zboson bosons. The large circles represent the active area boundaries of the \antikt\ $R=1.0$ jets, built from topological clusters. In the following, these jets are referred to as ungroomed jets. The clusters are represented by small solid squares with colours ranging from blue to red encoding low to high transverse energies. The grey regions indicate the active areas of the \kt\ $R=0.3$ subjets reconstructed from the constituents of the ungroomed jets. Only subjets with a \pt of at least $5\%$ of the ungroomed jet \pt are shown. Tracks associated with the jet and originating from the hard-scatter vertex (black open circles) or from \pileup vertices (black crosses) are also indicated. The violet ungroomed jet (with $\phi \approx 4.1$ and $y \approx -0.2$) has a \pt of $446 \GeV$ and is matched in $\Delta R$ to the truth \Zboson boson. While all three subjets have active areas overlapping with the $y$--$\phi$ positions of \pileup tracks, only two subjets have associated hard-scatter tracks. The invariant mass reconstructed from the two subjets with hard-scatter tracks is $89 \GeV$ and the one from all three subjets is $119 \GeV$. This event display shows that tracking information can provide information complementary to calorimeter-based trimming. Track-assisted trimming would allow the rejection of the third subjet, which is likely to originate from \pileup, while keeping the two subjets from the \Zboson boson.
 
\begin{figure}[!htbp]
  \centering
  \subfigure[]{
      \includegraphics[width= 0.49\textwidth]{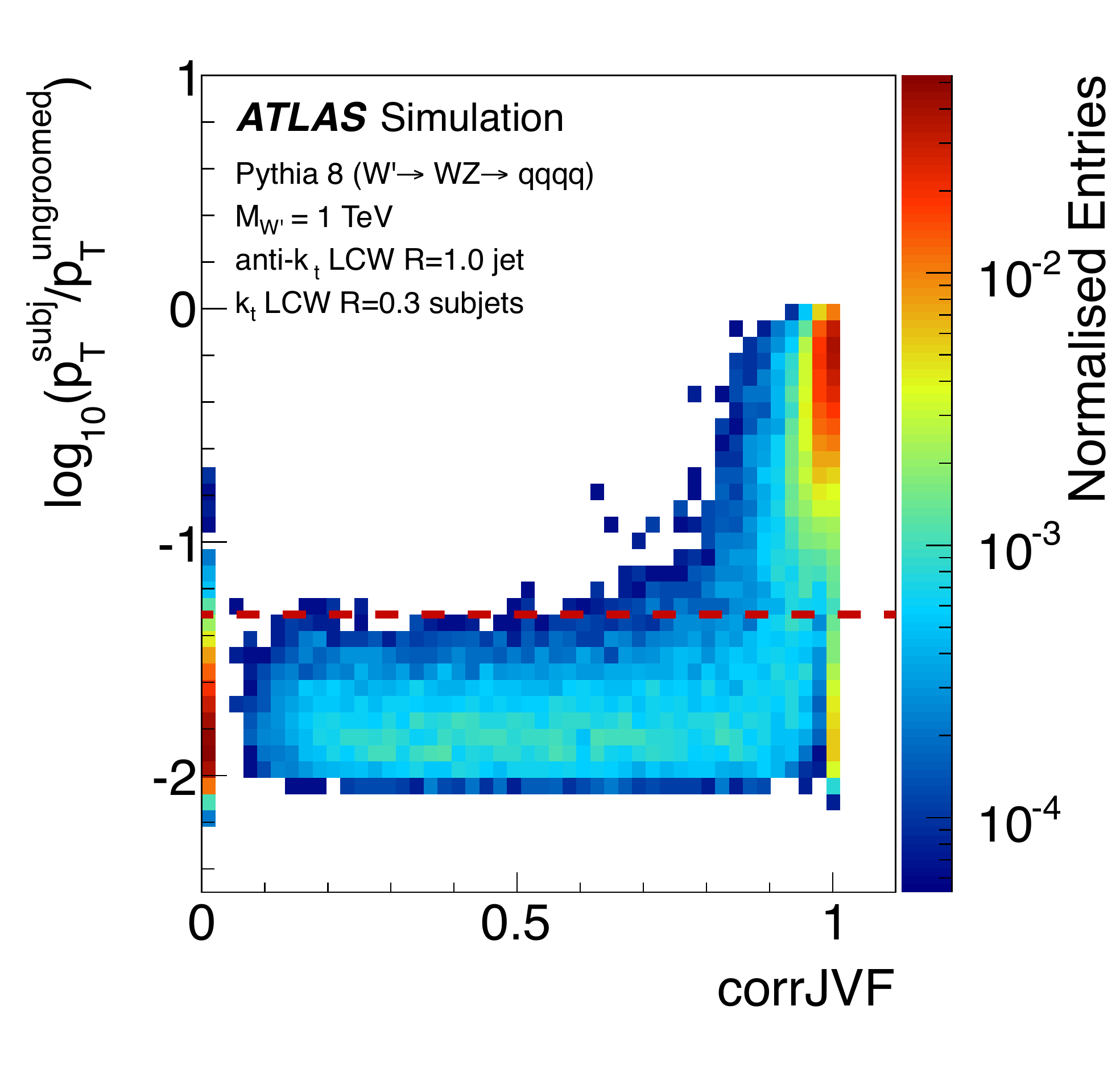}
    \label{fig:TBG:cJVFvsfCut}
  }
  \subfigure[]{
      \includegraphics[width=0.48\textwidth]{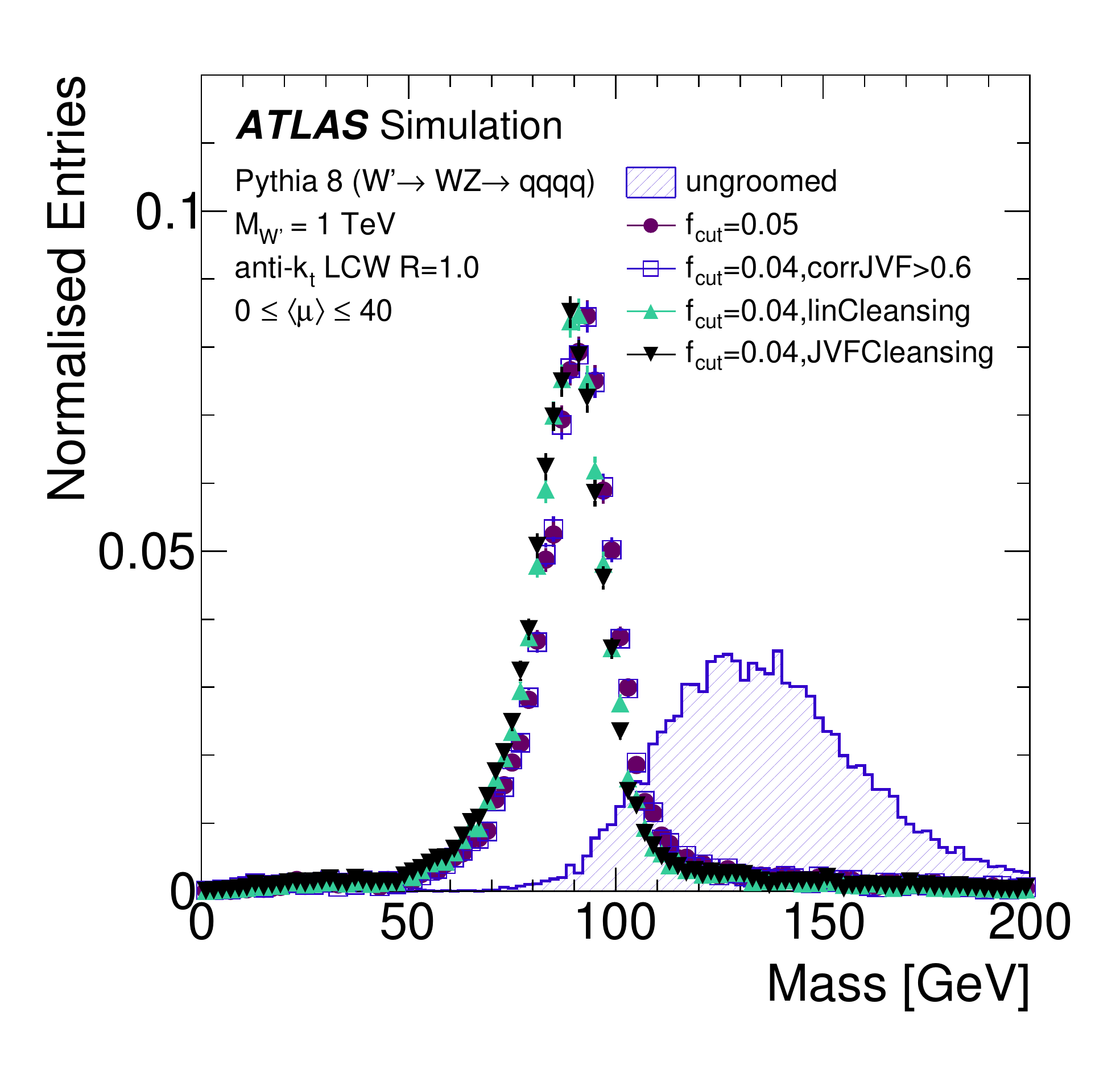}
    \label{fig:TBG:trackgrooming}
  }
 \caption{\subref{fig:TBG:cJVFvsfCut} Correlation of subjet \pt fraction, defined as the ratio of the subjet \pt to the ungroomed jet \pt, and subjet \cJVF for \antikt\ $R=1.0$ jets with $\pt>300 \GeV$ and $|\eta| <1.5$. The dotted line indicates the standard calorimeter-based trimming $\fcut$ of $5\%$. 
\subref{fig:TBG:trackgrooming} Distribution of jet mass for calorimeter- and track-based trimming configurations and jet cleansing. The default trimmed jet mass (purple filled circles) with $\fcut = 0.05$ is compared to calorimeter-based trimming with $(\fcut = 0.04)$ and $\cJVF>0.6$ (blue open squares), linear cleansing (green upward triangles) and \JVF cleansing (black downward triangles).  The dashed blue histogram is the mass distribution for ungroomed jets, with no \pileup subtraction applied. 
}
\label{fig:TBG}
\end{figure}

\Figref{TBG:cJVFvsfCut} shows the ratio of the subjet \pt to the ungroomed jet \pt 
on a logarithmic scale as a function of the subjet \cJVF in simulated $W^{\prime} \rightarrow W Z \rightarrow qqqq$ events. The subjet \pt is defined as the four-momentum sum of the constituents contained within the \kt jet that forms the subjet. The ungroomed jet \pt is defined as the \pt of the \largeR jet from which the subjets are then constructed. The two-dimensional distribution of this ratio is normalised to unit area. Approximately 4\% of subjets have no associated tracks ($\cJVF=-1$) and are omitted. Most subjets with significant \pt ratio also have large \cJVF, indicating that most of their charged \pt comes from the hard-scatter vertex. A large fraction of subjets with a low \pt  ratio $< 5\%$ $(\log_{10}[\pt^{\rm sub}/\pt^{\rm ungroomed}] < -1.3)$ and a few subjets with a significant \pt ratio, however, have small \cJVF values. Most such subjets are consistent with \pileup and are excluded by the track-based jet grooming procedure. Similarly, subjets with small \pt ratio and large \cJVF that would be removed by calorimeter-based trimming, are kept by the track-based trimming algorithm.

For the 2012 \pileup conditions with an average of about 21 \pp interactions per bunch crossing, an \fcut of 4\% in addition to the requirement of $\cJVF>0.6$ is found to be the optimal combination of trimming and \cJVF selection. A grooming configuration based solely on \cJVF (with no \fcut applied) is found to have a slightly worse mass resolution than trimming alone. 

The jet cleansing approach is implemented in two forms: \JVF cleansing and linear cleansing. In \JVF cleansing, the four-momentum of each subjet is scaled by the subjet \JVF, aiming to approximate the momentum of the subjet arising from neutral and charged particles from the hard-scatter vertex only. In linear cleansing, the subjet four-momentum from the hard-scatter vertex is approximated by scaling the reconstructed four-momentum based on the assumption that the ratio of charged to charged plus neutral \pileup \pt contributing to a subjet is 0.55~\cite{Krohn:2013lba}. Each \fcut used in these procedures is chosen to optimise the mass resolution. For 2012 \pileup conditions, the application of track-assisted grooming achieves a similar mass resolution to that of calorimeter-based trimming. 

\Figref{TBG:trackgrooming} compares the performance of the track-assisted grooming procedure with the variants of the jet cleansing concept. All of the methods studied show significant improvements in the jet mass resolution and stability with respect to \pileup. For the \pileup conditions expected during the LHC \RunOne and \RunTwo, studies using simulated data do not exhibit any significant difference between \cJVF and jet cleansing. However, for higher luminosity conditions expected beyond 2023 at the LHC the track-based grooming provides an alternative to calorimeter-only approaches.  Another advantage of track-based grooming over standard calorimeter-based grooming is that no \pt threshold is involved in the removal of subjets.  This means that in the limit of no \pileup, track-based grooming does not remove any signal, unlike for example trimming, which always rejects subjets that fall below the \fcut threshold.


\clearpage
\section{Conclusions}
\label{sec:Conclusions}

The presence of multiple simultaneous proton-proton interactions, known as \pileup, is one of the major challenges for jet reconstruction at the LHC.  ATLAS has implemented three main techniques to mitigate the effect of \pileup on jets and jet measurements: topological clustering, event-by-event jet \pileup subtraction, and jet-vertex tagging \pileup jet suppression. The first method reduces the impact of \pileup at the constituent level, whereas the latter two techniques are applied after jet reconstruction, to correct jet kinematic and substructure variables and to suppress jets induced by \pileup.  

Topological clustering partially suppresses the formation of calorimeter clusters from \pileup activity, before jet reconstruction, by considering \pileup as a form of noise in the definition of the energy significance thresholds for cells.  This acts as a constituent-level \pileup suppression and significantly reduces the contribution of \pileup to the inputs to jet reconstruction.  For the 20.3\invfb of \pp data collected at \sqseight, topological clustering used a fixed \pileup noise corresponding to $\avgmu=30$.  Fluctuations of \pileup due to different luminosity conditions as well as global and local event \pileup fluctuations can still affect the seeding and growth of clusters and require jet-level \pileup corrections. 

The jet-area \pileup subtraction method reduces global fluctuations of \pileup in jets and allows the correction of jet shape variables.  This method uses a direct measure of the \pileup activity in the calorimeter on an event-by-event basis (the \pt density $\rho$ in $\eta$--$\phi$ space), as well as a jet-by-jet measure of \pileup susceptibility (the jet area, \jetarea). A residual \pileup correction is necessary to fully accommodate the impact of \pileup on jet \pT\ as the high-occupancy jet core contributes some extra sensitivity to both in-time and out-of-time \pileup, and the effects of \pileup on forward jets are not fully described by the median \pT\ density as calculated from topological clusters in the central calorimeter.  The combination of $\rho\times\jetarea$ subtraction and residual correction results in a stable jet \pT\ response across the full range of \pileup conditions in 2012, and it significantly reduces the degradation in jet \pT\ resolution associated with fluctuations in \pileup. It also reduces the dependence of jet multiplicity on \pileup, shifting the majority of \pileup jets below the minimum jet-\pT\ threshold. For $\pT > 50 \GeV$, the \pileup subtraction procedure alone is sufficient to make the jet multiplicity stable as a function of \avgmu and \Npv within statistical errors.  Systematic uncertainties are typically below $2 \%$ for $R = 0.4$ \antikt\ jets with $\pT > 40 \GeV$ in the central region of the calorimeters; they reach up to $6 \%$ at low \pT\ and higher $\eta$.  Jet-area subtraction also significantly reduces the \pileup dependence of jet shape variables. 

Jet-vertex tagging enables the identification and rejection of \pileup jets arising from local fluctuations of \pileup within events, as well as from QCD jets originating from \pileup vertices.  A fundamental feature of the \JVT algorithm, introduced in this paper, is that its discrimination power is independent of the \pileup conditions, leading to hard-scatter jet selection efficiencies that are stable within $1\%$ for up to 35 interactions per bunch crossing.  This \pileup stability implies that there is no need to re-optimise selections based on \JVT as \pileup conditions change, even as the LHC transitions to \sqsthirt and 25~ns bunch spacing in \RunTwo.  The JVT selection efficiency, measured as a function of \pT\ and $\eta$, is found to agree between data and simulation within 1--2 $\%$.

Jet-vertex tagging has also been extended to the case of \largeR jets by introducing a track-based trimming algorithm at the subjet level.  The new track-based grooming achieves performance similar to that of calorimeter-based trimming, while using complementary tracking information.  In particular, track-based grooming does not need to rely on subjet \pT\ selection cuts as in the case of standard grooming methods.  Jet cleansing has also been studied and results in performance similar to that of all other methods considered. 

The suite of algorithms discussed in this paper has provided the capability to manage and suppress \pileup, both at the level already observed during the LHC \RunOne and at the level expected for \RunTwo. The impact on jet reconstruction and measurement is significant and has thus improved many aspects of the physics program in ATLAS.  \Pileup corrections and suppression algorithms both for small and large radius jets have enhanced the discovery potential of the ATLAS experiment and improved the precision for Standard Model measurements.  New and more advanced methods that are presented in this paper and developed towards the end of the LHC \RunOne will provide additional handles and improved precision for \pileup mitigation for the upcoming LHC \RunTwo and the future high-luminosity upgrades.

\section*{Acknowledgements}
\label{sec:Ack}

We thank CERN for the very successful operation of the LHC, as well as the
support staff from our institutions without whom ATLAS could not be
operated efficiently.

We acknowledge the support of ANPCyT, Argentina; YerPhI, Armenia; ARC, Australia; BMWFW and FWF, Austria; ANAS, Azerbaijan; SSTC, Belarus; CNPq and FAPESP, Brazil; NSERC, NRC and CFI, Canada; CERN; CONICYT, Chile; CAS, MOST and NSFC, China; COLCIENCIAS, Colombia; MSMT CR, MPO CR and VSC CR, Czech Republic; DNRF and DNSRC, Denmark; IN2P3-CNRS, CEA-DSM/IRFU, France; GNSF, Georgia; BMBF, HGF, and MPG, Germany; GSRT, Greece; RGC, Hong Kong SAR, China; ISF, I-CORE and Benoziyo Center, Israel; INFN, Italy; MEXT and JSPS, Japan; CNRST, Morocco; FOM and NWO, Netherlands; RCN, Norway; MNiSW and NCN, Poland; FCT, Portugal; MNE/IFA, Romania; MES of Russia and NRC KI, Russian Federation; JINR; MESTD, Serbia; MSSR, Slovakia; ARRS and MIZ\v{S}, Slovenia; DST/NRF, South Africa; MINECO, Spain; SRC and Wallenberg Foundation, Sweden; SERI, SNSF and Cantons of Bern and Geneva, Switzerland; MOST, Taiwan; TAEK, Turkey; STFC, United Kingdom; DOE and NSF, United States of America. In addition, individual groups and members have received support from BCKDF, the Canada Council, CANARIE, CRC, Compute Canada, FQRNT, and the Ontario Innovation Trust, Canada; EPLANET, ERC, FP7, Horizon 2020 and Marie Sk{\l}odowska-Curie Actions, European Union; Investissements d'Avenir Labex and Idex, ANR, R{\'e}gion Auvergne and Fondation Partager le Savoir, France; DFG and AvH Foundation, Germany; Herakleitos, Thales and Aristeia programmes co-financed by EU-ESF and the Greek NSRF; BSF, GIF and Minerva, Israel; BRF, Norway; the Royal Society and Leverhulme Trust, United Kingdom.

The crucial computing support from all WLCG partners is acknowledged
gratefully, in particular from CERN and the ATLAS Tier-1 facilities at
TRIUMF (Canada), NDGF (Denmark, Norway, Sweden), CC-IN2P3 (France),
KIT/GridKA (Germany), INFN-CNAF (Italy), NL-T1 (Netherlands), PIC (Spain),
ASGC (Taiwan), RAL (UK) and BNL (USA) and in the Tier-2 facilities
worldwide.


\clearpage


\bibliographystyle{atlasBibStyleWithTitle}
\bibliography{ATLAS}

\providecommand{\href}[2]{#2}\begingroup\raggedright\begin{thebibliography}{10}

\bibitem{Bruning:2004ej}
O.~S. Br{\"u}ning {et~al.}, {\em {LHC Design Report Vol.1: The LHC Main Ring}},
  \href{https://cdsweb.cern.ch/record/782076}{CERN-2004-003-V-1} (2004).
\url{https://cdsweb.cern.ch/record/782076}.

\bibitem{Cacciari:2007fd}
M.~Cacciari and G.~P. Salam, {\em {Pileup subtraction using jet areas}},
  \href{http://dx.doi.org/10.1016/j.physletb.2007.09.077}{Phys. Lett.
  {\bfseries B 659} (2008) 119},
\href{http://arxiv.org/abs/0707.1378}{{\ttfamily arXiv:0707.1378 [hep-ph]}}.

\bibitem{jespaper2010}
{ATLAS} Collaboration, {\em {Jet energy measurement with the ATLAS detector in
  proton-proton collisions at $\sqrt{s}=7$ TeV}},
  \href{http://dx.doi.org/10.1140/epjc/s10052-013-2304-2}{Eur. Phys. J.
  {\bfseries C 73} (2013) 2304},
\href{http://arxiv.org/abs/1112.6426}{{\ttfamily arXiv:1112.6426 [hep-ex]}}.

\bibitem{jespaper2011}
{ATLAS} Collaboration, {\em {Jet energy measurement and its systematic
  uncertainty in proton-proton collisions at $\sqrt{s}=7$ TeV with the ATLAS
  detector}}, \href{http://dx.doi.org/10.1140/epjc/s10052-014-3190-y}{Eur.
  Phys. J. {\bfseries C 75} (2015) 17},
\href{http://arxiv.org/abs/1406.0076}{{\ttfamily arXiv:1406.0076 [hep-ex]}}.

\bibitem{Aad:2014tca}
{ATLAS} Collaboration, {\em {Fiducial and differential cross sections of Higgs
  boson production measured in the four-lepton decay channel in $pp$ collisions
  at $\sqrt{s}$=8 TeV with the ATLAS detector}},
  \href{http://dx.doi.org/10.1016/j.physletb.2014.09.054}{Phys. Lett.
  {\bfseries B 738} (2014) 234},
\href{http://arxiv.org/abs/1408.3226}{{\ttfamily arXiv:1408.3226 [hep-ex]}}.

\bibitem{Aad:2014eva}
{ATLAS} Collaboration, {\em {Measurements of Higgs boson production and
  couplings in the four-lepton channel in pp collisions at center-of-mass
  energies of 7 and 8 TeV with the ATLAS detector}},
  \href{http://dx.doi.org/10.1103/PhysRevD.91.012006}{Phys. Rev. {\bfseries D
  91} (2015) 012006},
\href{http://arxiv.org/abs/1408.5191}{{\ttfamily arXiv:1408.5191 [hep-ex]}}.

\bibitem{Aad:2014eha}
{ATLAS} Collaboration, {\em {Measurement of Higgs boson production in the
  diphoton decay channel in pp collisions at center-of-mass energies of 7 and 8
  TeV with the ATLAS detector}},
  \href{http://dx.doi.org/10.1103/PhysRevD.90.112015}{Phys. Rev. {\bfseries D
  90} (2014) 112015},
\href{http://arxiv.org/abs/1408.7084}{{\ttfamily arXiv:1408.7084 [hep-ex]}}.

\bibitem{Aad:2014xzb}
{ATLAS} Collaboration, {\em {Search for the $b\bar{b}$ decay of the Standard
  Model Higgs boson in associated $(W/Z)H$ production with the ATLAS
  detector}}, \href{http://dx.doi.org/10.1007/JHEP01(2015)069}{JHEP {\bfseries
  1501} (2015) 069},
\href{http://arxiv.org/abs/1409.6212}{{\ttfamily arXiv:1409.6212 [hep-ex]}}.

\bibitem{ATLAS:2014aga}
{ATLAS Collaboration}, {\em {Observation and measurement of Higgs boson decays
  to WW$^*$ with the ATLAS detector}},
  \href{http://dx.doi.org/10.1103/PhysRevD.92.012006}{Phys. Rev. {\bfseries D
  92} (2015) 012006},
\href{http://arxiv.org/abs/1412.2641}{{\ttfamily arXiv:1412.2641 [hep-ex]}}.

\bibitem{detPaper}
{ATLAS Collaboration}, {\em {The ATLAS Experiment at the CERN Large Hadron
  Collider}},
\href{http://dx.doi.org/10.1088/1748-0221/3/08/S08003}{JINST {\bfseries 3}
  (2008) S08003}.

\bibitem{PerfWithData2010}
{ATLAS Collaboration}, {\em {Performance of the ATLAS Detector using First
  Collision Data}}, \href{http://dx.doi.org/10.1007/JHEP09(2010)056}{JHEP
  {\bfseries 09} (2010) 056},
\href{http://arxiv.org/abs/1005.5254}{{\ttfamily arXiv:1005.5254 [hep-ex]}}.

\bibitem{Aad:2010ac}
{ATLAS Collaboration}, {\em {Charged-particle multiplicities in pp interactions
  measured with the ATLAS detector at the LHC}},
  \href{http://dx.doi.org/10.1088/1367-2630/13/5/053033}{New J. Phys.
  {\bfseries 13} (2011) 053033},
\href{http://arxiv.org/abs/1012.5104}{{\ttfamily arXiv:1012.5104 [hep-ex]}}.

\bibitem{ATLAS-CONF-2012-042}
{ATLAS Collaboration}, {\em {Performance of the ATLAS Inner Detector Track and
  Vertex Reconstruction in the High Pile-Up LHC Environment}},
  \href{http://cdsweb.cern.ch/record/1435196}{ATLAS-CONF-2012-042} (2012).
  \url{http://cdsweb.cern.ch/record/1435196}.

\bibitem{ATLAS-CONF-2014-047}
{ATLAS Collaboration}, {\em {Alignment of the ATLAS Inner Detector and its
  Performance in 2012}},
  \href{https://cdsweb.cern.ch/record/1741021}{ATLAS-CONF-2014-047} (2014).
  \url{https://cdsweb.cern.ch/record/1741021}.

\bibitem{ATL-PHYS-PUB-2014-002}
{ATLAS Collaboration}, {\em {A measurement of single hadron response using data
  at $\sqrt{s}=8$~TeV with the ATLAS detector}},
  \href{http://cds.cern.ch/record/1668961}{ATL-PHYS-PUB-2014-002} (2014).
  \url{http://cds.cern.ch/record/1668961}.

\bibitem{Buchanan:2008zzc}
N.~Buchanan {et~al.}, {\em {ATLAS liquid argon calorimeter front end
  electronics}},
\href{http://dx.doi.org/10.1088/1748-0221/3/09/P09003}{JINST {\bfseries 3}
  (2008) P09003}.

\bibitem{Abreu:2010zzc}
H.~Abreu {et~al.}, {\em {Performance of the electronic readout of the ATLAS
  liquid argon calorimeters}},
\href{http://dx.doi.org/10.1088/1748-0221/5/09/P09003}{JINST {\bfseries 5}
  (2010) P09003}.

\bibitem{ClelandStern1994}
W.~E. Cleland and E.~G. Stern, {\em {Signal processing considerations for
  liquid ionization calorimeters in a high rate environment}},
  \href{http://dx.doi.org/10.1016/0168-9002(94)91332-3}{Nucl. Instrum. Meth.
  {\bfseries A 338} (1994) 467}.

\bibitem{TileReadiness}
{ATLAS Collaboration}, {\em {Readiness of the {ATLAS} Tile calorimeter for
  {LHC} collisions}},
  \href{http://dx.doi.org/10.1140/epjc/s10052-010-1508-y}{Eur. Phys. J.
  {\bfseries C 70} (2010) 1193},
  \href{http://arxiv.org/abs/1007.5423}{{\ttfamily arXiv:1007.5423
  [physics.ins-det]}}.

\bibitem{Aad:2014una}
{ATLAS} Collaboration, {\em {Monitoring and data quality assessment of the
  ATLAS liquid argon calorimeter}},
  \href{http://dx.doi.org/10.1088/1748-0221/9/07/P07024}{JINST {\bfseries 9}
  (2014) P07024},
\href{http://arxiv.org/abs/1405.3768}{{\ttfamily arXiv:1405.3768 [hep-ex]}}.

\bibitem{Aad:2013ucp}
{ATLAS} Collaboration, {\em {Improved luminosity determination in $pp$
  collisions at $\sqrt{s} = 7$~TeV using the ATLAS detector at the LHC}},
  \href{http://dx.doi.org/10.1140/epjc/s10052-013-2518-3}{Eur. Phys. J.
  {\bfseries C 73} (2013) 2518},
\href{http://arxiv.org/abs/1302.4393}{{\ttfamily arXiv:1302.4393 [hep-ex]}}.

\bibitem{Aad:2013zwa}
{ATLAS} Collaboration, {\em {Characterisation and mitigation of beam-induced
  backgrounds observed in the ATLAS detector during the 2011 proton-proton
  run}}, \href{http://dx.doi.org/10.1088/1748-0221/8/07/P07004}{JINST
  {\bfseries 8} (2013) P07004},
\href{http://arxiv.org/abs/1303.0223}{{\ttfamily arXiv:1303.0223 [hep-ex]}}.

\bibitem{Cacciari200657}
M.~Cacciari and G.~P. Salam, {\em {Dispelling the $N^{3}$ myth for the $k_t$
  jet-finder}}, \href{http://dx.doi.org/10.1016/j.physletb.2006.08.037}{Phys.
  Lett. {\bfseries B 641} (2006) 57},
  \href{http://arxiv.org/abs/hep-ph/0512210}{{\ttfamily arXiv:hep-ph/0512210
  [hep-ph]}}.

\bibitem{Cacciari:2008gp}
M.~Cacciari, G.~P. Salam, and G.~Soyez, {\em {The anti-$k_{t}$ jet clustering
  algorithm}}, \href{http://dx.doi.org/10.1088/1126-6708/2008/04/063}{JHEP
  {\bfseries 04} (2008) 063},
\href{http://arxiv.org/abs/0802.1189}{{\ttfamily arXiv:0802.1189 [hep-ph]}}.

\bibitem{TopoClusters}
W.~Lampl {et~al.}, {\em {Calorimeter clustering algorithms: description and
  performance}},
  \href{http://cdsweb.cern.ch/record/1099735}{ATL-LARG-PUB-2008-002} (2008).
  \url{http://cdsweb.cern.ch/record/1099735}.

\bibitem{Aad:2014vka}
{ATLAS} Collaboration, {\em {Search for dark matter in events with a $Z$ boson
  and missing transverse momentum in $pp$ collisions at $\sqrt{s}$=8 TeV with
  the ATLAS detector}},
  \href{http://dx.doi.org/10.1103/PhysRevD.90.012004}{Phys. Rev. {\bfseries D
  90} (2014) 012004},
\href{http://arxiv.org/abs/1404.0051}{{\ttfamily arXiv:1404.0051 [hep-ex]}}.

\bibitem{Aad:2014xra}
{ATLAS} Collaboration, {\em {Search for $W' \rightarrow tb \rightarrow qqbb$
  decays in $pp$ collisions at $\sqrt{s}$ = 8 TeV with the ATLAS detector}},
  \href{http://dx.doi.org/10.1140/epjc/s10052-015-3372-2}{Eur. Phys. J.
  {\bfseries C 75} (2015) 165},
\href{http://arxiv.org/abs/1408.0886}{{\ttfamily arXiv:1408.0886 [hep-ex]}}.

\bibitem{Krohn2010}
D.~Krohn, J.~Thaler, and L.-T. Wang, {\em {Jet trimming}},
  \href{http://dx.doi.org/10.1007/JHEP02(2010)084}{JHEP {\bfseries 2010} (2010)
  20}, \href{http://arxiv.org/abs/0912.1342}{{\ttfamily arXiv:0912.1342
  [hep-ph]}}.

\bibitem{Ellis1993}
S.~D. Ellis and D.~E. Soper, {\em {Successive combination jet algorithm for
  hadron collisions}}, \href{http://dx.doi.org/10.1103/PhysRevD.48.3160}{Phys.
  Rev. {\bfseries D 48} (1993) 3160},
\href{http://arxiv.org/abs/hep-ph/9305266}{{\ttfamily arXiv:hep-ph/9305266
  [hep-ph]}}.

\bibitem{Catani1993}
S.~Catani, Y.~L. Dokshitzer, M.~Seymour, and B.~Webber, {\em {Longitudinally
  invariant $K_t$ clustering algorithms for hadron hadron collisions}},
\href{http://dx.doi.org/10.1016/0550-3213(93)90166-M}{Nucl. Phys. {\bfseries B
  406} (1993) 187}.

\bibitem{Aad:2013gja}
{ATLAS Collaboration}, {\em {Performance of jet substructure techniques for
  large-$R$ jets in proton-proton collisions at $\sqrt{s}$ = 7 TeV using the
  ATLAS detector}}, \href{http://dx.doi.org/10.1007/JHEP09(2013)076}{JHEP
  {\bfseries 1309} (2013) 076},
\href{http://arxiv.org/abs/1306.4945}{{\ttfamily arXiv:1306.4945 [hep-ex]}}.

\bibitem{pythia8}
T.~Sj{\"o}strand, S.~Mrenna, and P.~Z. Skands, {\em {A Brief Introduction to
  PYTHIA 8.1}}, \href{http://dx.doi.org/10.1016/j.cpc.2008.01.036}{Comput.
  Phys. Commun. {\bfseries 178} (2008) 852--867},
\href{http://arxiv.org/abs/0710.3820}{{\ttfamily arXiv:0710.3820 [hep-ph]}}.

\bibitem{MC12AU2}
{ATLAS Collaboration}, {\em {Summary of ATLAS Pythia 8 tunes}},
  \href{http://cdsweb.cern.ch/record/1474107}{ATL-PHYS-PUB-2012-003} (2012).
\url{http://cdsweb.cern.ch/record/1474107}.

\bibitem{Lai:2010vv}
H.-L. Lai {et~al.}, {\em {New parton distributions for collider physics}},
  \href{http://dx.doi.org/10.1103/PhysRevD.82.074024}{Phys. Rev. {\bfseries
  D82} (2010) 074024},
\href{http://arxiv.org/abs/1007.2241}{{\ttfamily arXiv:1007.2241 [hep-ph]}}.

\bibitem{Herwigpp}
M.~B{\"a}hr {et~al.}, {\em Herwig++ physics and manual},
  \href{http://dx.doi.org/10.1140/epjc/s10052-008-0798-9}{Eur. Phys. J.
  {\bfseries C 58} (2008) 639--707},
  \href{http://arxiv.org/abs/0803.0883}{{\ttfamily arXiv:0803.0883 [hep-ph]}}.

\bibitem{PDF-CTEQ}
J.~Pumplin {et~al.}, {\em {New generation of parton distributions with
  uncertainties from global QCD analysis}},
  \href{http://dx.doi.org/10.1088/1126-6708/2002/07/012}{JHEP {\bfseries 07}
  (2002) 012},
\href{http://arxiv.org/abs/0201195}{{\ttfamily arXiv:0201195 [hep-ph]}}.

\bibitem{Gieseke:2012ft}
S.~Gieseke, C.~Rohr, and A.~Siodmok, {\em {Colour reconnections in Herwig++}},
  \href{http://dx.doi.org/10.1140/epjc/s10052-012-2225-5}{Eur. Phys. J.
  {\bfseries C 72} (2012) 2225},
\href{http://arxiv.org/abs/1206.0041}{{\ttfamily arXiv:1206.0041 [hep-ph]}}.

\bibitem{Geant4}
{GEANT4} Collaboration, S.~Agostinelli {et~al.}, {\em {GEANT4: A simulation
  toolkit}},
\href{http://dx.doi.org/10.1016/S0168-9002(03)01368-8}{Nucl. Instrum. Meth.
  {\bfseries A 506} (2003) 250}.

\bibitem{simulation}
{ATLAS Collaboration}, {\em {The {ATLAS} simulation infrastructure}},
  \href{http://dx.doi.org/10.1140/epjc/s10052-010-1429-9}{Eur. Phys. J.
  {\bfseries C 70} (2010) 823},
  \href{http://arxiv.org/abs/1005.4568}{{\ttfamily arXiv:1005.4568
  [physics.ins-det]}}.

\bibitem{Alioli:2010xd}
S.~Alioli, P.~Nason, C.~Oleari, and E.~Re, {\em {A general framework for
  implementing NLO calculations in shower Monte Carlo programs: the POWHEG
  BOX}}, \href{http://dx.doi.org/10.1007/JHEP06(2010)043}{JHEP {\bfseries 1006}
  (2010) 043}, \href{http://arxiv.org/abs/1002.2581}{{\ttfamily arXiv:1002.2581
  [hep-ph]}}.

\bibitem{Frixione:2007vw}
S.~Frixione, P.~Nason, and C.~Oleari, {\em {Matching NLO QCD computations with
  Parton Shower simulations: the POWHEG method}},
  \href{http://dx.doi.org/10.1088/1126-6708/2007/11/070}{JHEP {\bfseries 0711}
  (2007) 070}, \href{http://arxiv.org/abs/0709.2092}{{\ttfamily arXiv:0709.2092
  [hep-ph]}}.

\bibitem{Nason:2004rx}
P.~Nason, {\em {A New method for combining NLO QCD with shower Monte Carlo
  algorithms}}, \href{http://dx.doi.org/10.1088/1126-6708/2004/11/040}{JHEP
  {\bfseries 0411} (2004) 040},
  \href{http://arxiv.org/abs/hep-ph/0409146}{{\ttfamily arXiv:hep-ph/0409146
  [hep-ph]}}.

\bibitem{Gleisberg:2008ta}
T.~Gleisberg {et~al.}, {\em {Event generation with SHERPA 1.1}},
  \href{http://dx.doi.org/10.1088/1126-6708/2009/02/007}{JHEP {\bfseries 02}
  (2009) 007},
\href{http://arxiv.org/abs/0811.4622}{{\ttfamily arXiv:0811.4622 [hep-ph]}}.

\bibitem{Golonka:2005pn}
P.~Golonka and Z.~Was, {\em {PHOTOS Monte Carlo: A Precision tool for QED
  corrections in $Z$ and $W$ decays}},
  \href{http://dx.doi.org/10.1140/epjc/s2005-02396-4}{Eur. Phys. J. {\bfseries
  C45} (2006) 97--107},
\href{http://arxiv.org/abs/hep-ph/0506026}{{\ttfamily arXiv:hep-ph/0506026
  [hep-ph]}}.

\bibitem{PDF-MSTW2008}
G.~Watt and R.~Thorne, {\em {Study of Monte Carlo approach to experimental
  uncertainty propagation with MSTW 2008 PDFs}},
  \href{http://dx.doi.org/10.1007/JHEP08(2012)052}{JHEP {\bfseries 1208} (2012)
  052},
\href{http://arxiv.org/abs/1205.4024}{{\ttfamily arXiv:1205.4024 [hep-ph]}}.

\bibitem{Aad:2011dr}
{ATLAS Collaboration}, {\em {Luminosity determination in p-p collisions at
  $\sqrt{s}=7$~TeV using the {ATLAS} detector at the LHC}},
  \href{http://dx.doi.org/10.1140/epjc/s10052-011-1630-5}{Eur. Phys. J.
  {\bfseries C 71} (2011) 1630},
  \href{http://arxiv.org/abs/1101.2185}{{\ttfamily arXiv:1101.2185 [hep-ex]}}.

\bibitem{Chatrchyan:2011ds}
{CMS} Collaboration, {\em {Determination of Jet Energy Calibration and
  Transverse Momentum Resolution in CMS}},
  \href{http://dx.doi.org/10.1088/1748-0221/6/11/P11002}{JINST {\bfseries 6}
  (2011) P11002},
\href{http://arxiv.org/abs/1107.4277}{{\ttfamily arXiv:1107.4277
  [physics.ins-det]}}.

\bibitem{Berta:2014eza}
P.~Berta, M.~Spousta, D.~W. Miller, and R.~Leitner, {\em {Particle-level pileup
  subtraction for jets and jet shapes}},
  \href{http://dx.doi.org/10.1007/JHEP06(2014)092}{JHEP {\bfseries 06} (2014)
  092},
\href{http://arxiv.org/abs/1403.3108}{{\ttfamily arXiv:1403.3108 [hep-ex]}}.

\bibitem{Bertolini:2014bba}
D.~Bertolini, P.~Harris, M.~Low, and N.~Tran, {\em {Pileup Per Particle
  Identification}}, \href{http://dx.doi.org/10.1007/JHEP10(2014)059}{JHEP
  {\bfseries 10} (2014) 59},
\href{http://arxiv.org/abs/1407.6013}{{\ttfamily arXiv:1407.6013 [hep-ph]}}.

\bibitem{Cacciari2008}
M.~Cacciari and G.~P. Salam, {\em {The catchment area of jets}},
  \href{http://dx.doi.org/10.1088/1126-6708/2008/04/005}{JHEP {\bfseries 04}
  (2008) 005}, \href{http://arxiv.org/abs/0802.1188}{{\ttfamily arXiv:0802.1188
  [hep-ph]}}.

\bibitem{Soyez2012}
G.~Soyez, G.~P. Salam, J.~Kim, S.~Dutta, and M.~Cacciari, {\em {Pileup
  subtraction for jet shapes}},
  \href{http://dx.doi.org/10.1103/PhysRevLett.110.162001}{Phys. Rev. Lett.
  {\bfseries 110} (2013) 162001},
\href{http://arxiv.org/abs/1211.2811}{{\ttfamily arXiv:1211.2811 [hep-ph]}}.

\bibitem{Thaler:2010tr}
J.~Thaler and K.~Van~Tilburg, {\em {Identifying Boosted Objects with
  $N$-subjettiness}}, \href{http://dx.doi.org/10.1007/JHEP03(2011)015}{JHEP
  {\bfseries 1103} (2011) 015},
  \href{http://arxiv.org/abs/1011.2268}{{\ttfamily arXiv:1011.2268 [hep-ph]}}.

\bibitem{Thaler:2011gf}
J.~Thaler and K.~Van~Tilburg, {\em {Maximizing Boosted Top Identification by
  Minimizing N-subjettiness}},
  \href{http://dx.doi.org/10.1007/JHEP02(2012)093}{JHEP {\bfseries 1202} (2012)
  093},
\href{http://arxiv.org/abs/1108.2701}{{\ttfamily arXiv:1108.2701 [hep-ph]}}.

\bibitem{Thaler2008}
J.~Thaler and L.-T. Wang, {\em {Strategies to Identify Boosted Tops}},
  \href{http://dx.doi.org/10.1088/1126-6708/2008/07/092}{JHEP {\bfseries 07}
  (2008) 092},
\href{http://arxiv.org/abs/0806.0023}{{\ttfamily arXiv:0806.0023 [hep-ph]}}.

\bibitem{BOOST2011}
A.~Altheimer {et~al.}, {\em {Jet Substructure at the Tevatron and LHC: New
  results, new tools, new benchmarks (BOOST 2011 Working Group Report)}},
  \href{http://dx.doi.org/10.1088/0954-3899/39/6/063001}{J. Phys. {\bfseries
  G39} (2012) 063001},
\href{http://arxiv.org/abs/1201.0008}{{\ttfamily arXiv:1201.0008 [hep-ph]}}.

\bibitem{RPVGluino7TeV}
{ATLAS Collaboration}, {\em {Search for pair production of massive particles
  decaying into three quarks with the ATLAS detector in $\sqrt{s}=7$ TeV $pp$
  collisions at the LHC}},
  \href{http://dx.doi.org/10.1007/JHEP12(2012)086}{JHEP {\bfseries 1212} (2012)
  086},
\href{http://arxiv.org/abs/1210.4813}{{\ttfamily arXiv:1210.4813 [hep-ex]}}.

\bibitem{BOOST2012}
A.~Altheimer {et~al.}, {\em {Boosted objects and jet substructure at the LHC
  (BOOST 2012 Working Group Report)}},
  \href{http://dx.doi.org/10.1140/epjc/s10052-014-2792-8}{Eur. Phys. J.
  {\bfseries C 74} (2014) 2792},
\href{http://arxiv.org/abs/1311.2708}{{\ttfamily arXiv:1311.2708 [hep-ex]}}.

\bibitem{Hocker:2007ht}
A.~Hoecker {et~al.}, {\em {TMVA: Toolkit for Multivariate Data Analysis}}, PoS
  {\bfseries ACAT} (2007) 040,
\href{http://arxiv.org/abs/physics/0703039}{{\ttfamily arXiv:physics/0703039}}.

\bibitem{Aad:2014gea}
{ATLAS} Collaboration, {\em {Light-quark and gluon jet discrimination in $pp$
  collisions at $\sqrt{s}=7$~TeV with the ATLAS detector}},
  \href{http://dx.doi.org/10.1140/epjc/s10052-014-3023-z}{Eur. Phys. J.
  {\bfseries C 74} (2014) 3023},
\href{http://arxiv.org/abs/1405.6583}{{\ttfamily arXiv:1405.6583 [hep-ex]}}.

\bibitem{CMS-PAS-JME-13-005}
{CMS} Collaboration, {\em {Pileup Jet Identification}},
  \href{http://cdsweb.cern.ch/record/1581583}{CMS-PAS-JME-13-005} (2013).
  \url{http://cdsweb.cern.ch/record/1581583}.

\bibitem{Krohn:2013lba}
D.~Krohn, M.~D. Schwartz, M.~Low, and L.-T. Wang, {\em {Jet Cleansing: Pileup
  Removal at High Luminosity}},
  \href{http://dx.doi.org/10.1103/PhysRevD.90.065020}{Phys. Rev. {\bfseries D
  90} (2014) 065020},
\href{http://arxiv.org/abs/1309.4777}{{\ttfamily arXiv:1309.4777 [hep-ph]}}.

\end{thebibliography}\endgroup




\clearpage

\clearpage
\onecolumn
\begin{flushleft}
{\Large The ATLAS Collaboration}

\bigskip

G.~Aad$^{\rm 85}$,
B.~Abbott$^{\rm 113}$,
J.~Abdallah$^{\rm 151}$,
O.~Abdinov$^{\rm 11}$,
R.~Aben$^{\rm 107}$,
M.~Abolins$^{\rm 90}$,
O.S.~AbouZeid$^{\rm 158}$,
H.~Abramowicz$^{\rm 153}$,
H.~Abreu$^{\rm 152}$,
R.~Abreu$^{\rm 116}$
% ...
\bigskip
\\
% ...
$^{11}$ Institute of Physics, Azerbaijan Academy of Sciences, Baku, Azerbaijan\\
% ...
$^{85}$ CPPM, Aix-Marseille Universit{\'e} and CNRS/IN2P3, Marseille, France\\
% ...
$^{90}$ Department of Physics and Astronomy, Michigan State University, East Lansing MI, United States of America\\
% ...
$^{107}$ Nikhef National Institute for Subatomic Physics and University of Amsterdam, Amsterdam, Netherlands\\
% ...
$^{113}$ Homer L. Dodge Department of Physics and Astronomy, University of Oklahoma, Norman OK, United States of America\\
% ...
$^{116}$ Center for High Energy Physics, University of Oregon, Eugene OR, United States of America\\
% ...
$^{151}$ Institute of Physics, Academia Sinica, Taipei, Taiwan\\
$^{152}$ Department of Physics, Technion: Israel Institute of Technology, Haifa, Israel\\
$^{153}$ Raymond and Beverly Sackler School of Physics and Astronomy, Tel Aviv University, Tel Aviv, Israel\\
% ...
$^{158}$ Department of Physics, University of Toronto, Toronto ON, Canada\\
% ...
\end{flushleft}

\end{document}